\documentclass[12pt]{article}
\pdfoutput=1 
\usepackage{amssymb, amsthm,amsmath,amsfonts}
\usepackage{graphicx}
\usepackage{xcolor, colortbl}
\usepackage{empheq}
\usepackage[pdftex, bookmarks=true,colorlinks,linkcolor=red,urlcolor=blue, citecolor=blue]{hyperref}
\usepackage{cite}
\usepackage{subcaption}
\usepackage{slashed}

\usepackage{amsthm}
\usepackage{array}
\usepackage{tocloft} 
\usepackage{soul}
\usepackage{float}
\usepackage{bm}

\makeatletter\@addtoreset{equation}{section}\makeatother

\newcommand{\dd}{\mathop{}\!\mathrm{d}}

\newcommand{\AdS}{\mathrm{AdS}}
\newcommand{\BTZ}{\mathrm{BTZ}}
\newcommand{\cO}{\mathcal O}

\newcommand{\dT}{\Delta T}
\newcommand{\dOme}{\Delta \Omega}
\newcommand{\dF}{\Delta \mathcal{F}}

\newcommand{\dpt}{\Delta p_t}

\newcommand{\Fstar}{\mathcal{F}_\star}
\newcommand{\FAdS}{\mathcal{F}_\AdS}

\newcommand{\Fhair}{\mathcal{F}_{\rm hairy}}

\newcommand{\calB}{\mathcal{B}_-}

\def\bebal{\begin{equation}\begin{aligned}}
\def\eeeal{\end{aligned}\end{equation}}

\newcommand{\preprint}[1]{\begin{table}[t]  
             \begin{flushright}               
             {#1}                             
             \end{flushright}                 
             \end{table}}                     
\renewcommand{\title}[1]{\vbox{\center\LARGE{#1}}\vspace{5mm}}
\renewcommand{\author}[1]{\vbox{\center#1}\vspace{5mm}}
\newcommand{\address}[1]{\vbox{\center\em#1}}

\newcommand{\be}{\begin{equation}}
\newcommand{\ee}{\end{equation}}
\newcommand{\bea}{\begin{eqnarray}}
\newcommand{\eea}{\end{eqnarray}}
\newcommand{\bse}{\begin{subequations}}
\newcommand{\ese}{\end{subequations}}
\newcommand{\beqa}{\begin{eqnarray}}
\newcommand{\eeqa}{\end{eqnarray}}
\newcommand{\beqar}{\begin{eqnarray*}}
\newcommand{\eeqar}{\end{eqnarray*}}
\newcommand{\bi}{\begin{itemize}}
\newcommand{\ei}{\end{itemize}}
\newcommand{\bn}{\begin{enumerate}}
\newcommand{\en}{\end{enumerate}}

\newcommand{\ba}{\begin{array}}
\newcommand{\ea}{\end{array}}
\newcommand{\bc}{\begin{center}}
\newcommand{\ec}{\end{center}}

\newcommand{\al}{\alpha}

\newcommand{\la}{\lambda}

\newcommand{\De}{\Delta}

\newcommand{\vp}{\varphi}

\definecolor{darkgreen}{rgb}{0,0.3,0}
\definecolor{darkblue}{rgb}{0,0,0.3}
\definecolor{darkred}{rgb}{0.7,0,0}

\usepackage{tikz}
\usetikzlibrary{arrows.meta,positioning,calc}

\def\lae{\mathrel{\mathop{\smash{\lower .5 ex \hbox{$\stackrel<\sim$}}}}}
\def\lae{\mathrel{\mathop{\smash{\lower .5 ex \hbox{$\stackrel>\sim$}}}}}

\def\doi#1#2{\href{http://doi.org/#1}{#2}}

\begin{document}

\unitlength = .8mm

\begin{titlepage}
\preprint{}

\date{\today}
\begin{center}

\vspace{-1cm}
\title{\boldmath 
Critical Kasner scaling from Cauchy-horizon collapse
\\near holographic phase transitions}
\vskip 0.2cm
{Ling-Long Gao$^{\,a}$}\footnote{Email: {\tt linglonggao@buaa.edu.cn}},
{Yan Liu$^{\,a}$}\footnote{Email: {\tt yanliu@buaa.edu.cn}} and
{Hong-Da Lyu$^{\,b, c}$}\footnote{Email: {\tt hongdalyu@buaa.edu.cn}} 

\address{${}^{a}$Department of Space Science, \\ and Peng Huanwu Collaborative Center for Research and Education, \\Beihang University, Beijing 100191, China}
\vspace{-0.2cm}
\address{${}^{b}$School of Physics, Southeast University, Nanjing 211189, China}
\vspace{-0.2cm}
\address{${}^{c}$Key Laboratory of Particle Physics and Particle Irradiation (MOE), \\Institute of Frontier and Interdisciplinary Science, \\Shandong University, Qingdao, Shandong 266237, China}
\end{center}
\vskip 0.3cm

\vspace{-0.5cm}
\abstract{
We study continuous scalar-hair bifurcations from stationary black holes with a nonextremal Cauchy horizon.  Although the exterior deformation vanishes at criticality, the critical zero mode generically develops a logarithmic branch at the reference Cauchy horizon, producing a nonuniform critical limit. We show that the coefficient of this logarithm in the  
scalar perturbation, together with local Cauchy-horizon data, determines the exponential collapse rate of the Einstein-Rosen bridge. 
If the post-collapse geometry evolves into a Kasner regime,  
the boundary order-parameter scaling  
$O\propto|\lambda-\lambda_c|^\beta$ further implies 
$
 1-p_t\propto|\lambda-\lambda_c|^{2\beta}. 
$  Crucially, zero-mode matching fixes both the critical exponents and the associated amplitudes, without any fit to the nonlinear interior. 
We illustrate this mechanism in the transition from rotating BTZ to hairy rotating black holes. For scalarized RN-AdS$_4$ black branes, we quantitatively verify the predicted Kasner scaling using fully nonlinear numerical solutions. At low temperature, two charged models with identical linear critical data instead approach different infrared endpoints, yielding respectively power-law and inverse-logarithmic Kasner scaling. Thus interiors near a finite-temperature bifurcation are controlled by zero-mode transmission, whereas low temperature scaling is governed by the nonlinear infrared completion.
}

\vfill
\end{titlepage}

\begingroup 
\hypersetup{linkcolor=black}
\setlength{\cftbeforesecskip}{8pt}
 \setlength{\cftbeforesubsecskip}{3pt}
\tableofcontents
\endgroup

\newpage
\section{Introduction 
}
\label{sec:introduction}

Critical phenomena in AdS black-hole systems are usually characterized by quantities defined at the asymptotic boundary or in the exterior region,
such as an order parameter, a susceptibility, and the associated critical
exponents \cite{Zaanen:2015oix, Hartnoll:2016apf}. Yet the corresponding spacetime extends beyond the event
horizon. This raises a natural question: to
what extent can the critical data of an exterior phase transition be transmitted across the event horizon and control the scaling behavior of the black-hole interior?

For black holes with a single horizon, previous studies of
Einstein--scalar models have shown that the critical scaling of an
exterior order parameter can be inherited by the Kasner geometry near the spacelike singularity \cite{Liu:2021hap,Gao:2026tck}. In such geometries, the radial
evolution connects the event horizon directly to the
kinetic-dominated interior, allowing the critical perturbation to be
followed continuously into the Kasner regime. 

Black holes with an inner horizon present a qualitatively different
setting. Nonextremal rotating BTZ black holes and RN-AdS black
branes possess an outer event horizon and an inner Cauchy horizon.
Under generic perturbations, the strong blueshift near the Cauchy
horizon can drive mass inflation, placing its fate at the center of
strong cosmic censorship
\cite{Poisson:1989zz,Dafermos:2003vim, Shahbazi-Moghaddam:2024emr}. 
Stationary scalar deformations of charged AdS black holes are known
to eliminate the regular Cauchy horizon, producing an
Einstein--Rosen bridge collapse followed by a spacelike Kasner
singularity 
\cite{Hartnoll:2020rwq,Hartnoll:2020fhc}, 
see also e.g. \cite{Dias:2021afz, Sword:2021pfm, Mansoori:2021wxf,  Liu:2022rsy, Carballo:2024hem,  Arean:2024pzo, Zhang:2025hkb, Zhao:2025odj}. 
More recently, critical interior scaling has been numerically observed in both
dynamical collapse and stationary scalarized black holes
\cite{Shao:2025fki, Shao:2025apr, Li:2025mhy}. 


What remains missing is how the linear mode generating the exterior phase transition is converted into the nonlinear destruction of the Cauchy horizon, and whether this conversion determines not only critical exponents but also the corresponding interior
amplitudes. The primary purpose of this work is to study this missing analytic connection
and to determine precisely which parts of the interior scaling are fixed
by the critical zero mode. 

To study quantitatively how the exterior critical mode controls the inner-horizon geometry, we consider a stationary hairy branch
parametrized by a small amplitude $\epsilon$, which can be identified with the expectation value of the dual boundary operator, with 
\begin{equation}
 \phi(z)=\epsilon\Phi(z)+\cdots,
 \qquad
 |\epsilon|\simeq A_\epsilon|\lambda-\lambda_c|^\beta .
 \label{eq:intro1}
\end{equation}
Here $\Phi$ is the critical zero mode regular at the event horizon, and $\lambda$ denotes a boundary control parameter which parametrizes a one-dimensional
path crossing the critical surface at $\lambda=\lambda_c$, with all
other ensemble variables held fixed. Near
a simple Cauchy horizon of the reference solution, this mode generically
takes the form
\begin{equation}
 \Phi(z)
 =
 A_-+B_-\log|z-z_-|+\cdots .
\end{equation}
Although the scalar itself remains parametrically small as
$\epsilon\rightarrow0$, its radial derivative is enhanced as
$\phi'\simeq\epsilon B_-/(z-z_-)$.  The 
combination
$
 {\cal B}_-\equiv\epsilon B_-
$
provides the physical seed transmitted from the exterior
zero mode into the nonlinear Cauchy-horizon region.

The critical limit is therefore nonuniform.  As $\epsilon\to0$, the
hairy solutions approach the reference geometry outside its Cauchy horizon, while the logarithmically enhanced
gradient becomes nonperturbative in a shrinking region near that
horizon.  The rapid transition region shrinks to zero radial width, but
the geometry beyond the reference Cauchy horizon does not converge to
the corresponding analytic continuation of the two-horizon solution. 

By resolving this singular region through a matched boundary-layer
analysis, we find
\begin{equation}
\label{eq:deltaer-gammaer}
 \Gamma_{\rm ER}\propto{\cal B}_-^{-2},
 \qquad
 |\phi'|_{\rm aft}\propto|{\cal B}_-|^{-1},
\end{equation}
where 
$\Gamma_{\rm ER}
$ is 
the 
exponential 
decay rate of the
Einstein-Rosen bridge collapse, and
$|\phi'|_{\rm aft}$ is the scalar gradient after ER collapse.  If the
solution after ER collapse subsequently enters a scalar-kinetic-dominated
overlap region, the first Kasner epoch satisfies
\begin{equation}
 |v_1|
 \propto
 |\lambda-\lambda_c|^{-\beta},
 \qquad
 1-p_t^{(1)}
 \propto
 |\lambda-\lambda_c|^{2\beta},
 \qquad
 p_x^{(1)}
 \propto
 |\lambda-\lambda_c|^{2\beta},
 \qquad
 |p_\phi^{(1)}|
 \propto
|\lambda-\lambda_c|^\beta .
 \label{eq:intro_Kasner_scaling}
\end{equation}
Here $v_1$ is the scalar velocity in the first Kasner epoch, $p_t^{(1)}$ is the Kasner exponent associated with the direction inherited from the exterior time coordinate,
$p_x^{(1)}$ denotes each equivalent transverse Kasner exponent, and
$p_\phi^{(1)}$ is the scalar Kasner exponent. 
A central point is that these results go beyond matching critical exponents. Once the zero mode is normalized relative to the exterior order parameter, its transmission coefficient $B_-$, together with the local Cauchy-horizon data, fixes the leading coefficients of the ER collapse and, when a kinetic-dominated overlap exists, those of the first Kasner epoch. Converting these coefficients into functions of the boundary control parameter then requires only the boundary relation in \eqref{eq:intro1}; 
no independent parameter is fitted to the nonlinear interior. 
Later Kasner transitions and the final approach to the singularity can involve additional dynamical epochs, including behavior related to
cosmological billiards
\cite{Henneaux:2022ijt}. 

{We test the analytic matching and scaling 
laws in two complementary 
examples using fully nonlinear numerical solutions.  For the
bifurcation from rotating BTZ to hairy rotating black holes, the
critical zero mode is obtained analytically and yields an explicit
nonzero logarithmic coefficient $B_-$.  For scalarized
RN--AdS$_4$ black branes, a sign-definite integral identity instead
establishes $B_-\neq0$.  In both cases, the transmitted zero-mode
data predict the Kasner scalings \eqref{eq:intro_Kasner_scaling}, as well as the amplitudes, which can be compared directly with the nonlinear solutions.}

Finally, we also 
discuss the interior scaling law in 
the low-temperature limit which is different from the above near-bifurcation mechanism and is instead
controlled by the infrared geometry of the full nonlinear solution.
Two charged models with identical linear critical data approach
different zero-temperature geometries: a finite-scalar
AdS$_2\times\mathbb{R}^2$ throat gives
$
 1-p_t\propto T^{2\delta_{\rm IR}},
$
 where $\delta_{\rm IR}$ is the positive irrelevant exponent of the 
 throat, while a running-scalar endpoint gives
$
 1-p_t
 \simeq
 2/\log(1/T) 
$. 

The remainder of the paper is organized as follows. In 
section~\ref{sec:general_mechanism}, we study the matching from the
logarithmic channel 
through the Einstein-Rosen bridge collapse
to the first Kasner epoch. Section~\ref{sec:rotating_BTZ} applies this
framework to the continuous bifurcation from rotating BTZ to a hairy 
branch.
Section~\ref{sec:charged_RN} studies scalarized RN--AdS$_4$ black
branes. 
In section~\ref{sec:low_temperature_scaling}, we show how distinct infrared
geometries produce power-law and inverse-logarithmic Kasner scaling at
low temperature. In Section~\ref{sec:condis}, we discuss our study and the related open questions. 
Technical derivations are deferred to the 
Appendices \ref{app:ER_normal_form}-\ref{app:zero_temperature}.

\section{From critical zero mode to interior scaling}
\label{sec:general_mechanism}

Before applying the construction to the rotating and charged examples, we derive the matching procedure summarized in the Introduction.  We first continue the event-horizon-regular critical 
zero mode to the reference Cauchy horizon and extract its logarithmic transmission coefficient.  The corresponding physical amplitude
$\mathcal{B}_{-}$ is then matched to the finite canonical radial momentum $\Pi_{\star}$ entering the nonlinear regime containing Einstein--Rosen
 bridge collapse and its post-collapse scalar gradient.  When a scalar-kinetic overlap exists, the same post-collapse data fix the first Kasner epoch. The
subsequent interior evolution is generally model dependent.  The radial organization is summarized in Fig.~\ref{fig:cartoon}.

\usetikzlibrary{arrows.meta,patterns,decorations.pathmorphing}
\definecolor{ink}{RGB}{30,33,36}
\definecolor{muted}{RGB}{107,113,118}
\definecolor{accent}{RGB}{0,103,165}

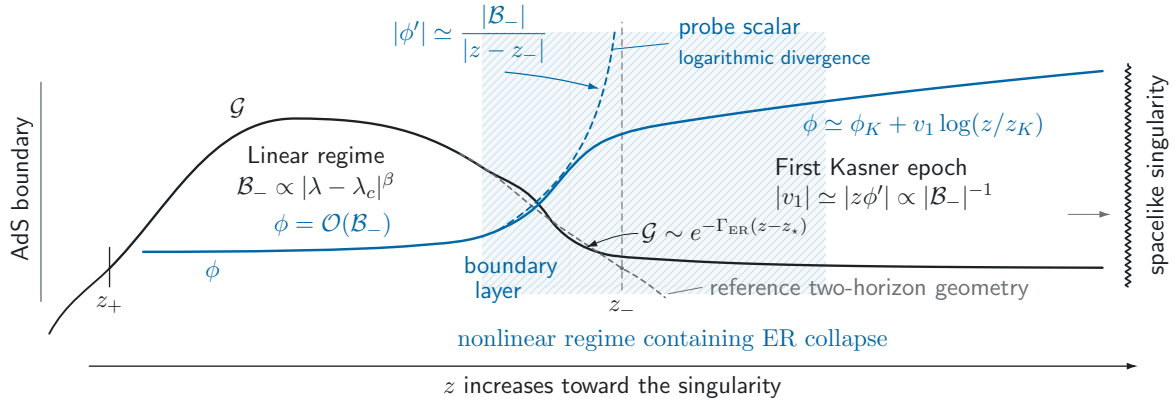
\begin{figure}[h!]
\begin{center}
\scalebox{0.84}{
\begin{tikzpicture}[
  x=1cm,y=1cm,line cap=round,line join=round,
  >={Latex[length=1.8mm,width=1.1mm]},
  every node/.style={text=ink,inner sep=1.5pt},
  region/.style={font=\sffamily\fontsize{11}{13}\selectfont},
  label/.style={font=\sffamily\fontsize{11}{13}\selectfont},
  small/.style={font=\sffamily\fontsize{11}{13}\selectfont},
  mathlabel/.style={font=\fontsize{11}{13}\selectfont},
  leader/.style={draw=muted,line width=.45pt}
]
\path[use as bounding box] (-.05,-.35) rectangle (18.10,5.75);
\def\erLeft{7.15}
\def\erRight{12.55}
\def\zMinus{9.35}

\fill[accent!4] (\erLeft,1.20) rectangle (\erRight,5.32);
\fill[pattern=north east lines,pattern color=accent!19]
  (\erLeft,1.20) rectangle (\erRight,5.32);

\draw[muted,line width=.6pt] (.2,1.08)--(.2,4.53);
\node[small,rotate=90] at (-.1,2.85) {AdS boundary};
\draw[ink,line width=.85pt]
  (.33,0.57) .. controls (.59,1.06) and (.96,1.28) .. (1.28,1.60);
\draw[ink,line width=.6pt] (1.28,1.31)--(1.28,1.93);
\node[mathlabel,anchor=north] at (1.28,1.21) {$z_+$};

\draw[ink,line width=1.05pt]
  (1.28,1.60)
  .. controls (2.20,2.52) and (2.87,3.96) .. (4.20,3.96)
  .. controls (5.51,3.96) and (6.26,3.77) .. (7.15,3.19)
  .. controls (7.55,{3.19-.40*.58/.89}) and (7.83,2.94) .. (8.20,2.40)
  .. controls (8.385,2.13) and (8.68,1.92) .. (9.10,1.82)
  .. controls (9.31,1.77) and (9.80,{1.73+.35*5*.13/10.15}) .. (10.15,1.73)
  -- plot[domain=10.15:16.92,samples=110]
    (\x,{1.60+.13*(\x/10.15)^(-5)});
\node[mathlabel,anchor=south] at (3.30,3.92) {$\mathcal G$};

\draw[muted,dash pattern=on 2.3pt off 2pt,line width=.75pt]
  (6.80,3.40)
  .. controls (7.16,3.23) and (7.44,2.99) .. (7.75,2.73)
  .. controls (8.33,2.24) and (8.98,1.86) .. (\zMinus,1.60)
  .. controls (9.61,1.47) and (9.81,1.32) .. (10.03,1.14);
\draw[muted,dash pattern=on 3pt off 2.2pt,line width=.5pt]
  (\zMinus,1.26)--(\zMinus,5.48);
\fill[muted] (\zMinus,1.60) circle (.9pt);
\node[small,anchor=north] at (\zMinus,1.15) {$z_-$};
\node[small,anchor=west,text=muted] at (10.67,1.28)
  {reference two-horizon geometry};
\draw[leader] (10.61,1.30)--(10.08,1.20);

\draw[accent,line width=1.10pt]
  (1.81,1.86)
  .. controls (3.33,1.86) and (5.20,1.88) .. (6.10,1.96)
  .. controls (6.55,2.00) and (6.96,2.00) .. (7.40,2.20)
  .. controls (7.62,2.30) and (7.90,2.43) .. (8.20,2.78)
  .. controls (8.50,3.13) and (8.74,3.46) .. (9.10,3.62)
  .. controls (9.46,3.78) and (9.80,{3.89-.35*1.60/10.15}) .. (10.15,3.89)
  -- plot[domain=10.15:16.92,samples=110]
    (\x,{3.89+1.60*ln(\x/10.15)});
\node[mathlabel,text=accent,anchor=north] at (2.90,1.79) {$\phi$};
\node[mathlabel,text=accent,anchor=south] at (4.80,2.03)
  {$\phi=\mathcal{O}(\mathcal B_-)$};

\draw[accent,dash pattern=on 3pt off 2.2pt,line width=.85pt]
  plot[domain=7.40:9.24,samples=150]
    (\x,{2.20+1.25*ln(2/(9.40-\x))});
\node[label,text=accent,anchor=west,align=left] at (10.18,5.16)
  {probe scalar\\[-1pt]{\fontsize{9}{11}\selectfont logarithmic divergence}};
\draw[accent,line width=.45pt] (10.08,5.14)--(9.20,5.05);

\node[region] at (4.53,3.35) {Linear regime};
\node[mathlabel] at (4.53,2.89)
  {$\mathcal B_-\propto|\lambda-\lambda_c|^\beta$};
\node[mathlabel,text=accent,anchor=south] at (6.96,4.78)
  {$|\phi'|\simeq\dfrac{|\mathcal B_-|}{|z-z_-|}$};
\draw[accent,line width=.55pt,->]
  (7.51,4.69) .. controls (8.12,4.63) and (8.45,4.56) .. (9.00,4.43);

\node[region,anchor=west] at (11.70,3.16) {First Kasner epoch};
\node[mathlabel,text=accent,anchor=north west] at (12.14,4.11)  {$\phi\simeq\phi_K+v_1\log(z/z_K)$};
\node[mathlabel,anchor=west] at (11.70,2.73)
  {$|v_1|\simeq|z\phi'|\propto |\mathcal B_-|^{-1}$};

\node[label,text=accent,anchor=south] at (7.60,1.40) {boundary};
\node[label,text=accent,anchor=south] at (7.40,1.00) {layer};
\node[mathlabel,anchor=west] at (9.6,2.2)
  {$\mathcal G\sim e^{-\Gamma_{\rm ER}(z-z_\star)}$};
\draw[leader,draw=ink,->] (9.6,2.15)
  .. controls (9.4,2.15) and (9.2,2.1) .. (8.78,1.93);

\node[mathlabel,text=accent,anchor=north] at (10.15,.7)
 {nonlinear regime containing ER collapse};

\draw[ink,line width=.8pt,decorate,
  decoration={zigzag,segment length=3pt,amplitude=1.2pt}]
  (17.32,1.31)--(17.32,4.82);
\node[small,rotate=90,anchor=north] at (17.55,3.07) {spacelike singularity};
\draw[leader,->] (16.38,2.45)--(17.05,2.45);

\draw[ink,line width=.6pt,->] (.96,.06)--(17.48,.06);
\node[label,anchor=north] at (9.20,-.04) {$z$ increases toward the singularity};
\end{tikzpicture}}
\end{center}
\vspace{-0.35cm}
\caption{\small Cartoon profile for the metric field and scalar field near a continuous scalar-hair bifurcation. The Cauchy horizon $z_-$ belongs to the reference solution; for any
nonzero ${\cal B}_-$ it is replaced by a ``narrow" nonlinear regime  ({\em shaded}) containing the Einstein-Rosen bridge collapse. The $\cal{O}$($\mathcal{B}_-^2$) pre-collapse layer and the wider
post-collapse window of order $o$($\mathcal{B}_-$), 
 are magnified in the plot for
clarity. The solution subsequently enters its first Kasner epoch.}
\label{fig:cartoon}
\end{figure}

\subsection{Logarithmic transmission at the Cauchy horizon}
\label{subsec:general_log_transmission}

We first describe the reference black hole in a form that applies to
both the static and rotating examples considered below.  We choose the
radial coordinate $z$ to increase toward the black-hole interior and
write
\begin{equation}
ds^2
=
\frac{1}{z^2}
\left[
-f(z)e^{-\chi(z)}dt^2
+
\frac{dz^2}{f(z)}
+
\sigma_{ij}(z)
\left(dx^i+N^i(z)dt\right)
\left(dx^j+N^j(z)dt\right)
\right],
\label{eq:general_stationary_metric}
\end{equation}
for $d+2$-dimensional asymptotic AdS spacetime, where $x^i$ are spatial coordinates of dual field theory. Here all metric functions depend only on $z$. 
The horizons are determined by the zeros of $f(z)$. The metric function 
$N^i(z)$ describes rotation.  The matrix
$\sigma_{ij}(z)$ describes the geometry of the spatial directions
along the horizon.  For an asymptotic AdS spacetime, we separate its overall scale into the prefactor
$z^{-2}$ and choose
$
\det\sigma_{ij}=1 .
$ 

The reference black hole is assumed to possess a nonextremal event
horizon at $z=z_+$ and a simple Cauchy horizon at $z=z_-$, satisfying $f(z_\pm)=0$ and $f'(z_+)<0, f'(z_-)>0$. 
We take $z$ to increase toward the black-hole interior, so that
$z_->z_+$.  Near the Cauchy horizon,
\begin{equation}
f(z)
=
f'_-(z-z_-)
+
\mathcal O\!\left((z-z_-)^2\right),
\qquad
f'_-\equiv f'(z_-)\neq0 .
\label{eq:general_F_CH}
\end{equation}
The Cauchy horizon of the reference solution is regular.  In
particular, $\chi$, $\sigma_{ij}$, $N^i$, and the background matter
fields remain finite at $z=z_-$. 

We now consider a stationary branch that bifurcates continuously from
this reference solution.  Close to the bifurcation point, the
deformation is controlled by a small amplitude $\epsilon$, and the
perturbing field can be written as
\begin{equation}
\phi(z)
=
\epsilon\Phi(z)
+
\mathcal O(\epsilon^2).
\label{eq:general_Psi_expansion}
\end{equation}
The linear mode $\Phi$ is regular at the event horizon.  We focus on perturbations that preserve all Killing symmetries of the reference background. 
Thus the perturbation 
depends only on the radial coordinate $z$, and the
linear problem reduces to an ordinary differential equation.
This includes both examples studied below: the symmetry-preserving
scalar mode of rotating BTZ and the homogeneous scalar mode of planar
RN--AdS$_4$, as well as the example of higher odd-dimensional rotating black holes with equal angular momenta.  More general rotating backgrounds may retain dependence
on non-Killing angular coordinates and do not admit the present one-dimensional reduction \cite{Dafermos:2017dbw}.

For a single scalar perturbation, the
quadratic action is 
\begin{equation}
S^{(2)}
=
-\frac{1}{2}
\int dz\,d^{d+1}x\,
\sqrt{-g}
\left[
Z(z)g^{zz}(z)(\Phi')^2
+
{\cal U}(z)\Phi^2
\right].
\label{eq:general_quadratic_action}
\end{equation}
Here and below, a prime denotes differentiation with respect to $z$.  $Z(z)$ is the radial kinetic coefficient and ${\cal U}(z)$
collects the terms without radial derivatives. 
We assume that both
are regular at the reference Cauchy horizon and that $Z(z_-)$ is
nonzero.

The equation for the zero mode $\Phi(z)$ takes the form
\begin{equation}
\left[
{\cal P}(z)\Phi'
\right]'
+
{\cal Q}(z)\Phi
=
0 ,
\label{eq:general_flux_equation}
\end{equation}
where
\begin{equation}
{\cal P}(z)
=
z^{-d}e^{-\chi/2}Z(z)f(z),
\qquad
{\cal Q}(z)
=
-z^{-(d+2)}e^{-\chi/2}{\cal U}(z).
\label{eq:general_PQ}
\end{equation}

Near the Cauchy horizon, \eqref{eq:general_PQ} takes the
form of a regular singular equation with\footnote{
The present argument requires ${\cal Q}(z)$ to remain finite at the
reference Cauchy horizon. For a charged perturbation, the coupling term $A_t^2/f$ can introduce additional inverse powers of
$f(z)$. The logarithmic channel must then be reconsidered; depending on the
effective horizon parameters, the local solutions may instead have
nonzero or complex indicial exponents.
}
\begin{equation}
{\cal P}(z)
=
{\cal P}_1(z-z_-)
+\cdots ,
\qquad
{\cal Q}(z)
=
{\cal Q}_-
+\cdots,
\qquad
{\cal P}_1
=
z_-^{-d}e^{-\chi_-/2}\,Z_-f'_- .\label{eq:general_P_near_CH}
\end{equation}
A Frobenius ansatz $\Phi\sim (z-z_-)^c$ gives
${\cal P}_1 c^2 (z-z_-)^{c-1}
+\mathcal O((z-z_-)^c)=0$, 
therefore 
$c^2=0 .$
The repeated root produces one regular and one logarithmic solution,
so that
\begin{equation}
\Phi(z)
=
A_-
+
B_-
\log
\left|
\frac{z-z_-}{\ell_-}
\right|
+\cdots .
\label{eq:general_log_mode}
\end{equation}
Here $\ell_-$ is a reference scale; changing it only shifts
$A_-$ and leaves $B_-$ unchanged.

The coefficient $B_-$ is fixed by the global linear solution obtained
by propagating the event-horizon-regular mode to the Cauchy horizon.
From \eqref{eq:general_log_mode}, we have 
\begin{equation}
\Phi'(z)
\simeq
\frac{B_-}{z-z_-}.
\label{eq:general_log_derivative}
\end{equation}
Thus the logarithmic mode produces a divergent radial gradient at the
reference Cauchy horizon. In the examples below, $B_-\neq0$ is obtained
analytically for rotating BTZ and proved by an integral
identity for the Einstein--Maxwell--scalar model.\footnote{ 
The special case $B_-=0$ corresponds to a mode that is regular at
both horizons.  Whether this occurs is a global property of the linear
solution and cannot be determined from the local Cauchy-horizon analysis
alone.}  

Finally, the normalization of $\Phi$ is arbitrary, so $B_-$ alone is
not a physical amplitude.  
In the full
perturbation \eqref{eq:general_Psi_expansion}, i.e. $\phi=\epsilon\Phi+\cdots$, close to  the Cauchy horizon, before ER collapse,  
we have \be\phi'=\epsilon\Phi'+\cdots
\simeq \frac{{\cal B}_-}{z-z_-}\,,~~~~~~~~ {\cal B}_-
\equiv
\epsilon B_- .
\label{eq:general_Bphysical}
\ee 
It is this physical logarithmic
amplitude ${\cal B}_-$ that is passed
from the linear problem to the nonlinear Cauchy-horizon region.


\subsection{Einstein-Rosen bridge collapse}
\label{subsec:erc}

From \eqref{eq:general_Bphysical}, we know that an arbitrarily small perturbation develops a large radial
gradient as the reference Cauchy horizon is approached.
This enhancement is 
visible in the scalar kinetic term,
\begin{equation}
(\nabla\phi)^2
=
g^{zz}(\phi')^2
\simeq
z_-^2 f'_-
\frac{{\cal B}_-^2}{z-z_-}.
\label{eq:general_kinetic_divergence}
\end{equation}
Although $\phi$ itself is small, its kinetic contribution 
diverges in a formal continuation to
$z=z_-$.  The linear expansion must break down before this point is
reached.

The backreaction becomes of order unity when 
\begin{equation}
 \frac{(z_--z)_{\rm bl}}{z_-}
 =
 {\cal O}({\cal B}_-^2),
\label{eq:general_breakdown_scale}
\end{equation}
where we use the subscript to refer it as ``boundary layer", as shown as the left boundary of shaded regime in Fig. \ref{fig:cartoon}, where the nonlinear effect starts to become important.  
If $B_-$ approaches a finite nonzero value at the
bifurcation, then
$
(z_--z)_{\rm bl}
=
\mathcal O(\epsilon^2). 
$ 
This sets the scale at which the Einstein-Rosen bridge starts to collapse. 

An important feature of the boundary layer region is that the field amplitude
remains small even though its gradient is large.  
Consequently, smooth matter terms that depend on the field amplitude
remain perturbatively close to their reference values across the
boundary layer.  The derivative terms are different: the large radial 
gradient produces order-one stress energy and therefore requires a
nonlinear treatment of both the scalar gradient and the metric.

The scale \eqref{eq:general_breakdown_scale} tells us where the linear solution ceases to be
valid. 
A natural matching quantity is the canonical
radial momentum which can be passed into the nonlinear region,
\begin{equation}
\Pi_\phi
\equiv
\sqrt{-g}\,Zg^{zz}\phi'
=
{\cal P}(z)\phi' .
\label{eq:general_radial_momentum}
\end{equation}
In the outer overlap region ${\cal B}_-^2\ll (z_--z)/z_-\ll1, $ the full geometry is still asymptotic to the reference solution,
whereas the logarithmic part of the scalar mode dominates its radial derivative. Using Eqs.~\eqref{eq:general_P_near_CH} and
\eqref{eq:general_Bphysical}, we obtain the matched incoming flux
\begin{equation}
\Pi_\star
=
{\cal P}_1{\cal B}_-
+
o({\cal B}_-).
\label{eq:general_momentum_matching}
\end{equation}  
The symbol $\star$ labels the datum entering the nonlinear layer, evaluated at $z_\star$ within the boundary layer \eqref{eq:general_breakdown_scale}, whose characteristic location will be specified later. 
We now show how its backreaction produces the Einstein-Rosen bridge collapse.

\subsubsection{Einstein--Rosen nonlinear completion}
\label{subsec:general_ER}

For stationary geometries it is useful to isolate the longitudinal
metric coefficient by removing the contribution of the rotational
shift.  Using the metric field introduced in 
\eqref{eq:general_stationary_metric}, we define
\begin{equation}
{\cal G}
\equiv
g_{tt}
-z^2 g_{ti}\sigma^{ij}g_{jt}
=
-\frac{f e^{-\chi}}{z^2}.
\label{eq:general_ER_geometry}
\end{equation}
For a static solution, ${\cal G}=g_{tt}$, while for a rotating
solution the shift contribution is projected out.

For the class of Einstein--matter systems specified  in
Appendix~\ref{app:ER_normal_form}, the scalar flux is constant to
leading order across the boundary layer, i.e. $\Pi_\phi
=
\Pi_\star\left(1+o(1)\right)$ from \eqref{eq:general_momentum_matching}. 
And as in \eqref{eq:app_ER_normal_form}, the   equations for metric and scalar fields reduce to
\begin{align}
{\cal G}'
&\simeq
-\Lambda_-
\frac{{\cal G}}{{\cal G}+{\cal G}_c},
~~~~~~
\phi'
\simeq
-{\cal K}_-
\frac{\Pi_\star}{{\cal G}+{\cal G}_c},
\label{eq:ER_Psi_equation}
\end{align}
where
\begin{equation}
{\cal G}_c
=
{\cal C}_-\Pi_\star^2 .
\label{eq:ER_Gc}
\end{equation}
The coefficients $\Lambda_-$, ${\cal C}_-$, and ${\cal K}_-$ are finite and
nonzero in the critical limit and are fixed by the reference
Cauchy-horizon geometry.  Their explicit expressions are derived in
Appendix~\ref{app:ER_normal_form}.

The equation for ${\cal G}$ in ~\eqref{eq:ER_Psi_equation} admits two different asymptotic behaviors, 
\be 
\mathcal{G} \gg \mathcal{G}_c:
\quad 
\text{linear fall-off};
\qquad \qquad
\mathcal{G} \ll \mathcal{G}_c:
\quad 
\text{exponential fall-off}.
\ee 
We thus choose the ``crossover
position" $z_\star$ in the boundary layer by ${\cal G}(z_\star)={\cal G}_c$, which provides a  boundary condition to integrate the equation for ${\cal G}$.
We obtain
\begin{equation}
{\cal G}
-
{\cal G}_c
+
{\cal G}_c
\log\left(
\frac{{\cal G}}{{\cal G}_c}
\right)
=
-\Lambda_-(z-z_\star).
\label{eq:ER_profile}
\end{equation}
This implicit solution describes the complete leading crossover
through the nonlinear layer.

Its two asymptotic limits make the physical behavior transparent.
On the outer side, ${\cal G}\gg{\cal G}_c$, Eq. 
\eqref{eq:ER_profile} reduces to
\be 
{\cal G}\simeq-\Lambda_-(z-z_\star),
\ee
reproducing the simple zero of the
reference Cauchy horizon.  At the same time,
Eq.~\eqref{eq:ER_Psi_equation} gives
$\phi'\propto\Pi_\star/{\cal G}$, matching the logarithmically
enhanced gradient found in 
Sec.~\ref{subsec:general_log_transmission}.

After the collapse, ${\cal G}\ll{\cal G}_c$, one instead has
\begin{equation}
{\cal G}(z)
\simeq
\,{\cal G}_c
e^{1-\Gamma_{\rm ER}(z-z_\star)}, 
\qquad
\Gamma_{\rm ER}
\equiv
\frac{\Lambda_-}{{\cal G}_c}; 
\qquad
\phi'_{\rm aft}\simeq -{\cal K}_-
\frac{\Pi_\star}{{\cal G}_c}. 
\label{eq:ER_exponential}
\end{equation}
The longitudinal metric therefore changes from the linear
Cauchy-horizon behavior to an exponential collapse once
${\cal G}$ becomes of order ${\cal G}_c$. This is known as the Einstein--Rosen bridge collapse \cite{Hartnoll:2020rwq, Hartnoll:2020fhc}. Note that after collapse, we still have $\Pi_\phi
=\Pi_\star\left(1+o(1)\right)$.

With the expressions \eqref{eq:general_momentum_matching}, \eqref{eq:ER_Gc} and \eqref{eq:app_ER_dictionary}, one immediately obtains
\begin{align}
\Gamma_{\rm ER}
\simeq 
\frac{2d}{
Z_-z_-{\cal B}_-^2
},
~~~~~~~~~
|\phi'|_{\rm aft}
\simeq 
\frac{2d}{
Z_-z_-|{\cal B}_-|
}, 
\label{eq:general_ER_amplitudes}
\end{align}
which has been announced in \eqref{eq:deltaer-gammaer} of Sec. \ref{sec:introduction}, and derived in \eqref{eq:app_ER_final_amplitudes} of Appendix~\ref{app:ER_normal_form}. 
These relations are the nonlinear completion of the matching derived in Sec.~\ref{subsec:general_log_transmission}.

The corresponding characteristic width scale of the Einstein-Rosen bridge collapse is
\be \Delta z_{\rm ER}=\Gamma_{\rm ER}^{-1}\simeq \frac{Z_-z_-}{2d} {\cal B}_-^2,\ee in the radial gauge of Eq.~\eqref{eq:general_stationary_metric}, and we will not treat this
coordinate-dependent quantity as an independent observable. This characteristic width is at the same order as \eqref{eq:general_breakdown_scale}. 

\subsection{Critical Kasner scaling}
\label{subsec:general_ER_to_Kasner}


On the collapsed side, define 
\begin{equation}
\delta z
\equiv
\frac{z-z_\star}{z_-}>0.
\end{equation}
Consider the asymptotic window
\begin{equation}
\frac{\Delta z_{\rm ER}}{z_-}
\ll
\delta z
\ll
|{\cal B}_-|.
\label{eq:general_kinetic_window}
\end{equation}
The lower inequality implies
$\Gamma_{\rm ER}(z-z_\star)\gg1$, and therefore
${\cal G}/{\cal G}_c\ll1$. Eq. \eqref{eq:ER_Psi_equation} then gives the
after collapse plateau
\begin{equation}
\phi'
=
\phi'_{\rm aft}[1+o(1)],
\qquad
|\phi'_{\rm aft}|
=
{\cal O}\!\left(\frac{1}{z_-|{\cal B}_-|}\right).
\end{equation}
The upper inequality ensures
\begin{equation}
|\Delta\phi|
\sim
|\phi'_{\rm aft}|(z-z_\star)
=
{\cal O}\!\left(\frac{\delta z}{|{\cal B}_-|}\right)
=o(1).
\end{equation}
Therefore smooth nondifferential matter terms remain finite, whereas
the scalar kinetic contribution is parametrically enhanced.
In particular, 
$
 Z(\phi)=Z_-+o(1).
$
Provided that no  additional matter 
develops the same enhancement, the leading 
equations reduce to those of Einstein gravity coupled to a scalar
with constant kinetic normalization $Z_-$. 

For the isotropic sector relevant to the examples in next two sections, 
the corresponding Kasner solution can be written as
\begin{align}
\phi(z)
\simeq
v\log z+\phi_K,
~~~~~
\chi(z)
\simeq
\frac{Z_-v^2}{d}\log z+\chi_K,
~~~~~
f(z)
\simeq
-f_K z^{\,d+1+Z_-v^2/(2d)},
\label{eq:general_Kasner_radial}
\end{align}
where $f_K>0$ and $v\equiv z\phi'$ is approximately constant during a Kasner
epoch. 

Equivalently, in terms of proper time $\tau$ toward the spacelike
singularity,
\begin{align}
ds^2
\simeq
-d\tau^2
+
\tau^{2p_t}d\widetilde t^{\,2}
+
\tau^{2p_x}
\sum_{i=1}^{d}d\widetilde x_i^{\,2},
~~~~~~~~
\phi
\simeq
-\sqrt{\frac{2}{Z_-}}\,
p_\phi\log\tau+\phi_0 ,
\label{eq:general_Kasner_proper_time}
\end{align}
where
\begin{equation}
p_t+d\, p_x=1,
\qquad
p_t^2+d\, p_x^2+p_\phi^2=1.
\label{eq:general_Kasner_relations}
\end{equation}
The two parameterizations are related by
\begin{equation}
p_t
=
\frac{
Z_-v^2-2d(d-1)
}{
Z_-v^2+2d(d+1)
},
\qquad
p_x
=
\frac{
4d
}{
Z_-v^2+2d(d+1)
},
\qquad
p_\phi
=
\frac{
2\sqrt{2Z_-}\,d\, v
}{
Z_-v^2+2d(d+1)
}.
\label{eq:general_Kasner_exponents}
\end{equation}

In the overlap between the ER collapse and first  Kasner regimes, $z=z_-+o(1)$, one has
$|v_1|=z_-|\phi'|_{\rm aft}[1+o(1)]$.
Using Eq.~\eqref{eq:general_ER_amplitudes}, this gives
\begin{equation}
|v_1|
=
\frac{2d}{
Z_-|{\cal B}_-|
}
\left[1+o(1)\right].
\label{eq:general_v1_scaling}
\end{equation}
Hence $|v_1|$ diverges as ${\cal B}_-\rightarrow0$, and the first
Kasner epoch approaches the large-$|v|$ regime. 
The matching is also manifest in the metric sector: Eq.~\eqref{eq:general_Kasner_radial} gives ${\cal G}_K\propto z^{\,d-1-Z_-v_1^2/(2d)}$. Hence, in the overlap region,  \eqref{eq:general_v1_scaling} and \eqref{eq:general_ER_amplitudes} yield $-\partial_z\log{\cal G}_K\simeq \Gamma_{\rm ER}$. Thus the scalar gradient and the longitudinal metric decay match onto the same first Kasner solution.

Expanding Eq.~\eqref{eq:general_Kasner_exponents} for
$|v_1|\gg1$, we obtain
\begin{align}
p_t^{(1)}
=
1-
\frac{4d^2}{Z_-v_1^2}
+
{\cal O}(v_1^{-4}),
~~~~~
p_x^{(1)}
=
\frac{4d}{Z_-v_1^2}
+
{\cal O}(v_1^{-4}),
~~~~~
p_\phi^{(1)}
=
\frac{2\sqrt{2}\,d}{
\sqrt{Z_-}\,v_1
}
+
{\cal O}(v_1^{-3}).
\label{eq:general_large_v_Kasner}
\end{align}
Here we included a superscript to indicate that they are the first Kasner exponents. 
Combining this expansion with
Eq.~\eqref{eq:general_v1_scaling} gives
\begin{equation}
1-p_t^{(1)}
\simeq 
Z_-{\cal B}_-^2
,
\qquad
p_x^{(1)}
\simeq 
\frac{Z_-}{d}{\cal B}_-^2
,
\qquad
|p_\phi^{(1)}|
\simeq 
\sqrt{2Z_-}\,|{\cal B}_-|
.
\label{eq:general_first_Kasner_scaling}
\end{equation}
Here $p_x$ denotes the exponent of each of the $d$ equivalent
transverse directions.  Thus, as ${\cal B}_-\rightarrow0$, the first
Kasner epoch approaches $p_t^{(1)}\rightarrow1$, while the transverse
and scalar Kasner exponents vanish.

This first-Kasner matching follows whenever the post-collapse region
remains dominated by the scalar kinetic term, with no competing
matter or shear channel acquiring the same parametric enhancement.
The subsequent evolution may not remain universal: once the scalar
moves away from the small-field regime, additional Kasner
transitions can depend on the large-field form of the matter couplings.

\subsubsection{Transmission of critical scaling}
\label{subsec:general_critical_scaling}

We now translate the matching results above into scaling with the
boundary control parameter.  We use $\lambda$ to denote the parameter that
drives a continuous bifurcation. 
If the order-parameter
amplitude behaves as
$
|\epsilon|
\simeq 
A_\epsilon\,
|\lambda-\lambda_c|^\beta,
$ 
and the logarithmic transmission coefficient has a finite nonzero
critical limit,
$
B_-
\rightarrow
B_-^{(c)}
\neq0,
$
then the physical Cauchy-horizon amplitude introduced in
Eq.~\eqref{eq:general_Bphysical} satisfies
\begin{equation}
|{\cal B}_-|
\simeq
A_{\cal B}\,
|\lambda-\lambda_c|^\beta
,
\qquad
A_{\cal B}
=
A_\epsilon |B_-^{(c)}|.
\label{eq:general_Bcal_scaling}
\end{equation}

Substituting Eq.~\eqref{eq:general_Bcal_scaling} into the ER matching
relations \eqref{eq:general_ER_amplitudes} in  Sec.~\ref{subsec:general_ER} gives
\begin{equation}
\Gamma_{\rm ER}
\propto
|\lambda-\lambda_c|^{-2\beta},
\qquad
|\phi'|_{\rm aft}
\propto
|\lambda-\lambda_c|^{-\beta}.
\label{eq:general_ER_critical_scaling}
\end{equation}
Thus the nonlinear Cauchy-horizon scales inherit their critical powers
directly from the exterior order parameter. 

When the first-Kasner matching \eqref{eq:general_first_Kasner_scaling} applies, the same critical
amplitude also gives
\begin{equation}
|v_1|
\propto
|\lambda-\lambda_c|^{-\beta},
\qquad
1-p_t^{(1)}
\propto
|\lambda-\lambda_c|^{2\beta},
\qquad
p_x^{(1)}
\propto
|\lambda-\lambda_c|^{2\beta},
\qquad
|p_\phi^{(1)}|
\propto
|\lambda-\lambda_c|^\beta ,
\label{eq:general_Kasner_critical_scaling}
\end{equation}
This establishes the scaling anticipated in \eqref{eq:intro_Kasner_scaling}. 
For an ordinary mean-field bifurcation, $\beta=1/2$, so that
$1-p_t^{(1)}$ are linear in
$|\lambda-\lambda_c|$, while $\Gamma_{\rm ER}$ diverges as
$|\lambda-\lambda_c|^{-1}$ and $|v_1|$ as
$|\lambda-\lambda_c|^{-1/2}$.  More generally, a multicritical
bifurcation with a different value of $\beta$ changes these powers
without changing the Cauchy-horizon matching itself. 
If the scalar potential is tuned to realize a multicritical point with $\beta=1/(2r)$, the exponent $\beta$ should be replaced by the corresponding multicritical exponent, in analogy with the one-horizon case \cite{Gao:2026tck}. 

\subsection{Comments on validity}
\label{subsec:general_scope}

The transmission mechanism involves three logically distinct steps, with following assumptions. 

First, logarithmic transmission requires a single regular two-derivative radial channel at a simple nonextremal Cauchy horizon.  Locally this produces a repeated indicial root, while the coefficient
$B_-$ is fixed globally by propagating the event-horizon-regular zero
mode. If the horizon is degenerate, if the radial channel is singular,
or if $B_-=0$,\footnote{If $B_-$ itself scales with the control parameter, that behavior must be included together with $\epsilon$ in the physical amplitude ${\cal B}_-=\epsilon B_-$.} the leading logarithmic matching must be replaced by
the next nonvanishing channel, and the resulting interior critical powers
are modified. 


Second, the Einstein--Rosen completion requires
$Z_->0$ and a positive finite regular source, ${\cal C}_->0$ in \eqref{eq:app_ER_constants}, 
together with the uniform remainder conditions stated in
Appendix~\ref{app:ER_normal_form}. These conditions ensure that no
matter or shear contribution competes with the scalar-gradient term
on a radial scale larger than ${\cal O}({\cal B}_-^2)$.


Third, matching to a first Kasner epoch requires the
post-collapse solution to remain scalar-kinetic dominated beyond the
local overlap region. The Einstein--Rosen collapse can therefore
remain valid even when a competing potential, gauge, or shear channel
prevents the formation of an extended Kasner plateau. The explicit
Kasner formulae used here additionally assume an isotropic transverse sector. 

In the following, we illustrate the generic mechanism discussed here to the rotating and charged examples. 

\section{Critical interior scaling in rotating black holes}
\label{sec:rotating_BTZ}

We now test the interior scaling from Cauchy-horizon matching of
Sec.~\ref{sec:general_mechanism} in a rotating setting.
Hairy rotating interiors with quadratic and super-exponential scalar
potentials were studied in~\cite{Gao:2023rqc,Gao:2025qqa},
while the double-trace instability and exterior phase structure of
hairy BTZ black holes were analyzed in~\cite{Dias:2025uyk,Dias:2026xuy}.
Here we connect these two directions: a mixed boundary condition
produces a continuous BTZ bifurcation, whose critical zero mode is
continued to the Cauchy horizon and matched to the near-critical
Einstein-Rosen bridge collapse.

\subsection{Einstein--scalar model} 

We begin by specifying the rotating model, the boundary conditions, 
and the conventions needed for the critical and interior analysis. 

The three dimensional bulk action for Einstein-scalar system is 
\begin{equation}
 S_{\rm bulk}=\frac{1}{16\pi G_3}\int_{\mathcal M}\dd^3x\sqrt{-g}
 \left[R-\frac12(\partial\phi)^2-V(\phi)\right].
 \label{eq:action-rotating}
\end{equation}
We work in units
$
 16\pi G_3=1$ and $L=1.
$
The scalar is real and its potential is
\begin{equation}
V(\phi)=
-2+\frac12m^2\phi^2+\frac{\lambda_4}{4}\phi^4.
 \label{eq:potential}
\end{equation}
The constant term supports an 
$\AdS_3$ solution at $\phi=0$.
The quartic coupling $\lambda_4$ does not qualitatively affect the black hole interior near phase transition, but determines the scaling relations of thermodynamic quantities and Kasner exponents.

The metric and scalar ansatz are
\begin{equation}
 \dd s^2=\frac{1}{z^2}\left[
 -f(z)e^{-\chi(z)}\dd t^2+\frac{\dd z^2}{f(z)}
 +\bigl(N(z)\dd t+\dd x\bigr)^2
 \right],
 \qquad \phi=\phi(z)\,,
 \label{eq:ansatz-rotating}
\end{equation}
where the spatial coordinate  is compact,
$ x\sim x+2\pi l\,,~ l=1$.\footnote{$l=1$ sets the unit of the dimensional quantities throughout Sec.~\ref{sec:rotating_BTZ}. The ensemble parameters of the system are $\Omega, T, \kappa$. }
The action \eqref{eq:action-rotating} yields independent equations for $f,\chi,\phi$ and $N$,
in which the equation for $N$ has the first integral
\begin{equation}
 J=-e^{\chi/2}\frac{N'}{z}\,.
 \label{eq:N}
\end{equation}
The constant $J$ fixes the angular momentum. 

The system \eqref{eq:ansatz-rotating} admits two hairless solutions: the rotating BTZ black hole and thermal AdS. The BTZ black hole has
\begin{equation}
 f_{\BTZ}=\left(1-\frac{z^2}{z_+^2}\right)
 \left(1-\frac{z^2}{z_-^2}\right),
 \qquad
 N=-\frac{z^2}{z_+z_-}\,,\qquad
 \label{eq:BTZ-metric-func}
\end{equation}
where $z_+$ and $z_-$ denote the outer and inner horizon.
The temperature, angular velocity, and entropy density along the $x$ direction with $x\sim x+2\pi$ are
\be 
T_{\BTZ}=\frac{1-\Omega^2}{2\pi z_+}\,,
\qquad
\Omega=\frac{z_+}{z_-}\,,
\qquad
s_{\BTZ}=\frac{4\pi}{z_+}\,.
\label{eq:btz-horizon-quantities}
\ee 
The thermal AdS has
\be 
f_{\AdS}=1+z^2\,,
\qquad
N=0\,,
\label{eq:ads-metric}
\ee 
with vanishing entropy.
The Hawking-Page transition occurs at the critical line
\begin{equation}
 T_{\rm HP}(\Omega)=\frac{\sqrt{1-\Omega^2}}{2\pi}.
 \label{eq:HP-temperature}
\end{equation}
The thermodynamics of the two solutions are given in Appendix \ref{sec:thermo-rotating}. 

The system \eqref{eq:ansatz-rotating} also admits two hairy solutions: hairy black hole and boson star, with different IR conditions. For a black hole, the outer event horizon is at $z=z_h$, where $f(z_h)=0.$ The regular expansion at a non-extremal outer horizon is
\bebal
f\to(z-z_h)\left(
\frac{J^2 z_h^3}{2}
+\frac{V(\phi_h)}{z_h}
\right)\,,~~~
\chi\to \chi_h\,,~~~
\phi\to \phi_h\,,~~~
N\to N_h\,.
\eeeal
The Hawking temperature, angular velocity, and entropy density along $x\sim x+2\pi$ are
\begin{equation}
 T=\frac{e^{-\chi(z_h)/2}|f'(z_h)|}{4\pi}=-\frac{1}{4\pi}e^{-\chi_h/2}\left(
\frac{J^2 z_h^3}{2}
+\frac{V(\phi_h)}{z_h}
\right)\,,
~~~
\Omega=-N_h\,,
~~
s=\frac{4\pi}{z_h}\,.
 \label{eq:rota-horizon-quantities}
\end{equation}
We do not aim to consider the extremal case $T=0$, since the following boson star solution could dominate the phase diagram at low temperature.

For a horizonless boson star, its center is at $z=\infty$. The regular center gives $J=0$ from \eqref{eq:N}, and thus $N'=0$. The constant $N$ is then fixed to zero by choosing the non-rotating boundary frame $N(0)=0$. 
The regular expansions for other functions are 
\bebal
f(z)\to z^2\,,\qquad
\chi(z)\to \chi_0\,,\qquad
\phi(z)\to \phi_0\,.
\eeeal
The leading behavior $f=z^2$ ensures that there is no conical singularity at $z=\infty$.  

The two hairy solutions share the same asymptotic expansions near the AdS boundary $z\to 0$. We choose $m^2=-15/16$ to avoid logarithmic terms for simplicity, and we have
\bebal
 f(z)\to 1+\frac38\alpha^2z^{3/2}+f_{(2)}z^2
 \,,~~~
 \chi(z)\to 0\,,~~~
 \phi(z)\to \alpha z^{3/4}+\eta z^{5/4}\,,~~~
 N(z)\to -\frac{J}{2} z^2\,.
 \label{eq:UV-expansion}
\eeeal
Note that the boundary time is normalized by $\chi(0)=0$, and the non-rotating boundary frame by $N(0)=0$. This can be implemented by
the two shift symmetries,
\begin{align}
    \chi&\rightarrow \chi-\chi_b\,,\qquad t\rightarrow e^{-\frac{\chi_b}{2}} t\,,\qquad N\rightarrow e^{\frac{\chi_b}{2}} N\,,\\
    N&\rightarrow N-N_b\,,\qquad x\rightarrow x+N_b t\,,
\end{align}
where $\chi_b$ and $N_b$ are arbitrary constants.

BTZ black hole is stable against scalar perturbation with Dirichlet or Neumann boundary conditions. To realize the phase transition between the rotating BTZ and hairy black hole, we consider the double trace deformation,
\be
S\to S-{\kappa} \int \dd^2 x \,{O}^2\,.
\ee
We use the alternative quantization, which is allowed for the scalar mass $m^2=-15/16$ in the range $-1<m^2<0$.
The sourceless condition is $\eta-\kappa \al=0$, and the scalar condensate is 
$O=-\alpha/2\,,$ 
which serves as the order parameter.

We consider the grand canonical ensemble, where the temperature $T$ and angular velocity $\Omega$ control the phase diagram at fixed double-trace coupling $\kappa$. \footnote{Our goal here is to realize a phase transition, at least locally, between the rotating BTZ and hairy black hole. One may also consider the microcanonical or canonical ensemble, provided that such a phase transition exists.}
The corresponding holographic renormalization of the Euclidean bulk action gives the thermodynamic densities along the $x$ direction with $x\sim x+2\pi$,
\begin{equation}
 \mathcal E
 =-f_{(2)}+\frac54\alpha\eta-\frac14\kappa\alpha^2\,,\qquad
 \mathcal J=J\,,
 \label{eq:energyFromUVData}
\end{equation}
where $\mathcal E$ and $\mathcal J$ are the energy and angular momentum. The variables on the right-hand sides are defined in the UV expansions \eqref{eq:UV-expansion}.
The thermodynamic potential is the free energy density,
\begin{equation}
 \mathcal F=
 \mathcal E-Ts-\Omega \mathcal J\,,
 \qquad
\dd\mathcal F =
-s\,\dd T-\mathcal J\,\dd \Omega\,, 
\end{equation}
where the thermodynamic quantities $T,s,\Omega$ are given in \eqref{eq:rota-horizon-quantities}.
At fixed $\kappa$, the preferred saddle at $(T,\Omega)$ minimizes $\mathcal F$. The details are shown in Appendix \ref{sec:thermo-rotating}.

\subsection{Phase diagram}
\label{sec:phase-double-trace}
In the previous section, we analyzed the full nonlinear boson star and hairy black hole solutions, and derived their thermodynamics in the grand canonical ensemble at fixed double-trace coupling $\kappa$. In this section, we further investigate their existence perturbatively, which helps us to get the full phase diagram.

The existence of boson star originates from the instability of AdS vacuum when $\kappa<\kappa_b$. This is independent of $T$ and $\Omega$, since the bulk solutions for boson star and thermal AdS are independent of them. Hairy black hole originates from the instability of BTZ black hole upon crossing the critical line $T(\Omega)$ with fixed $\kappa$. The following perturbative analysis of the onset scalar determines $\kappa_b$ and $T(\Omega)$.

We first discuss the scalar perturbation on $\AdS_3$, which is regular at the center $z=\infty$.  
The linearized scalar equation is
\begin{equation}
\vp''+\left(\frac{f_{\AdS}'}{f_{\AdS}}-\frac1z\right)\vp'
 -\frac{m^2}{z^2 f_{\AdS}}\vp=0,
 \qquad m^2=-\frac{15}{16}\,,
\end{equation}
with $f_\AdS$ given by \eqref{eq:ads-metric}.
Imposing the double trace boundary condition $\eta=\kappa \alpha$ at $z=0$ and regularity at $z=\infty$, the onset scalar satisfies 
\be
\label{eq:kappab}
\kappa=\kappa_b=-
 \frac{\Gamma(3/4)\Gamma(5/8)^2}
 {\Gamma(3/8)^2\Gamma(5/4)}
 \simeq -0.49513\,,
\ee
which signals the instability of $\AdS_3$. When $\kappa<\kappa_b$, the free energy of boson star is always lower than that of thermal $\AdS_3$ \cite{Dias:2026xuy}, and boson star is thus thermodynamically more preferred.

For the hairy black hole originating from the instability of BTZ black hole, the onset scalar perturbation is
\begin{equation}
\Phi''+\left(\frac{f_{\BTZ}'}{f_{\BTZ}}-\frac1z\right)\Phi'
 -\frac{m^2}{z^2 f_\BTZ}\Phi=0,
 \qquad m^2=-\frac{15}{16}\,, \label{eq:rotatingZeroMode}
\end{equation}
with $f_\BTZ$ given by \eqref{eq:BTZ-metric-func}.
A basis normalized to the two UV falloffs is
\begin{align}
 \Phi_-(z)&=\left(\frac{u(z)}{D}\right)^{3/8}
 {}_2F_1\left(\frac38,\frac38;\frac34;u(z)\right),
 \qquad
 \Phi_+(z)=\left(\frac{u(z)}{D}\right)^{5/8}
 {}_2F_1\left(\frac58,\frac58;\frac54;u(z)\right)\,,
 \label{eq:rotatingBasis}
\end{align}
where
\begin{equation}
 u(z)\equiv\frac{Dz^2}{1-z^2/z_-^2}\,,\qquad D\equiv\frac1{z_+^2}-\frac1{z_-^2}\,.
 \label{eq:rota-uz}
\end{equation}
We therefore write
\begin{equation}
 \Phi(z)\simeq \Phi_-(z)+\kappa\Phi_+(z),
 \qquad
 \label{eq:rota-onset}
\end{equation}
Both $\Phi_-$ and $\Phi_+$ contain a logarithmic branch at the outer horizon $z_+$. Regularity of $\Phi$ fixes the double-trace coupling to
\begin{equation}
\kappa=\kappa_b\left(\frac1{z_+^2}-\frac1{z_-^2}\right)^{1/4}\,.
\label{eq:rotatingRegularityRatio}
\end{equation}
The right-hand side is negative, consistently selecting $\kappa<0$.

At fixed $\kappa$, Eq.~\eqref{eq:rotatingRegularityRatio} is a discrete eigenvalue condition
for the horizon scale.  The lowest, nodeless zero mode gives the critical outer horizon,
\begin{equation}
 z_{+,c}=\frac{\kappa_b^2}{\kappa^2}{\sqrt{1-\Omega^2}}.
 \label{eq:criticalRadii}
\end{equation}
Substitution into the BTZ thermodynamics \eqref{eq:btz-horizon-quantities} and \eqref{eq:BTZ-thermodynamics} gives the critical line and free energy,
\begin{equation}
 T(\Omega;\kappa)
 =\frac{\kappa^2}{2\pi \kappa_b^2}\sqrt{1-\Omega^2}\,,\qquad \mathcal{F}_c(\kappa)=-\frac{\kappa^4}{\kappa_b^4}\,.
\label{eq:rota-TOmega-Fc}
\end{equation}
We consider phase transition of non-extremal black hole, which requires $T(\Omega;\kappa)>0$.

\begin{figure}[h!]
\begin{center}
\includegraphics[width=0.5\textwidth]{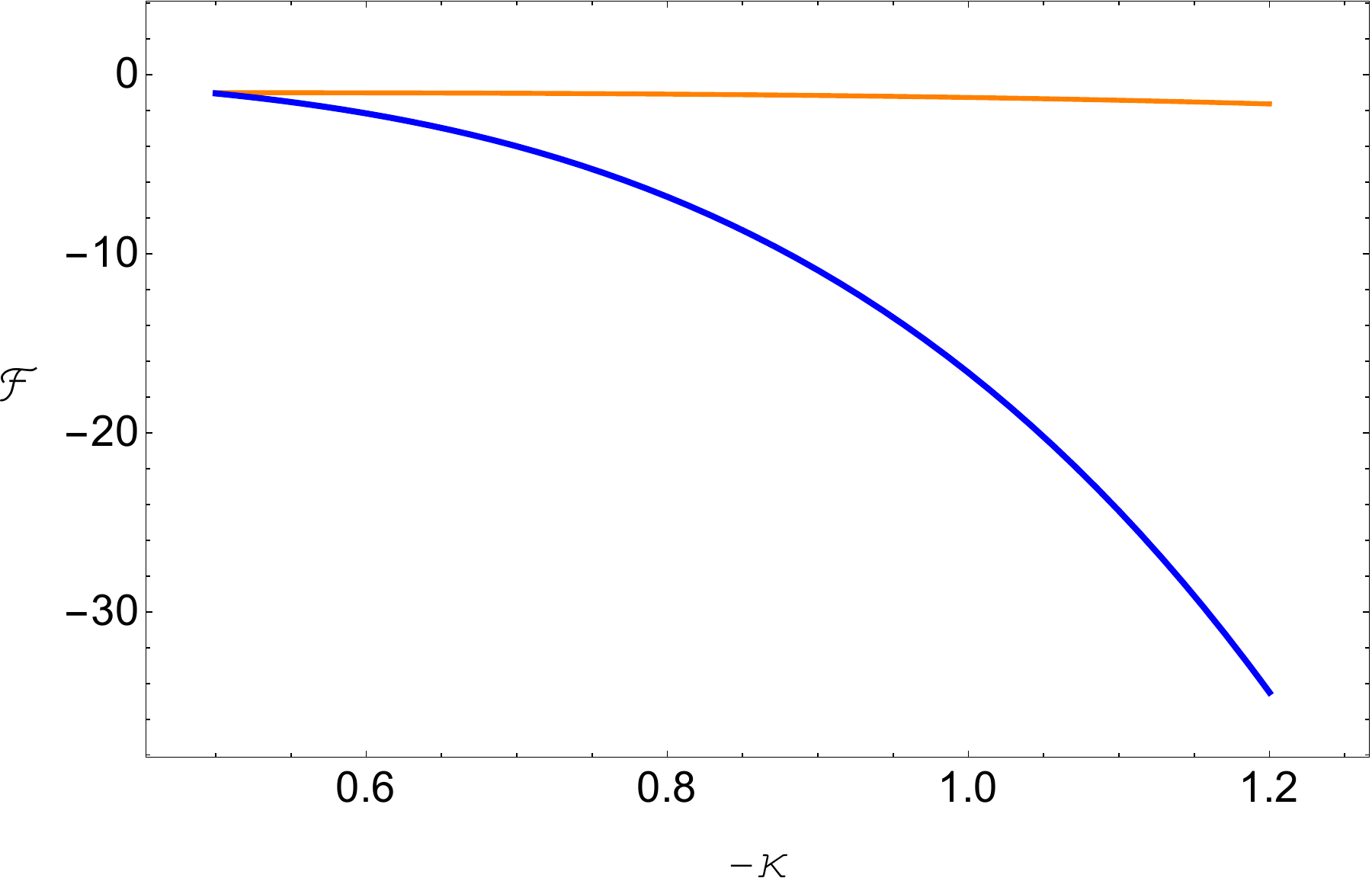}
\end{center}
\vspace{-0.5cm}
\caption{\small The free energy as a function of $-\kappa$. {\em Orange}: boson star. {\em Blue}: hairy black hole at the critical line. $\Fstar$ depends on the quartic coupling $\lambda_4$ in the potential. We choose $\lambda_4=1/5$ for definiteness. }
\label{fig:Fstar-kappa}
\end{figure}

To realize the phase transition from BTZ to hairy black hole, thermal AdS must be thermodynamically disfavored, i.e., $\FAdS>\mathcal{F}_c(\kappa)$ or equivalently $T(\Omega;\kappa)>T_{\rm HP}(\Omega)$, which requires $\kappa<\kappa_b$.
Within this range, however, boson star beats thermal AdS at any $T(\Omega)$ as discussed around \eqref{eq:kappab}. Therefore, boson star must also satisfy 
$\Fstar(\kappa)>\mathcal{F}_c(\kappa)$. Fig.~\ref{fig:Fstar-kappa} shows the numerical result of $\Fstar(\kappa)$, where we fix the quartic coupling $\lambda_4=1/5$ for definiteness. One can select $\kappa$ in this range
\be 
-1.2<\kappa<\kappa_b\,,
\ee 
where $\mathcal{F}_c$ is clearly lower than $\Fstar$, indicating there exists an intermediate region for $(T,\Omega)$, where hairy black hole is thermodynamically preferred.
Boson star could be thermodynamically preferred far away below $T(\Omega)$, since $\Fhair$ increases as $T$ or $\Omega$ decreases while $\Fstar$ keeps constant once $\kappa$ is fixed.

In the following we study the phase diagram, with the double trace coupling $\kappa$ fixed, in two directions: 
(i) varying temperature $T$ with fixed angular velocity $\Omega$, and (ii) varying $\Omega$ with fixed $T$.

\subsubsection{Scaling with fixed $\Omega$ and $\kappa$}

\label{sec:scalings of thermodynamic quantities}
We now study the high-order deviation away from the critical point $T_c(\Omega)$. The double-trace coupling $\kappa$, angular velocity $\Omega$, and bulk quartic coupling $\lambda_4$ are held fixed. The temperature $T$ is not fixed: it is the control variable that shifts away from its critical value $T_c$.
It is very convenient to choose $\al$ in the UV asymptotics $\phi=\alpha z^{3/4}+\eta z^{5/4}$ as the small parameter for the perturbative expansion.

The weakly nonlinear expansions in terms of $\alpha$ are
\begin{align}
 \phi&=\al\Phi+\al^3\phi_3+\cO(\al^5),&
 f&=f_0+\al^2 f_2+\cO(\al^4),&
 \chi&=\al^2\chi_2+\cO(\al^4),\nonumber\\
 N&=N_0+\al^2N_2+\cO(\al^4),&
 z_h&=z_{+,c}+\al^2z_{h,2}+\cO(\al^4).
 \label{eq:weakExpansion}
\end{align}
The onset mode at $\cO(\al)$ is already derived in \eqref{eq:rota-onset}. At $\cO(\al^2)$ its stress tensor sources
$(f_2,\chi_2,N_2)$ and the horizon displacement $z_{h,2}$, which obey the asymptotic AdS conditions, regularity at the outer horizon $z_h$, and fixed angular velocity
$N(z_h)=-\Omega$.  At $\cO(\al^3)$ the equation for $\phi_3$ is sourced by $(f_2,\chi_2,N_2)$ and by
$\lambda_4\Phi^3$.  Its UV expansion is,
\begin{equation}
\phi_3(z)=\alpha_3z^{3/4}+\eta_3z^{5/4}+\cO(z^{9/4})\,,
 \label{eq:cubicNoLog}
\end{equation}
where the amplitude convention fixes $\alpha_3=0$. Together with the outer-horizon regularity and the double trace boundary condition $\eta-\kappa\alpha=0$, one can get the solution for $\phi_3(z)$.

Solving the equations for $f_2, \chi_2$ and $\phi_3$, and then replacing $\al$ with the order parameter $O=-\al/2$ gives, 
\begin{align}
O
 &= -\frac{1}{2}\left[
 \frac{-\kappa}{\sqrt{1-\Omega^2}\,\mathcal A(\lambda_4)}
 \right]^{1/2}(T_c-T)^{1/2}+\cO\left((T_c-T)^{3/2}\right)\,,
\label{eq:rota-O-scaling}\\
 \mathcal F_{\rm BTZ}-\mathcal F_{\rm hairy}
 &=\frac{\mathcal \pi \kappa_b^2}{
 8(1-\Omega^2)\mathcal A(\lambda_4)}(T_c-T)^2+\cO\left((T_c-T)^{3}\right)\,.
 \label{eq:rota-F-scaling}
\end{align}
The coefficient functions are
\begin{align}
 \mathcal A(\lambda_4)&=0.030738+0.16802\,\lambda_4,
\end{align}
The first term is generated by gravitational backreaction, while the term linear in $\lambda_4$ comes directly from the quartic interaction. 
A locally preferred hairy branch
has $\mathcal A>0$, which requires
\be 
\la_4>\la_4^t=-0.18294 \,.
\ee 
In this range, the order parameter has the mean-field exponent $1/2$, while the free-energy
difference begins quadratically in $T_c-T$.\footnote{When $\lambda_4<\lambda_4^t$, there exists a first-order phase transition without obvious scaling relations. We do not consider the case.}

\subsubsection{Scaling with fixed $T$ and $\kappa$}
We now keep $(T,\kappa,\lambda_4)$ fixed and vary the angular velocity
$\Omega$.  Its critical value $\Omega_c$ is defined by the critical line $T(\Omega)$ in \eqref{eq:rota-TOmega-Fc} with fixed $T$. 
We expand $\Omega$ as
\begin{equation}
\Omega=\Omega_c+\Omega_2\al^2+O(\al^4)\,,\qquad
\Omega_c=\sqrt{1-\left(\frac{2\pi \kappa_b^2T}{\kappa^2}\right)^2}\,.
 \label{eq:rota-Omega-expan}
\end{equation}
Start from the local hairy solution \eqref{eq:rota-O-scaling} in terms of $\al$,
\begin{equation}
 T_{\rm hairy}(\Omega,\al)
 =T_c(\Omega)-C_T(\Omega)\al^2+O(\al^4),
 \qquad
 C_T(\Omega)=\frac{\sqrt{1-\Omega^2}}{-\kappa}
 \mathcal A(\lambda_4).
 \label{eq:rota-T-scaling-alpha}
\end{equation}
Substituting 
\eqref{eq:rota-Omega-expan} into \eqref{eq:rota-T-scaling-alpha}, while imposing
$T_{\rm hairy}=T$, gives at $O(\al^2)$
\begin{equation}
T'(\Omega_c)\Omega_2=C_T(\Omega_c)\,,\qquad 
T'(\Omega_c)
 =-\frac{\kappa^2}{2\pi \kappa_b^2}
 \frac{\Omega_c}{\sqrt{1-\Omega_c^2}}<0\,,
\end{equation}
we obtain the simple scaling
\begin{equation}
 \Omega_c-\Omega=C_\Omega\al^2+O(\al^4),
 \qquad
 C_\Omega=\frac{2\pi \kappa_b^2(1-\Omega_c^2)}
 {|\kappa|^3\Omega_c}\mathcal A(\lambda_4)>0.
\label{eq:rota-Omega-scaling-alpha}
\end{equation}
Thus decreasing $\Omega$ through $\Omega_c$ moves the system onto the hairy side;
increasing it through $\Omega_c$ moves the system onto the BTZ side.

Eliminating $\al$ gives the scaling relations,
\begin{align}
 O
 &\simeq-\frac{1}{2\sqrt{C_\Omega}}
 (\Omega_c-\Omega)^{1/2}\,,
~~~~~~~~~
 \mathcal F_{\rm BTZ}-
 \mathcal F_{\rm hairy}
 \simeq\frac{\mathcal \pi \kappa_b^2 \mathcal{A}(\lambda_4)}{8\kappa^2C_\Omega^2}
 (\Omega_c-\Omega)^2\,.
\label{eq:rota-OF-omega}
\end{align}
The critical exponents are again $1/2$ for the order parameter and $2$ for the
free-energy difference.  
This derivation assumes an ordinary rotating hairy black hole,
$0<\Omega_c<1$.  The limit $\Omega_c\to1$ is the zero-temperature extremal limit; it is not considered in this paper. The endpoint $\Omega_c=0$ is also special because $T_c'(0)=0$. 

\subsection{Scaling of the Kasner epoch}

In the previous section we study the scaling relations of thermodynamic quantities of the boundary field theory. In this section we show the onset scalar can transmit into the black hole interior, analytically, and thus produces the exact scaling relation between the Kasner exponents and boundary quantities.

The onset mode \eqref{eq:rota-onset} derived in the previous section is regular at the outer horizon $z=z_+$.
It can be analytically continued through the black-hole interior, $z_+<z<z_-$, toward the inner Cauchy horizon $z=z_-$. 
The relevant leading continuation is
\begin{equation}
 {}_2F_1\left(a,a;2a;u(z)\right)
 \sim
 \frac{\Gamma(2a)}{\Gamma(a)^2}(-u(z))^{-a}
 \left[\log(-u(z))+2\frac{\Gamma'(1)}{\Gamma(1)}-2\frac{\Gamma'(a)}{\Gamma(a)}\right]\,,
 \qquad z\rightarrow z_-\,,
\label{eq:hypergeometricInfinity}
\end{equation}
with $a=3/8$ or $5/8$ and $u(z)$ defined in \eqref{eq:rota-uz}.
After using
the outer-horizon regularity relation in \eqref{eq:rotatingRegularityRatio}, the 
solution has the simple asymptotic form 
\begin{equation}
 \Phi(z)\simeq B_-\log\left(1-\frac{z}{z_-}\right)\,,\qquad B_-=-2\cos\!\left(\frac{3\pi}{8}\right)
 \frac{\Gamma(3/4)}{\Gamma(3/8)^2}(\frac{\kappa}{\kappa_b})^{-3/2}\,,
 \qquad z\rightarrow z_-,
 \label{eq:rota-Phi-z-}
\end{equation}
Note that the value of $B_-$ depends on the normalization of the onset mode, which we fix by setting $\Phi \sim z^{3/4}+\kappa z^{5/4}$ near $z\to0$. The physical scalar perturbation is $\phi=\alpha\Phi$.

The behavior of onset scalar \eqref{eq:rota-Phi-z-} near the inner horizon $z_-$ overlaps with the ER collapse, and thus gives the data before ER collapse,
\be 
|\phi'|_{\rm bef}=\al |\Phi'| \simeq \frac{\al |B_-|}{|z_--z|} \,.
\label{eq:rota-phi'}
\ee 
Comparing the dynamical equations with the general results in Sec.~\ref{sec:general_mechanism}, besides \eqref{eq:relation1} and \eqref{eq:app_ER_explicit_sources}, we have 
\begin{align}
\begin{split}
d&=1, \qquad
Z(z)=1,
\qquad
{\cal U}(z)=m^2,
\qquad
{\cal B}_-=\alpha B_-
\\
{\cal G}
&=
-\frac{f e^{-\chi}}{z^2}\,,
\qquad
{\cal Y}=e^{-\chi/2},
~~~~~~~~
\Pi_\phi
=
\frac{f e^{-\chi/2}}{z}\phi',
\\
{\cal R}_\chi&={\cal R}_f=0\,,
\qquad
{\cal S}
=
\frac{V(\phi)}{z}
+
\frac{J^2z^3}{2}.
\label{eq:BTZ_general_dictionary}
\end{split}
\end{align}
One can check that 
the assumptions \eqref{eq:app_ER_regular_sources} and \eqref{eq:app_ER_remainders} apply. The rotating model therefore satisfies
the conditions required for the standard ER completion discussed in
Sec.~\ref{subsec:general_ER}.
Thus we get the scalar data after the ER bridge collapse directly from \eqref{eq:general_ER_amplitudes}, 
\be 
|\phi'|_{\rm aft}
\simeq
\frac{1}{\al}\frac{2}{|B_-| z_-} \,.
\label{eq:rota-phi'-after}
\ee 

It is well known that the interior of hairy rotating black hole enters a Kasner epoch after the ER collapse, with the metric and scalar following the Kasner form,  
\be 
\dd s^2=-\dd \tau^2+ \tau^{2p_t}\dd t^2+\tau^{2p_x}(N_K\dd t+\dd x)^2\,,\qquad
\phi=v \log z=-\sqrt{2}p_\phi \log \tau\,.
\label{eq:Kas-rota}
\ee 
The constant $N=N_K$ can be locally absorbed into the redefinition $\tilde{x}=x+N_K t$, reducing to the standard Kasner form.\footnote{In \cite{Gao:2023rqc} we numerically find that $N_K$ obeys a relation with $\Omega$, 
$N_K=-{1}/{\Omega}\,,$
which could establish a simple connection between the horizon and singularity. In this paper, we give an analytical proof in Appendix \ref{app:shift_relation}.} The Kasner exponents are
\be 
p_t=\frac{v^2}{v^2+4}\,,\qquad
p_x=\frac{4}{v^2+4}\,,\qquad
p_\phi=\frac{2\sqrt{2}v}{v^2+4}\,,
\label{eq:rota-Kas-expo}
\ee 
which follows the Kasner relation $p_t+p_x=1\,,p_t^2+p_x^2+p_\phi^2=1$. The only independent degree of freedom is characterized by the scalar velocity $v$.

As discussed in Sec.~\ref{sec:general_mechanism}, close to the critical line $T(\Omega)$, 
the interior enters the Kasner regime \eqref{eq:Kas-rota} immediately after the ER collapse. The scalar data \eqref{eq:rota-phi'-after} thus serves as the input of the Kasner epoch, which gives
\be 
|v|\simeq
z_- |\phi'|_{\rm aft}
\simeq
\frac{1}{\al}\frac{2}{|B_-|}\,.
\ee 
Substitution into \eqref{eq:rota-Kas-expo} gives the scaling relations of Kasner exponents,
\be 
1-p_t= p_x \simeq B_-^2 \al^2=4 B_-^2 O^2\,,\qquad
|p_\phi|\simeq \sqrt{2}|B_-| \al=2\sqrt{2}|B_- O|\,,
\label{eq:rota-Kasner-scaling}
\ee 
where the equality uses $\al=-2 O$.
Combining the temperature scaling \eqref{eq:rota-T-scaling-alpha} and the angular velocity scaling \eqref{eq:rota-Omega-scaling-alpha} gives
\begin{align}
 1-p_t = p_x \simeq \frac{|p_\phi|^2}{2}
 &\simeq
 \frac{|\kappa| B_-^2}
 {\sqrt{1-\Omega^2}\,\mathcal A(\lambda_4)}
 (T_c-T),
 \qquad ~~~~~~~~\text{fixed } \Omega,
 \\
 1-p_t = p_x \simeq \frac{|p_\phi|^2}{2}
 &\simeq
 \frac{|\kappa|^3\Omega_c B_-^2}
 {2\pi\kappa_b^2(1-\Omega_c^2)\,\mathcal A(\lambda_4)}
 (\Omega_c-\Omega),
 \qquad \text{fixed } T.
\end{align}
The scaling relations with respect to the order parameter $O$ are universal, regardless of the path of approach to the critical line $T(\Omega)$, while the scaling relations with phase parameters $T$ and $\Omega$ are different.


\subsection{Numerical verification}
\label{sec:numerical-verification}

We now examine our analytical results in the two models: (i), fixed $\Omega$ and $\kappa$ with varying $T$, and (ii), fixed $T$ and $\kappa$ with varying $\Omega$.
The model parameters are chosen to be
\be 
m^2=-15/16\,,\qquad
\kappa=-1\,,\qquad
\lambda_4=1/5\,.
\label{eq:rota-para-model}
\ee 
Actually, we have chosen $m^2=-15/16$ to simplify the analytical analysis by avoiding logarithmic terms at the boundary
$z=0$.
The choice of $\kappa$ satisfies  $-1.2<\kappa<\kappa_b$, allowing for a local phase transition from BTZ to hairy rotating black hole, as discussed in Sec.~\ref{sec:phase-double-trace}. 
The quartic coupling in the potential is chosen to satisfy $\lambda_4>\lambda_4^t$, allowing for a second-order phase transition, as discussed in Sec.~\ref{sec:scalings of thermodynamic quantities}.

We consider non-extremal rotating black holes. The critical point is chosen to be
\be 
\qquad
T_c=T_c(\Omega_c)\simeq 0.56223\,,\qquad
\Omega_c=\frac{1}{2}\,.
\label{eq:rota-para-phase}
\ee
We denote the difference between a quantity $x$ and its critical value $x_c$ by $\Delta x\equiv x_c-x$, while for the free energy we define $\dF \equiv \mathcal{F_{\BTZ}}-\mathcal{F}$.
The corresponding analytical scaling relations of the order parameter $O$ and free energy $\mathcal{F}$ in model (i) and (ii) are
\begin{align}
 (\rm i)&:\quad {-O}\simeq 2.1182\,\dT^{1/2}\,,
 \qquad
 \Delta\mathcal F\simeq 1.9950\,\dT^2
 \label{eq:OF-T-analy}\\
 (\rm ii)&:\quad {-O}\simeq 1.2968\,\dOme^{1/2},\qquad
 \Delta\mathcal F \simeq 0.28028\,\dOme^2\,.
 \label{eq:OF-omega-analy}
\end{align}
Both of them show again that the hairy branches exist at $T<T_c$ or $\Omega<\Omega_c$ and are thermodynamically preferred, at least near the critical point.

\begin{figure}[h!]
\begin{center}
\includegraphics[width=0.43\textwidth]{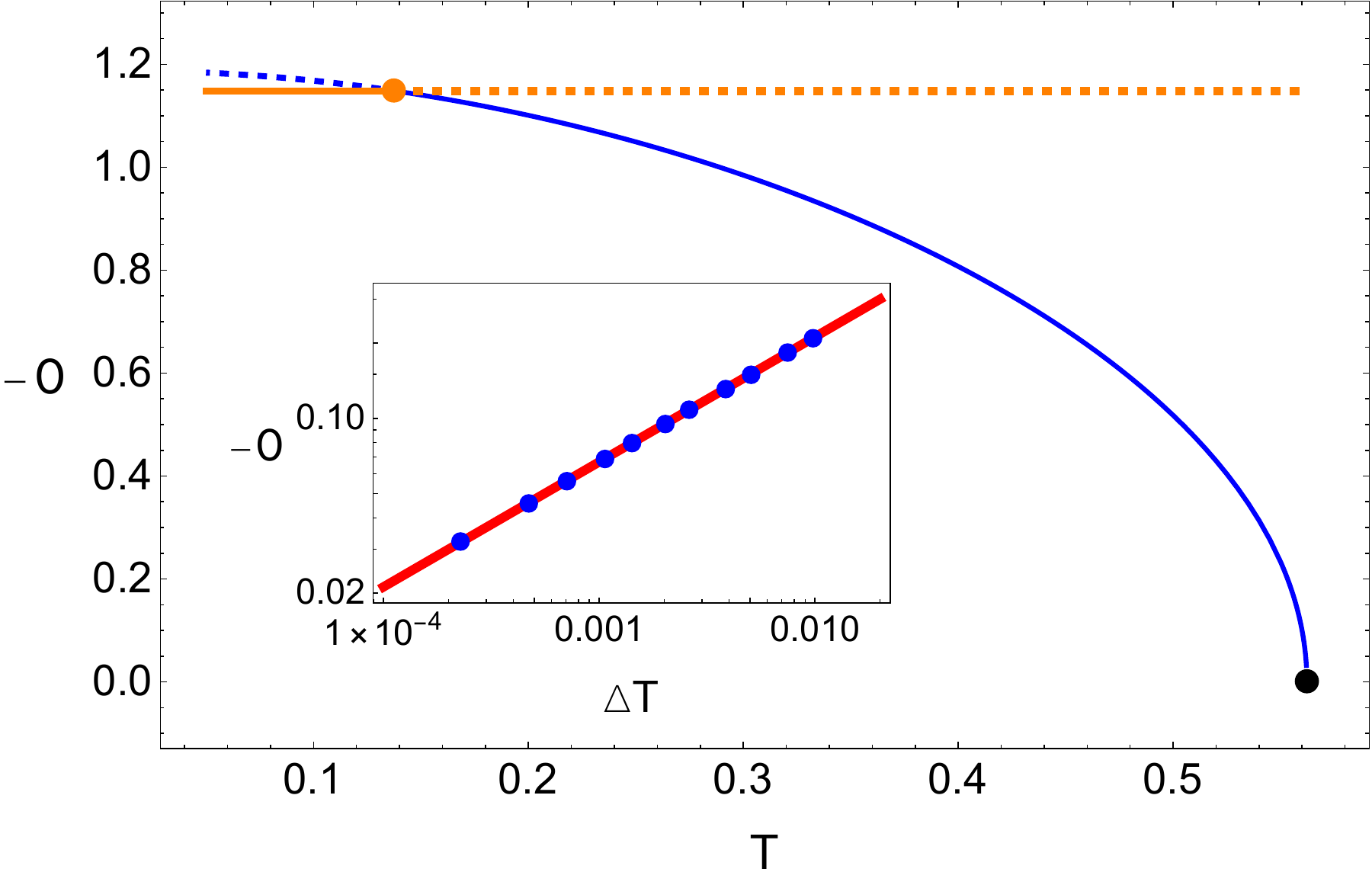}
\hspace{0.04\textwidth}
\includegraphics[width=0.43\textwidth]{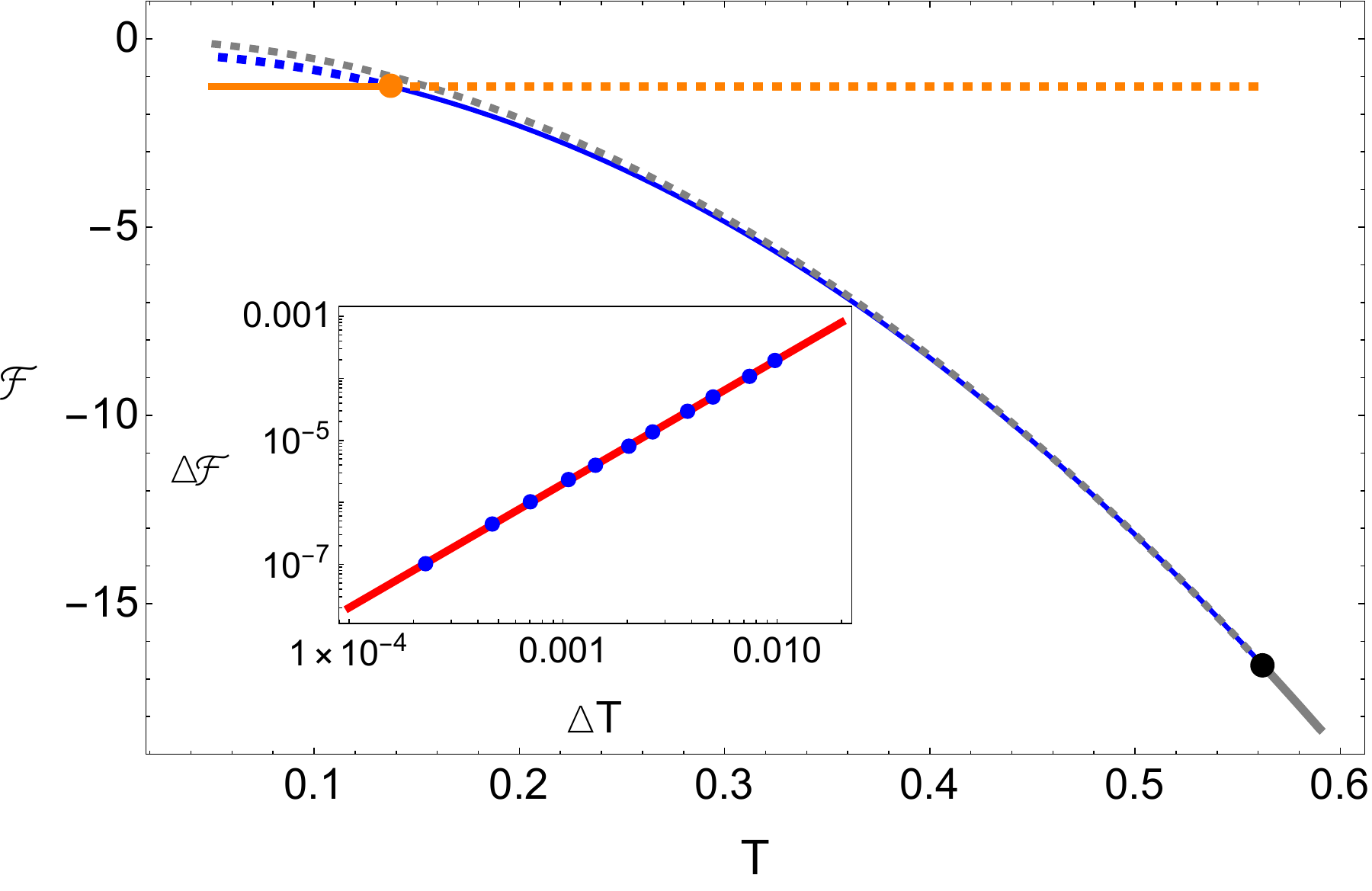}
\end{center}
\vspace{-0.5cm}
\caption{\small
Thermodynamic phase diagram in Model (i) at fixed angular velocity $\Omega=\Omega_c$ and $\kappa=-1$  with varying temperature $T$.
{\em Left}: condensate ${O}$ as a function of $T$ for different phases.  
{\em Right}: grand-potential density $\mathcal{F}$ versus $T$.
In both panels, hairy black hole, boson star, and $\BTZ$ black hole are shown in blue, orange, and gray lines, respectively.
Solid lines denote the thermodynamically preferred phase, while dashed lines denote subdominant branches. The insets highlight the continuous approach of the hairy black hole to $\BTZ$ branch at $T_c$ (black dot), with analytical predictions \eqref{eq:OF-T-analy} (red), and numerical results (blue dots). The orange dot denotes the first-order transition  at $T_\star$ between the hairy black hole and boson star.
}
\label{fig:rota-scaling-num}
\end{figure}

Fig.~\ref{fig:rota-scaling-num} shows the thermodynamic diagrams in model (i). There are three different phases: $\BTZ$ black hole at high temperature above $T_c$, hairy rotating black hole at intermediate temperature, $T_\star<T<T_c$, and boson star at low temperature below $T_\star$. We focus on the continuous phase transition between the hairy and $\BTZ$ black hole at $T_c$. The numerical results of scalings of thermodynamic quantities are,
\be 
{-O} \simeq 2.1126\, \dT^{\,0.49967}\,,\qquad
\Delta{\cal F}\simeq1.9857\,\dT^{\,1.9994}\,,
\ee 
which are in good agreement with the analytical results \eqref{eq:OF-T-analy}.\footnote{As one approaches the critical point $T_c$, the relative error becomes increasingly small, further supporting the validity of our analytical predictions.}
We do not focus on the first-order phase transition between the boson star and hairy black hole at $T_\star$, whose value is determined only numerically.

\begin{figure}[h!]
\begin{center}
\includegraphics[width=0.43\textwidth]{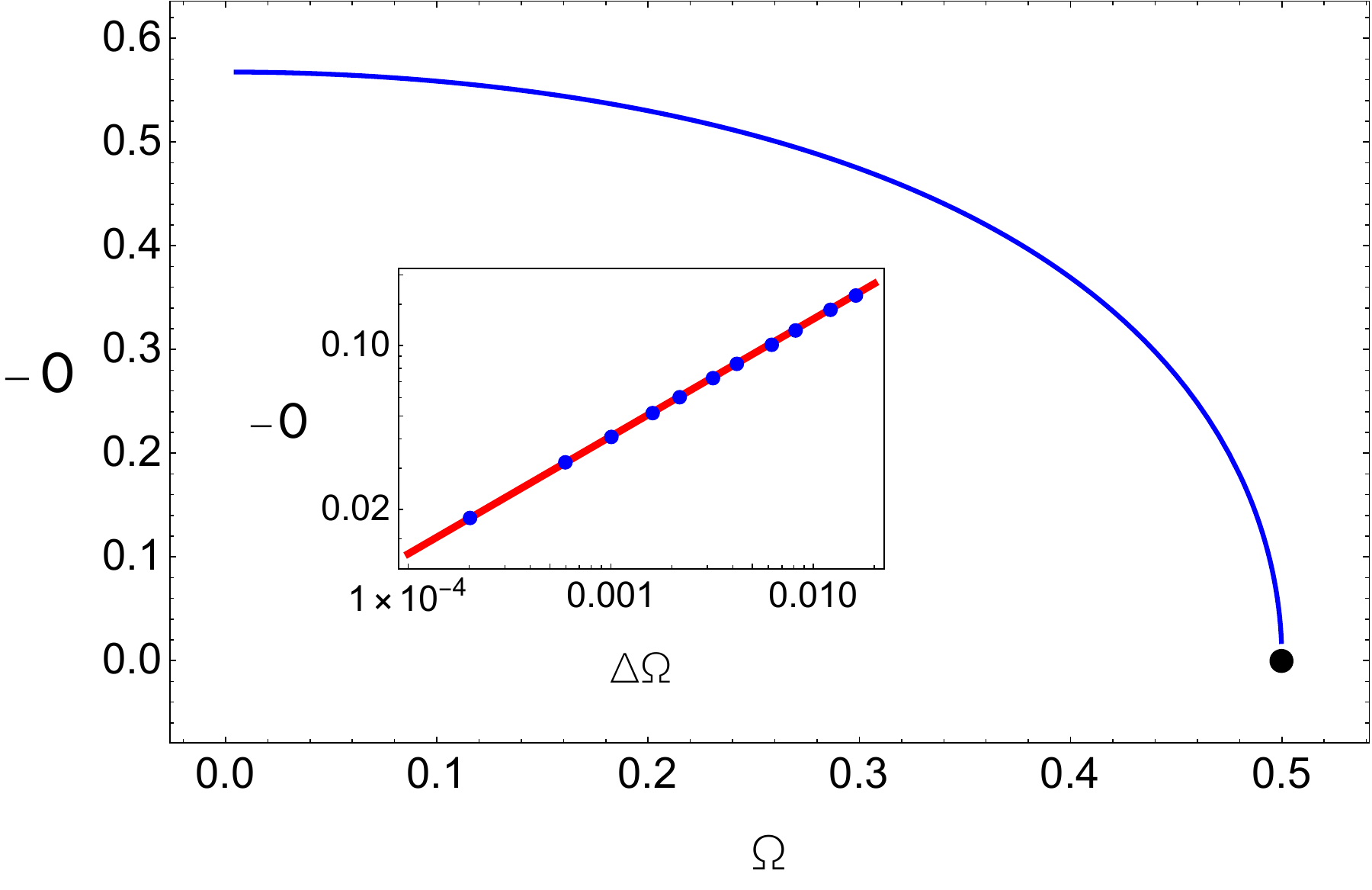}
\hspace{0.04\textwidth}
\includegraphics[width=0.43\textwidth]{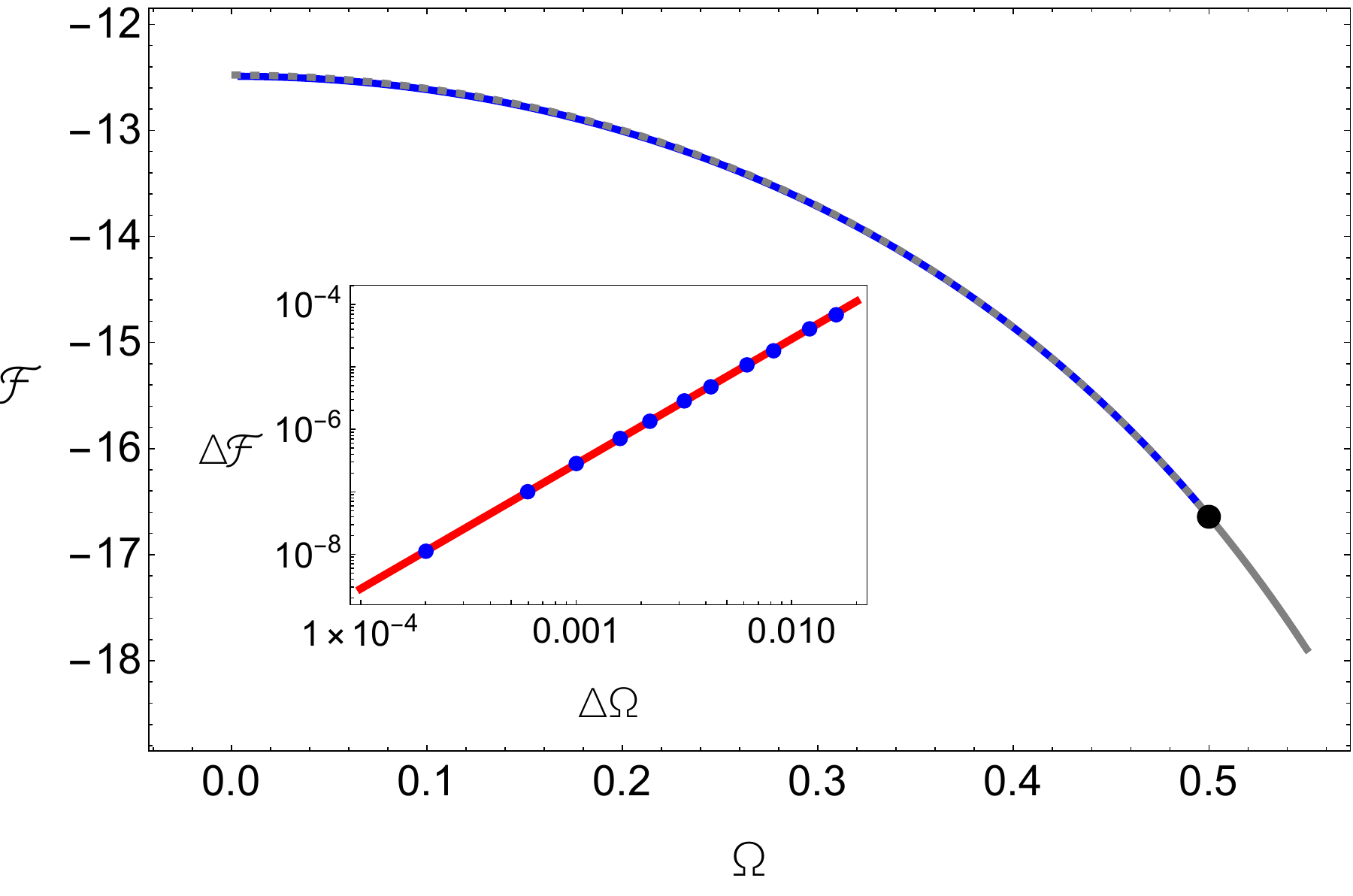}
\end{center}
\vspace{-0.5cm}
\caption{\small
Thermodynamic phase diagram in Model (ii) at fixed $T=T_c$ and $\kappa=-1$ with varying $\Omega$.
{\em Left}: condensate ${O}$ versus $\Omega$.
{\em Right}: grand-potential density $\mathcal{F}$ versus $\Omega$.
In both panels, hairy black hole and $\BTZ$ black hole are shown in blue and gray lines, respectively.
Solid and dashed lines denote the thermodynamically preferred and subdominant phases, respectively. The insets show the approach of hairy black hole to $\BTZ$ branch at $\Omega_c$ (black dot), with analytical predictions \eqref{eq:OF-omega-analy} (red), and numerical results (blue dots).
}
\label{fig:rota-scaling-fixT-num}
\end{figure}

Fig.~\ref{fig:rota-scaling-fixT-num} shows the thermodynamic diagrams in model (ii). There are two phases: $\BTZ$ black hole at high angular velocity $\Omega>\Omega_c$, and hairy rotating black hole at  $\Omega<\Omega_c$. 
The numerical results of scaling relations are
\be 
{-O} \simeq 1.2854\, \dOme^{\, 0.49891}\,,\qquad
\Delta{\cal F}\simeq 0.27768\,\dOme^{1.9990}\,,
\ee 
which are again in good agreement with the analytical results \eqref{eq:OF-omega-analy}.
We do not show the boson star branch, since it is always thermodynamically unstable compared to the two phases.

We now turn to the scaling of black hole singularity, which is characterized by the independent Kasner exponent $p_t$.\footnote{Note that near criticality, the shrinking radial scale of the ER collapse and the rapid suppression of the longitudinal metric coefficient make a reliable numerical extraction of this rate challenging, therefore, the ER collapse rate is derived analytically but is not extracted independently from the nonlinear numerical solutions.} Under the choice of the model parameters \eqref{eq:rota-para-model}, and the critical point \eqref{eq:rota-para-phase}, 
the analytical predictions are
\begin{align}
({\rm i}):&\quad
    \dpt\simeq 0.013527\, O^2\,,\qquad \dpt\simeq 0.060692\,\dT\,,\qquad 
    \text{fixed $\Omega$}
    \label{eq:rota-dpt-fixed-omega-analy}
    \\
({\rm ii}):&\quad
    \dpt\simeq 0.013527\, O^2\,,\qquad \dpt\simeq  0.022748\, \Delta \Omega\,,\qquad
    \text{fixed $T$}
    \label{eq:rota-dpt-fixedT-analy}
\end{align}
Note that the scaling relation with the order parameter $O$ are universal, regardless of the path of approach to the critical point,  while the scaling with the varying phase parameters, such as $T$ or $\Omega$, are path-dependent.

\begin{figure}[h!]
\begin{center}
\includegraphics[width=0.435\textwidth]{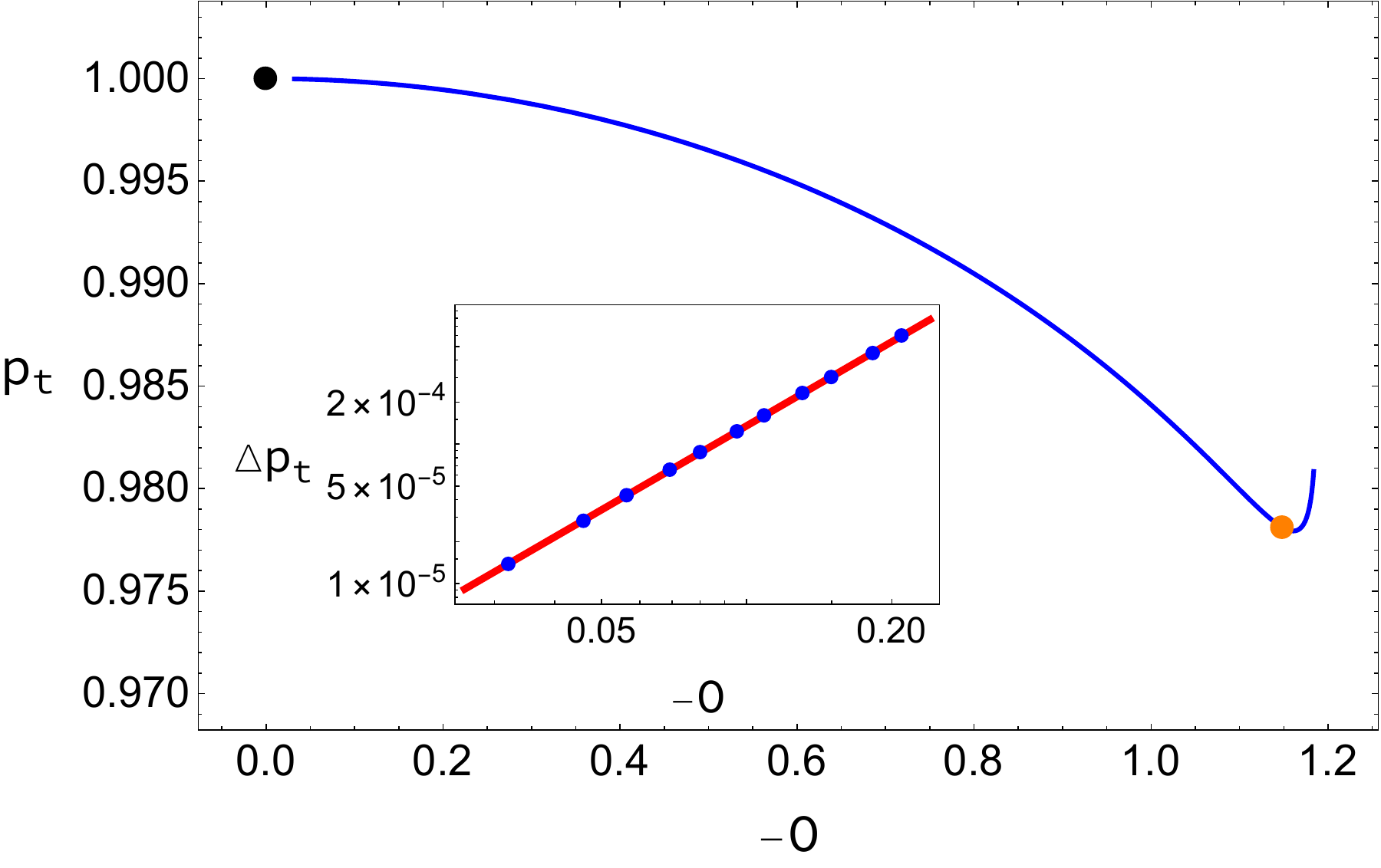}
\hspace{0.04\textwidth}
\includegraphics[width=0.43\textwidth]{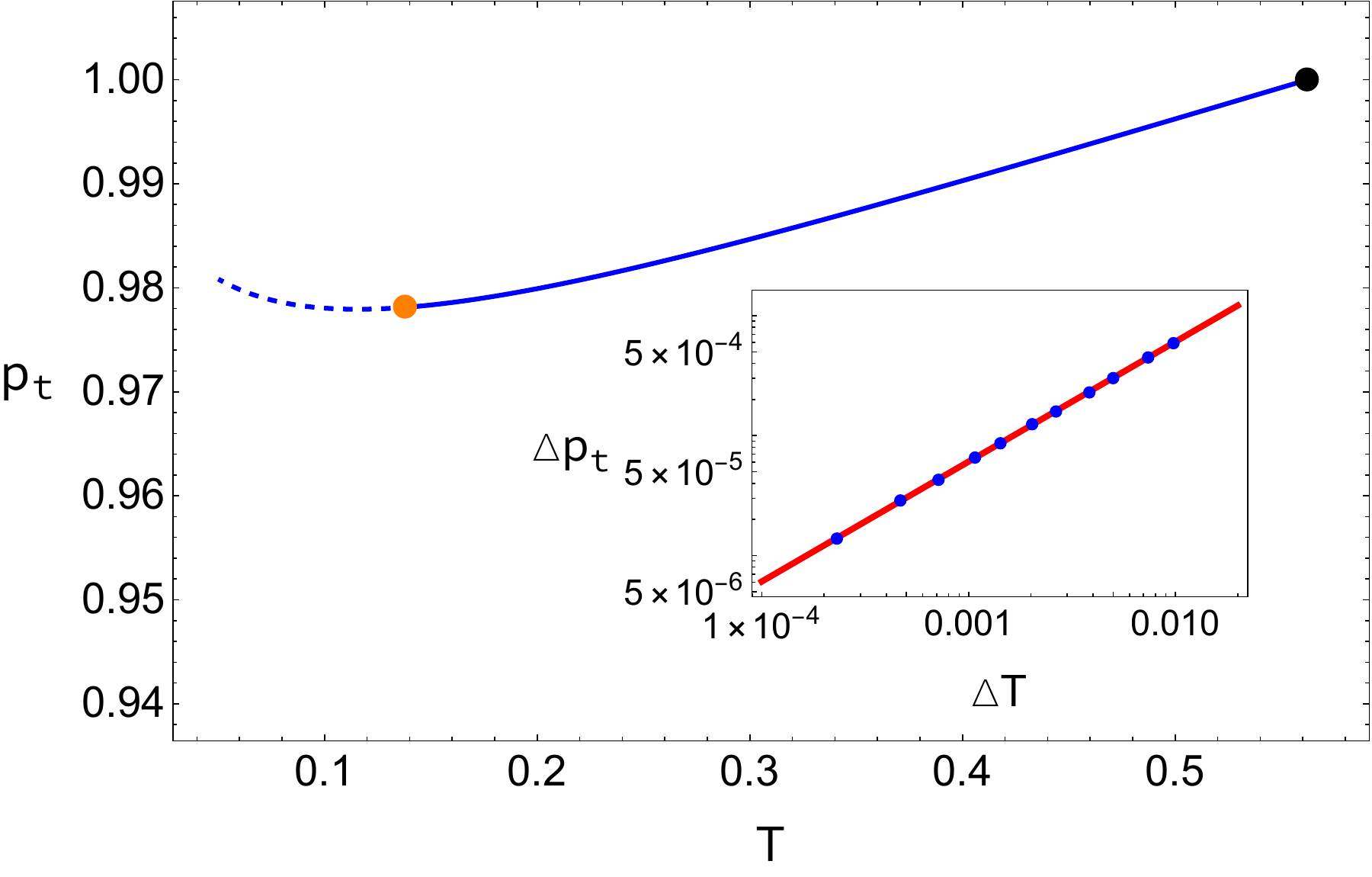}
\end{center}
\vspace{-0.5cm}
\caption{\small The plots of Kasner exponent $p_t$ (blue lines) in the phase diagram of model (i). {\em Left}: $p_t$ versus $O$. {\em Right}: $p_t$ versus $T$.  The black dot signals the phase transition at $T_c$ between BTZ and hairy black hole. The orange dot signals the phase transition at $T_\star$ between hairy branch and boson star. The insets show the scaling behavior near $T_c$, with analytical predictions \eqref{eq:rota-dpt-fixed-omega-analy}(red lines), and numerical results (blue dots).
}
\label{fig:rota-pt-fix-omega}
\end{figure}

\begin{figure}[h!]
\begin{center}
\includegraphics[width=0.432\textwidth]{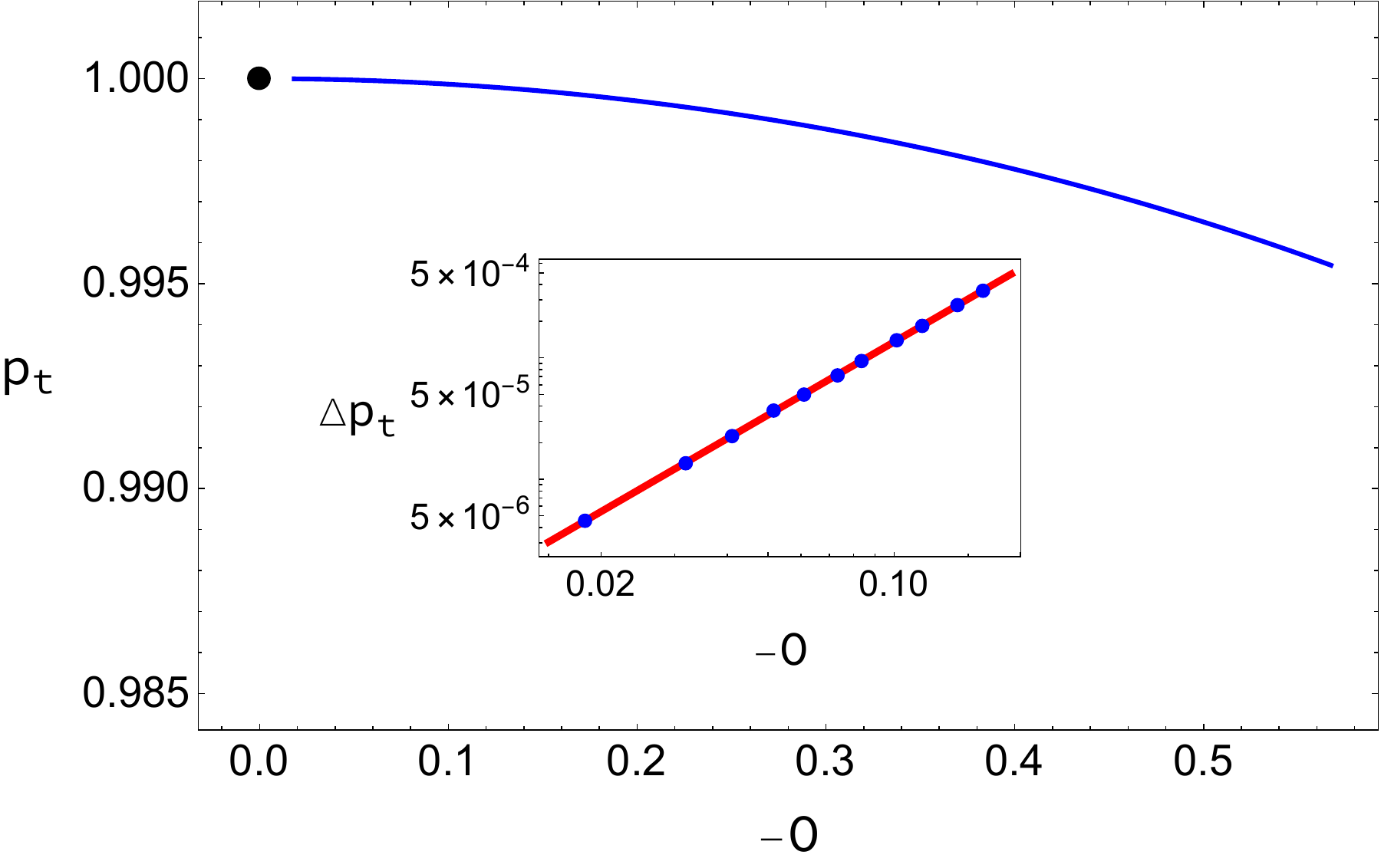}
\hspace{0.04\textwidth}
\includegraphics[width=0.43\textwidth]{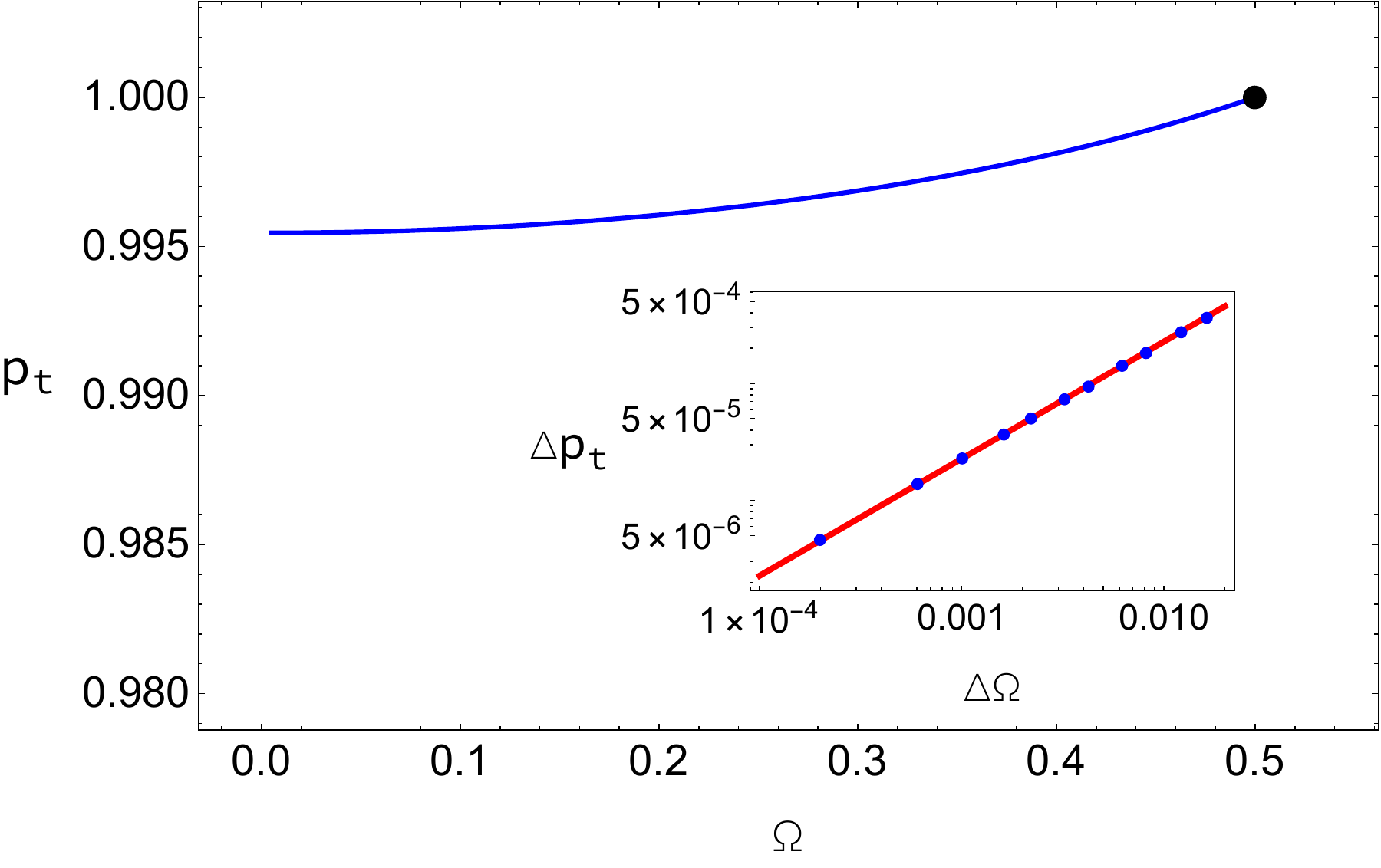}
\end{center}
\vspace{-0.5cm}
\caption{\small The plots of Kasner exponent $p_t$ (blue lines) in the phase diagram of model (ii). {\em Left}: $p_t$ versus $O$. {\em Right}: $p_t$ versus $\Omega$. 
The black dot signals the phase transition at $\Omega_c$ between BTZ and hairy black hole. The insets show the scaling behavior with analytical predictions \eqref{eq:rota-dpt-fixedT-analy}(red lines), and numerical results (blue dots).
}
\label{fig:rota-pt-fix-T}
\end{figure}

Figs.~\ref{fig:rota-pt-fix-omega} and \ref{fig:rota-pt-fix-T} show the plot of $p_t$ with respect to $O$ and $T$ at fixed $\Omega_c$, and  with $O$ and $\Omega$ at fixed $T_c$, in models (i) and (ii), respectively. There are clear scaling relations with the numerical fits
\begin{align}
({\rm i}):&\quad
    \dpt\simeq 0.013568\, |O|^{2.0009}\,,\qquad \dpt\simeq 0.060674\,\dT^{0.99996}\,,\qquad 
    \text{fixed $\Omega$}
    \label{eq:rota-dpt-fixed-omega-num}\\
({\rm ii}):&\quad
    \dpt\simeq 0.013552\, |O|^{2.0005}\,,\qquad \dpt\simeq  0.022329\, \Delta \Omega^{0.99763}\,,\qquad
    \text{fixed $T$}
\end{align}
These numerical results are again in good agreement with the analytical predictions \eqref{eq:rota-dpt-fixed-omega-analy} and \eqref{eq:rota-dpt-fixedT-analy}. In particular, they confirm the universal scaling relation between the Kasner exponent $p_t$ and the order parameter $O$.

\section{Critical interior scaling in scalarized charged black holes}
\label{sec:charged_RN}

Having illustrated the interior scaling from Cauchy-horizon matching mechanism in rotating
BTZ, we now turn to a four-dimensional charged realization based on
RN-AdS$_4$.  This example further confirms that the critical zero
mode near phase transition has a nonvanishing Cauchy-horizon coefficient 
$B_-$ and provides
a setting in which the resulting Einstein--Rosen and first-Kasner
predictions can be tested through fully nonlinear scalarized black
branes.

\subsection{Einstein--Maxwell--real scalar model}
We consider the generalized holographic model \cite{Aprile:2010yb} which is an Einstein-Maxwell real-scalar system 
\begin{align}
S=\frac{1}{16\pi G_N}\int d^4x\sqrt{-g}
\left[
R-\frac{1}{4}G(\phi)F^{ab}F_{ab}
-\frac{1}{2}(\nabla\phi)^2
-V(\phi)
\right],
\label{eq:action}
\end{align}
where $G(\phi)$, $V(\phi)$ describe the gauge kinetic coupling, the scalar potential respectively. 
We again set $16\pi G_N=L=1$ for convenience. The AdS$_4$ vacuum satisfies 
$V(0)=-6$.

The scalar is real and  neutral under the bulk gauge field, therefore the
Maxwell equation admits a conserved electric flux. 
We focus on the following particular model\footnote{More general choices may include nonanalytic powers such as
$|\phi|^q$ \cite{Aprile:2010yb}, or exponential and super-exponential
large-field behavior in $G(\phi)$ and $V(\phi)$
\cite{Cai:2023igv}. Provided that their expansions agree through
quadratic order at $\phi=0$, these modifications leave the linearized
instability unchanged but can affect both the nonlinear critical
behavior and the deep-interior dynamics once the scalar becomes large;
see Appendix~\ref{subsec:Kasner_dynamics}.} 
\begin{align}
G(\phi)
&=1+g_2\phi^2+g_4\phi^4\,,
~~~~~~~~~~
V(\phi)
=-6+\frac{1}{2}m^2\phi^2+\lambda_4\phi^4\,.
\label{eq:model1}
\end{align}
Both functions $G(\phi)$ and $V(\phi)$ are analytic even functions around $\phi=0$. 
We choose
$m^2=-2, g_2=0.6$, thus the effective mass of the scalar perturbation violates the zero temperature near-horizon AdS$_2$ BF bound, triggering the scalarization instability  \cite{Hartnoll:2008kx}. 

We consider the static and homogeneous black brane ansatz
\begin{align}
\begin{split}
ds^2
&=
\frac{1}{z^2}
\left[
-f(z)e^{-\chi(z)}dt^2
+\frac{dz^2}{f(z)}
+dx^2+dy^2
\right],
\\
A&=\psi(z)dt,
\qquad
\phi=\phi(z).
\label{eq:ansatz}
\end{split}
\end{align}
The asymptotic AdS boundary is located at $z=0$, and the event
horizon at $z=z_h$, where $f(z_h)=0$.

For the ansatz~\eqref{eq:ansatz}, the Maxwell equation can be integrated once,
\begin{equation}
\psi'
=
\frac{E}{G(\phi)}e^{-\chi/2},
\label{eq:maxwell_first_integral}
\end{equation}
where $E$ is the conserved electric flux.
The remaining equations of motion can be written in a
convenient form as 
\begin{align}
\chi'
&=
\frac{z}{2}\phi'^2,
\label{eq:chi_eom}
\\
\left(
\frac{e^{-\chi/2}f}{z^3}
\right)'
&=
\frac{e^{-\chi/2}}{4z^4}
\left[
\frac{z^4E^2}{G(\phi)}
+2V(\phi)
\right],
\label{eq:f_eom}
\\
\left(
\frac{e^{-\chi/2}f\phi'}{z^2}
\right)'
&=
\frac{e^{-\chi/2}}{z^4}
\left[
\partial_\phi V
-\frac{z^4E^2}{2}
\frac{\partial_\phi G}{G(\phi)^2}
\right].
\label{eq:phi_eom}
\end{align}
These equations, together with
Eq.~\eqref{eq:maxwell_first_integral}, are radial equations that
we use to construct the black-hole solutions and follow their evolution across the horizon.

Near the AdS boundary $z\to 0$, the fields have the asymptotic expansions 
\begin{align}
\begin{split}
\phi
&=
\phi_1 z+\phi_2 z^2+\cdots ,
~~~~~~~~~~~~~~~~~~~~~~
\chi
=
\chi_0+\frac{\phi_1^2}{4}z^2+\cdots ,
\\
f
&=
1+\frac{\phi_1^2}{4}z^2+f_3 z^3+\cdots ,
~~~~~~~~~~~~~
\psi
=
\mu-\rho z+\cdots .
\label{eq:UV_expansion}
\end{split}
\end{align}
Here $\mu$ and $\rho$ are the chemical potential and charge density,
and Eq.~\eqref{eq:maxwell_first_integral} gives
$\rho=-E e^{-\chi_0/2}$.  The residual rescaling of the time
coordinate will be used to set $\chi_0=0$. Since $m^2=-2$, both standard and alternative quantizations are allowed.
Throughout this work, we adopt the standard quantization and impose the
source-free boundary condition $\phi_1=0$, so that
$\phi_2\propto O$.

At a regular nonextremal event horizon $z=z_h$, regularity requires
$f(z_h)=0$ and $\psi(z_h)=0$.  Writing
$\phi_h\equiv\phi(z_h)$ and $\chi_h\equiv\chi(z_h)$, the leading
near-horizon expansion is fixed by  
\begin{align}
\begin{split}
f'(z_h)
&=
\frac{1}{4z_h}
\left(
\frac{z_h^4E^2}{G_h}
+
2V_h
\right),
~~~~~~~
\phi'(z_h)
=
\frac{1}{z_h^2 f'(z_h)}
\left(
\partial_\phi V (z_h)
-
\frac{z_h^4E^2\partial_\phi G (z_h)}{2G_h^2}
\right),
\\
\chi'(z_h)
&=
\frac{z_h}{2}\phi'(z_h)^2,
\qquad
~~~~~~~~~~~~~~~~
\psi'(z_h)
=
\frac{E}{G_h}e^{-\chi_h/2},
\label{eq:horizon_data}
\end{split}
\end{align}
where all matter couplings carrying a subscript $h$ are evaluated at
$\phi=\phi_h$.  The Hawking temperature is
\begin{equation}
T
=
-\frac{e^{-\chi_h/2}}{4\pi}f'(z_h).
\label{eq:temperature}
\end{equation}

The equations possess two residual scaling symmetries.  A constant
rescaling of the time coordinate is used to impose $\chi(0)=0$, while
the overall AdS scaling
$(t,x,y,z)\rightarrow b(t,x,y,z)$ may be used to fix one
dimensionful scale, for example the ensemble parameter $\mu=1$, or $z_h=1$ in the numerical construction.

The reduced radial system also admits the conserved quantity
\begin{equation}
{\cal Q}
=
\frac{e^{-\chi/2}}{z^2}
\left(
f'-f\chi'
\right)
-
E\psi,
\qquad
{\cal Q}'=0 .
\label{eq:radial_charge}
\end{equation}
Evaluating it at the AdS boundary and at the event horizon gives
\begin{equation}
e^{-\chi_0/2}
\left(
3f_3-2\phi_1\phi_2
\right)
-
E\mu
=
-Ts,
\qquad
s=\frac{4\pi}{z_h^2}.
\label{eq:radial_charge_relation}
\end{equation}
Further discussion on holographic renormalization and thermodynamics can be found in Appendix~\ref{app:holo-ren}. 

\subsection{Perturbative analysis near the critical point}
\label{sec:critical_perturbation}

We now analyze the bifurcation of the nodeless scalar-haired branch from RN-AdS$_4$.
Throughout this section we set $
g_2=0.6,
$ and focus on the grand canonical ensemble, i.e. $\mu=1$
. We consider the following two models used later in the numerical analysis, 
\begin{align}
\begin{split}
{\cal M}_{\rm I}:&\qquad
G(\phi)=1+0.6\phi^2,
\qquad
V(\phi)=-6-\phi^2+10\phi^4,
\\
{\cal M}_{\rm II}:&\qquad
G(\phi)=1+0.6\phi^2,
\qquad
V(\phi)=-6-\phi^2 .
\label{eq:critical_models}
\end{split}
\end{align}
The two theories have identical quadratic expansions around
$\phi=0$.  They therefore share the same linear critical mode and
the same critical RN-AdS$_4$ geometry, while their nonlinear
departures from the critical point are different.

\subsubsection{Expansion around the critical RN solution}
\label{subsec:critical_expansion}

To keep the event horizon at a fixed coordinate position, in this subsection we use
\begin{equation}
u=\frac{z}{z_h},
\label{eq:u_coordinate}
\end{equation}
where $z_h$ is the outer horizon of hairy black hole. Now the AdS boundary and the event horizon are at $u=0$ and
$u=1$, respectively. 
We work in 
$\psi(0)=1,
~
\psi(1)=0 .
$

At the bifurcation point the scalar vanishes and the background is
RN-AdS$_4$, then $z_h=z_c$ with $z_c$ the event horizon position of this
critical RN solution, 
\begin{align}
\begin{split}
\phi_0=0,
~~~~
\chi_0=0,
~~~~~
\psi_0=1-u,
~~~~
f_0(u)
=
1-
\left(
1+\frac{z_c^2}{4}
\right)u^3
+
\frac{z_c^2}{4}u^4\,,
\label{eq:critical_RN}
\end{split}
\end{align}
Its temperature is
\begin{equation}
T_c
=
\frac{1}{4\pi z_c}
\left(
3-\frac{z_c^2}{4}
\right).
\label{eq:Tc_RN}
\end{equation}

In standard quantization the normalizable scalar mode behaves as
\begin{equation}
\phi(u)
=
\epsilon u^2+\cdots ,
\qquad
u\rightarrow0,
\end{equation}
which defines the perturbative amplitude $\epsilon$.  Because the
theory is invariant under $\phi\rightarrow-\phi$, the weakly
nonlinear expansion takes the form
\begin{align}
\begin{split}
\phi
&=
\epsilon\Phi
+
\epsilon^3\phi_3
+
\mathcal O(\epsilon^5),
~~~~
f
=
f_0
+
\epsilon^2f_2
+
\mathcal O(\epsilon^4),
~~~~~
\chi
=
\epsilon^2\chi_2
+
\mathcal O(\epsilon^4),
\\
\psi
&=
\psi_0
+
\epsilon^2\psi_2
+
\mathcal O(\epsilon^4),
~~~~~
z_h
=
z_c
+
\epsilon^2z_2
+
\mathcal O(\epsilon^4).
\label{eq:critical_expansions}
\end{split}
\end{align}
We normalize the zero mode by
\begin{equation}
\Phi(u)
=
u^2+\mathcal O(u^3),
\qquad
\phi_3=o(u^2),
\qquad
u\rightarrow0 .
\label{eq:critical_zero_mode_norm}
\end{equation}

\subsubsection{Critical zero mode}
\label{subsec:linear_zero_mode}

At order $\mathcal O(\epsilon)$ the scalar decouples from the metric
and Maxwell perturbations.  Its equation is
\begin{equation}
\left(
\frac{f_0}{u^2}\Phi'
\right)'
+
\left(
\frac{2}{u^4}
+
g_2z_c^2
\right)\Phi =0 .
\label{eq:linear_zero_mode}
\end{equation}
Here the prime denotes the derivatives with respect to 
$u$. Regularity at $u=1$ gives
\begin{equation}
\Phi'(1)
=
\frac{
2+g_2z_c^2
}{
3-z_c^2/4
}
\Phi(1).
\label{eq:critical_horizon_bc}
\end{equation}

Together with the UV normalization in
Eq.~\eqref{eq:critical_zero_mode_norm}, this is an eigenvalue problem
for $z_c$.  For $g_2=0.6$, the nodeless solution which has the lowest free energy \cite{Horowitz:2010gk} gives
\begin{equation}
z_c
=
2.3589,
\qquad
\Phi(1)
=
1.5813.
\label{eq:critical_linear_numeric}
\end{equation}
Equation~\eqref{eq:Tc_RN} then gives
\begin{equation}
T_c
=
0.054273.
\label{eq:critical_Tc_numeric}
\end{equation}

Note that the quartic term in $V(\phi)$ does not enter
Eq.~\eqref{eq:linear_zero_mode}.  Consequently
${\cal M}_{\rm I}$ and ${\cal M}_{\rm II}$ in \eqref{eq:critical_models} have exactly the same
critical temperature and the same linear zero mode.

\subsubsection{Quadratic backreaction and cubic solvability}
\label{subsec:quadratic_backreaction}

The first backreaction appears at order $\epsilon^2$.  The equation \eqref{eq:chi_eom} gives 
\begin{equation}
\chi_2'
=
\frac{u}{2}\Phi'^2,
\qquad
\chi_2(u)
=
\frac12
\int_0^u d\sigma\,\sigma\,\Phi'(\sigma)^2 .
\label{eq:chi2_solution}
\end{equation}
The Maxwell equation gives
\begin{equation}
\psi_2'
=
C_E
+
\frac{\chi_2}{2}
+
g_2\Phi^2 .
\label{eq:psi2_prime}
\end{equation}
Imposing $\psi_2(0)=\psi_2(1)=0$ fixes
\begin{equation}
C_E
=
-
\int_0^1du
\left(
\frac{\chi_2}{2}
+
g_2\Phi^2
\right).
\label{eq:CE}
\end{equation}
For the zero mode above,
\begin{equation}
\chi_2(1)
=
1.7940,
\qquad
C_E
=
-0.34043.
\label{eq:quadratic_numeric}
\end{equation}
These numbers depend only on the common linear zero mode and are
therefore identical for ${\cal M}_{\rm I}$ and ${\cal M}_{\rm II}$ in \eqref{eq:critical_models}.

The remaining equation is
\begin{align}
f_2'
-\frac{3}{u}f_2
={}&
\frac12 f_0\chi_2'
-\frac{\Phi^2}{2u}
+
\frac{u^3}{4}
\left[
2z_cz_2
+
z_c^2
\left(
\chi_2
+
g_2\Phi^2
-
2\psi_2'
\right)
\right].
\label{eq:f2_equation}
\end{align}
With
$
f_2(0)=f_2(1)=0,
$ we write
\begin{equation}
f_2
=
f_2^{(0)}
+
z_2 f_2^{(h)} .
\label{eq:f2_split}
\end{equation}
Using Eqs.~\eqref{eq:chi2_solution},
\eqref{eq:psi2_prime}, and \eqref{eq:linear_zero_mode},
the part independent of $z_2$ can be integrated explicitly,
\begin{equation}
f_2^{(0)}(u)
=
\frac{u f_0(u)}{4}\Phi(u)\Phi'(u)
+
\frac{z_c^2 C_E}{2}u^3(1-u).
\end{equation}
The part proportional to $z_2$ is,
\begin{equation}
f_2^{(h)}(u)
=
\frac{z_c}{2}u^3(u-1).
\label{eq:f2h_explicit}
\end{equation}

The value $z_2$ in \eqref{eq:critical_expansions} is fixed by the scalar equation at order
$\epsilon^3$.  
The cubic equation takes the form
\begin{equation}
\left(
\frac{f_0}{u^2}\phi_3'
\right)'
+
\left(
\frac{2}{u^4}
+
g_2 z_c^2
\right)\phi_3
=
R_3 ,
\end{equation}
with
\begin{align}
R_3
={}&
\frac{4\lambda_4}{u^4}\Phi^3
+
\frac{\chi_2}{u^4}\Phi
-
g_2z_c^2
\left(
\frac{2z_2}{z_c}
+
\frac{\chi_2}{2}
-
2\psi_2'
\right)\Phi
-
\left(\frac{1}{u^2}
\left(
f_2-\frac{\chi_2}{2}f_0
\right) \Phi'\right)'.
\label{eq:R3_source}
\end{align}

Since ${\cal L}$ possesses the zero mode $\Phi$, regularity of
$\phi_3$ requires the Fredholm condition
\begin{equation}
\int_0^1du\,\Phi R_3=0 .
\label{eq:solvability_condition}
\end{equation}
Writing
\begin{equation}
R_3
=
R_3^{(0)}
+
z_2R_3^{(h)},
\end{equation}
one obtains
\begin{equation}
z_2
=
-
\frac{{\cal I}^{(0)}}{{\cal I}^{(h)}},
\qquad
{\cal I}^{(0)}
\equiv
\int_0^1du\,
\Phi R_3^{(0)}\,, \qquad {\cal I}^{(h)}
\equiv
\int_0^1du\,
\Phi R_3^{(h)} .
\label{eq:zh2_solution}
\end{equation}

For $g_2=0.6$, the part associated with the horizon displacement is
common to the two models,
\begin{equation}
{\cal I}^{(h)}
=
-1.6061.
\end{equation}
The quartic potential changes only ${\cal I}^{(0)}$.  Numerically, 
\begin{equation}
\begin{array}{c|cc}
\text{model}
&
{\cal I}^{(0)}
&
z_2
\\
\hline
{\cal M}_{\rm I}
&
30.441~~~
&
18.953~~~
\\
{\cal M}_{\rm II}
&
1.9795~~~
&
1.2325~~~
\end{array}
\label{eq:cubic_numeric_models}
\end{equation}
and therefore
\begin{align}
z_h^{\rm I}
&=
2.3589
+
18.953\,\epsilon^2
+
\mathcal O(\epsilon^4),
~~~~~
z_h^{\rm II}
=
2.3589
+
1.2325\,\epsilon^2
+
\mathcal O(\epsilon^4).
\label{eq:zh_branches_numeric}
\end{align}

\subsubsection{Critical scaling of the condensate}
\label{subsec:condensate_scaling}

The Hawking temperature \eqref{eq:temperature} in $u$ coordinate is
\begin{equation}
T
=
-
\frac{
e^{-\chi(1)/2}f'(1)
}{
4\pi z_h
}.
\end{equation}
Using Eq.~\eqref{eq:critical_expansions}, we find
\begin{equation}
T
=
T_c
-
T_2\epsilon^2
+
\mathcal O(\epsilon^4),
\label{eq:T_epsilon_expansion}
\end{equation}
where
\begin{equation}
T_2
=
\frac{1}{4\pi z_c}
\left[
f_2'(1)
-\frac12 f_0'(1)\chi_2(1)
-\frac{z_2}{z_c}f_0'(1)
\right].
\label{eq:T2}
\end{equation}
For the two models this gives
\begin{align}
{\cal M}_{\rm I}:\qquad
T
&=
0.054273
-
1.1582\,\epsilon^2
+
\mathcal O(\epsilon^4),
\nonumber\\
{\cal M}_{\rm II}:\qquad
T
&=
0.054273
-
0.045441\,\epsilon^2
+
\mathcal O(\epsilon^4).
\label{eq:T_branches_numeric}
\end{align}
Thus both branches are supercritical and extend toward
$T<T_c$.

Since
\begin{equation}
T_c-T
=
T_2\epsilon^2
+
\mathcal O(\epsilon^4),
\end{equation}
the order parameter has the usual mean-field scaling
\begin{equation}
|\epsilon|
=
\frac{1}{\sqrt{T_2}}
(T_c-T)^{1/2}
+\cdots .
\label{eq:epsilon_mean_field}
\end{equation}
Using
$\epsilon=z_h^2\phi_{(2)}$ and
$O=\phi_{(2)}$,
we obtain
\begin{equation}
O
=
\frac{1}{z_c^2\sqrt{T_2}}
(T_c-T)^{1/2}
+\cdots .
\label{eq:beta_model1}
\end{equation}
Hence
\begin{align}
\begin{split}
{\cal M}_{\rm I}:~~~
O
\simeq
0.16698\,
(T_c-T)^{1/2};
~~~~~~~~~
{\cal M}_{\rm II}:~~~~
O
\simeq
0.84302\,
(T_c-T)^{1/2}.
\label{eq:condensate_numeric_models}
\end{split}
\end{align}

The two models therefore have the same critical temperature and the
same mean-field exponent,
$
\beta=1/2,
$
but different amplitudes.  This distinction will be directly tested
using the fully nonlinear solutions in
Sec.~\ref{sec:num}.

\subsection{Critical interior  scaling}
\label{sec:inte}

Section~\ref{sec:general_mechanism} gives the general matching from a
critical zero mode to the nonlinear Cauchy-horizon region.  We now
specialize those results to the charged case.  

\subsubsection{Cauchy-horizon transmission of the critical zero mode}
\label{subsec:critical_inner_breakdown}

At the critical point the background is the RN-AdS$_4$ solution
\eqref{eq:critical_RN}. Let $z_->z_c$ denote its inner Cauchy horizon, i.e. 
$f_0(z_-)=0$.
The horizon is simple for the nonextremal solutions considered here,
and $f_0<0$ between the two horizons.

The critical scalar mode obeys
Eq.~\eqref{eq:linear_zero_mode}.  We now show that the solution
regular at the event horizon cannot also be regular at $z=z_-$, in the similar strategy of \cite{Hartnoll:2020rwq, Hartnoll:2020fhc, Cai:2020wrp}.
Multiplying the zero-mode equation by $\Phi$ and integrating between
the two horizons gives
\begin{equation}
\left[
\frac{f_0}{z^2}
\Phi\,\Phi'
\right]_{z_c}^{z_-}
-
\int_{z_c}^{z_-}
\frac{f_0}{z^2}
(\Phi')^2\,dz
+
\int_{z_c}^{z_-}
\left(
\frac{2}{z^4}
+
\frac{g_2}{z_c^2}
\right)
\Phi^2\,dz
=
0\,.
\label{eq:zero_mode_integral_identity}
\end{equation}
If $\Phi$ were regular at both horizons, the boundary term would
vanish.  Since $f_0<0$ for $z_c<z<z_-$ and $g_2>0$, the remaining two
terms are nonnegative and can vanish simultaneously only for the
trivial solution.  Therefore a nontrivial zero mode regular at the
event horizon cannot be regular at the Cauchy horizon.

The integral identity therefore excludes the exceptional regular
branch at the Cauchy horizon.  According to the local analysis of
Sec.~\ref{subsec:general_log_transmission}, the critical zero mode
must consequently have a nonvanishing logarithmic transmission
coefficient,
$B_-\neq0$.

\subsubsection{ER collapse and first-Kasner scaling}
\label{subsec:critical_interior_matching}

We now specialize the general results of
Sec.~\ref{sec:general_mechanism} and Appendix \ref{app:ER_normal_form} to the Einstein--Maxwell--scalar model.  Comparing the charged equations
\eqref{eq:chi_eom}--\eqref{eq:phi_eom} with the general results, besides \eqref{eq:relation1} and \eqref{eq:app_ER_explicit_sources}, the relevant identifications are 
\begin{align}
d&=2, \qquad
Z(z)=1,
\qquad
{\cal U}(z)=\partial_\phi^2 V(0)-\frac{\partial_\phi^2 G(0)}{2G(0)^2}z^4 E^2,
\nonumber\\
{\cal G}
&=
-\frac{f e^{-\chi}}{z^2}\,,
\qquad
{\cal Y}=e^{-\chi/2},
~~~~~~~~
\Pi_\phi
=
\frac{f e^{-\chi/2}}{z^2}\phi',
\nonumber\\
{\cal R}_\chi&={\cal R}_f=0\,,
\qquad
{\cal S}
=
\frac{z^3E^2}{4G(\phi)}
+
\frac{V(\phi)}{2z}.
\label{eq:RN_general_dictionary}
\end{align}
For the critical RN solution,
$
\phi=0,
\chi=0,$
we have $
{\cal Y}_-=1,
Z_-=1, {\cal S}_->0
$, 
thus 
the assumptions \eqref{eq:app_ER_regular_sources} and \eqref{eq:app_ER_remainders} apply. The charged model therefore satisfies
the conditions required for the standard ER completion discussed in
Sec.~\ref{subsec:general_ER}.

The only information transmitted from the critical linear mode is
then its physical logarithmic amplitude
${\cal B}_-=\epsilon B_-$.  Substituting $d=2$ and $Z_-=1$ into 
Eqs.~\eqref{eq:general_ER_amplitudes},
\eqref{eq:general_v1_scaling}, and
\eqref{eq:general_first_Kasner_scaling}, we immediately obtain
\begin{align}
\begin{split}
\Gamma_{\rm ER}
\simeq
\frac{4}{
z_-{\cal B}_-^2
},
~~~~~
|\phi'|_{\rm aft}
\simeq
\frac{4}{
z_-|{\cal B}_-|
},
~~~~~~
|v_1|
\simeq
\frac{4}{|{\cal B}_-|},
\label{eq:RN_ER_matching}
\end{split}
\end{align}
together with
\begin{align}
1-p_t
\simeq 
{\cal B}_-^2,
~~~~~~~~~
p_x
=
p_y
&\simeq
\frac12{\cal B}_-^2, 
~~~~~~~~~
|p_\phi|
\simeq 
\sqrt{2}\,|{\cal B}_-|.
\label{eq:RN_first_Kasner_matching}
\end{align}
Thus no additional nonlinear interior parameter is required: the
leading ER collapse and the first Kasner epoch are fixed completely
by the critical RN geometry and the transmission coefficient $B_-$.

To express these results in terms of the boundary control parameter, we use the equation obtained from \eqref{eq:T_epsilon_expansion},
\begin{equation}
\begin{split}
\Delta T
\equiv
T_c-T
=
T_2\epsilon^2
+
{\cal O}(\epsilon^4).
\label{eq:RN_deltaT}
\end{split}
\end{equation}
Since ${\cal B}_-=\epsilon B_-$, Eqs.~\eqref{eq:RN_ER_matching}
and \eqref{eq:RN_first_Kasner_matching} give
\begin{align}
\begin{split}
\Gamma_{\rm ER}
\simeq
\frac{
4T_2
}{
z_-B_-^2
}
\Delta T^{-1},
~~~~
|\phi'|_{\rm aft}
\simeq
\frac{
4\sqrt{T_2}
}{
z_-|B_-|
}
\Delta T^{-1/2},
~~~~
|v_1|
\simeq 
\frac{
4\sqrt{T_2}
}{
|B_-|
}
\Delta T^{-1/2},
\label{eq:RN_ER_temperature_scaling}
\end{split}
\end{align}
and
\begin{align}
\begin{split}
\dpt \equiv 1-p_t
&\simeq 
\frac{B_-^2}{T_2}
\Delta T,
~~~~~~~
p_x
=
p_y
\simeq 
\frac{B_-^2}{2T_2}
\Delta T,
~~~~~~~
|p_\phi|
\simeq 
\frac{\sqrt{2}|B_-|}{\sqrt{T_2}}
\Delta T^{1/2}.
\label{eq:RN_Kasner_temperature_scaling}
\end{split}
\end{align}

The two models ${\cal M}_{\rm I}$ and ${\cal M}_{\rm II}$ share the
same critical RN geometry and the same critical zero mode.  They
therefore have the same $z_-$ and $B_-$, and hence the same
matching amplitudes when expressed in terms of $\epsilon$.  Their
temperature-space amplitudes differ only through the coefficient
$T_2$, whereas all critical powers are identical.  These predictions
will be tested directly in Sec.~\ref{sec:num}.

Close to the phase transition, the first Kasner epoch is not followed by an additional Kasner inversion in the class of models considered here, as discussed in Appendix~\ref{subsec:Kasner_dynamics}. Away from this regime, for models in Appendix~\ref{subsec:Kasner_dynamics},  the subsequent interior evolution probes the large-field behavior of the matter couplings and becomes model dependent. The corresponding transition laws between successive Kasner epochs are derived in Appendix~\ref{subsec:Kasner_dynamics}.

\subsection{Numerical verification}
\label{sec:num}

We now compare the analytic predictions of
Secs.~\ref{subsec:critical_expansion} and
\ref{subsec:critical_interior_matching} with the fully nonlinear 
hairy black-hole solutions.  
The two
models ${\cal M}_{\rm I}$ and ${\cal M}_{\rm II}$ share the same critical
temperature and the same Cauchy-horizon transmission data,
\begin{equation}
T_c=0.054273,
\qquad
z_-=3.5622,
\qquad
B_-=-10.619.
\label{eq:num_common_critical_data}
\end{equation}
Their distinction first appears in the nonlinear relation between the
bifurcation amplitude $\epsilon$ and the temperature distance from
criticality.

\subsubsection{Exterior critical behavior}
\label{subsec:num_exterior}

We first verify the exterior critical behavior obtained in
Sec.~\ref{subsec:critical_expansion}. 
Numerically, we find
\begin{align}
\begin{split}
{\cal M}_{\rm I}:~~~
{O}
\simeq
0.16701\,\Delta T^{1/2};
~~~~~~~~~
{\cal M}_{\rm II}:~~~
{O}
\simeq 
0.84314\,\Delta T^{1/2}
\,,
\label{eq:num_condensate_prediction}
\end{split}
\end{align}
where $\Delta T\equiv T_c-T>0$. These results are in good agreement with the perturbative expansion gives \eqref{eq:condensate_numeric_models}.

\begin{figure}[h!]
\begin{center}
\includegraphics[width=0.43\textwidth]{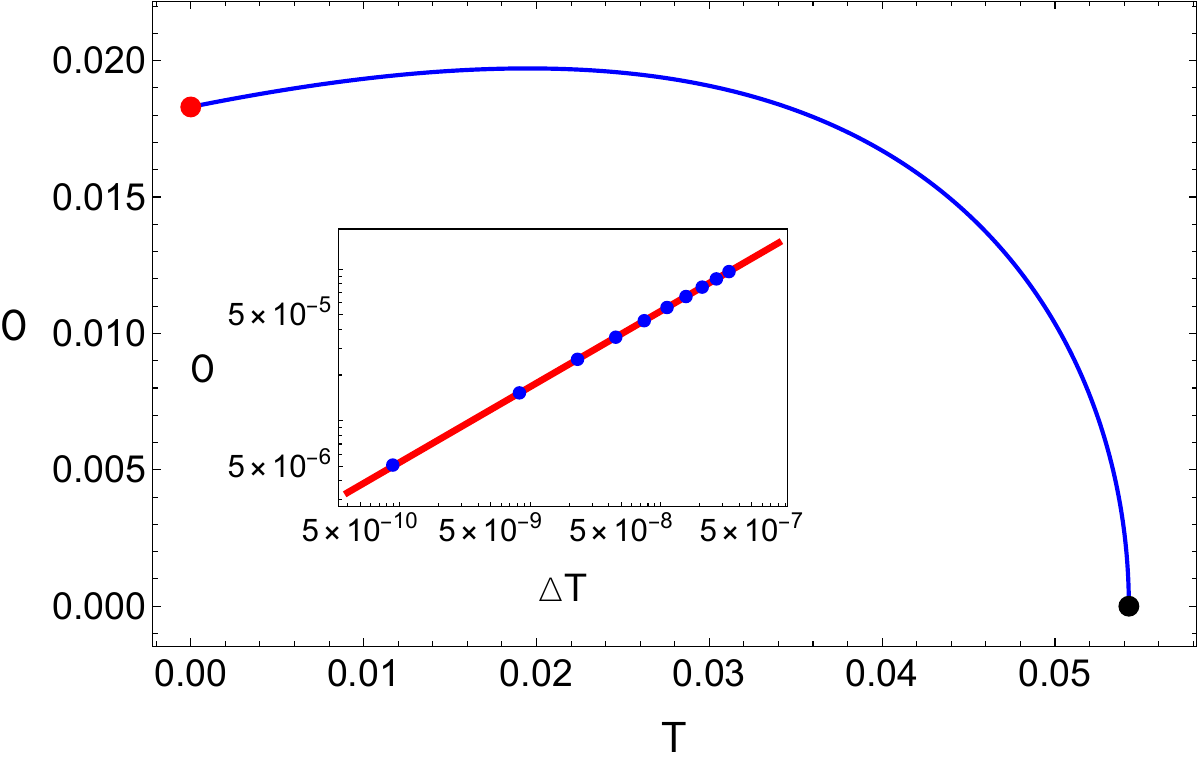}
\hspace{0.04\textwidth}
\includegraphics[width=0.43\textwidth]{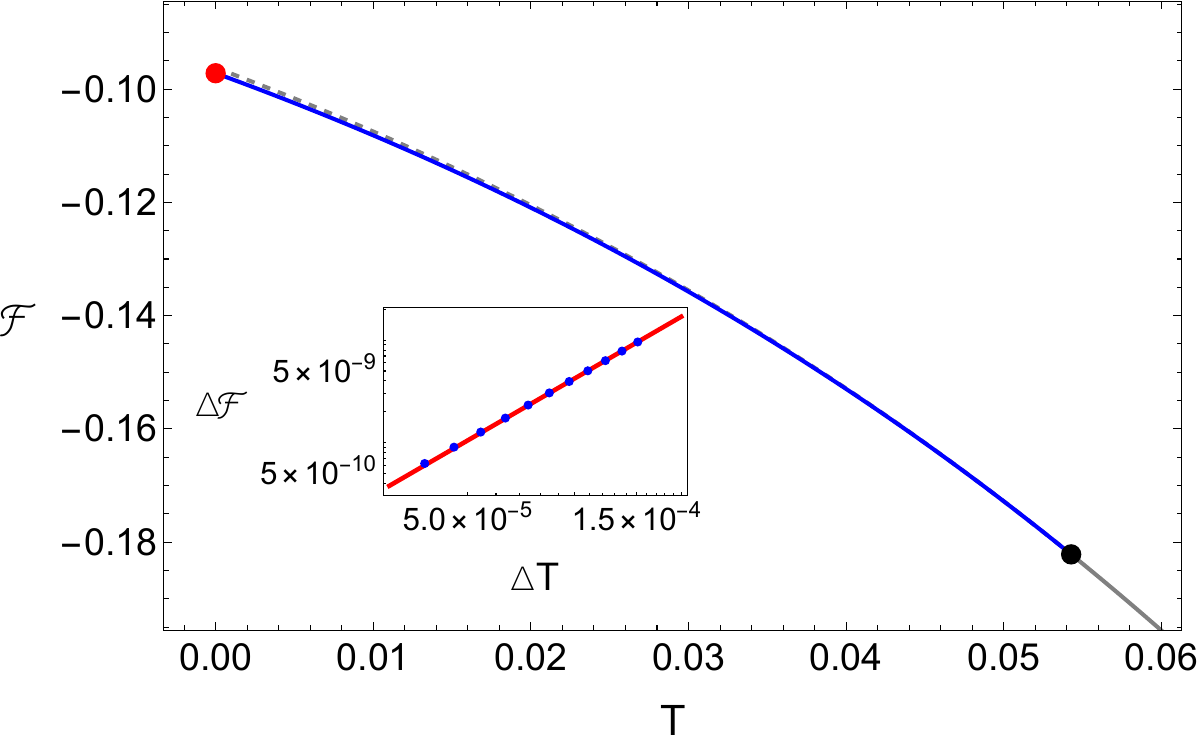}
\end{center}
\vspace{-0.5cm}
\caption{\small
Near-critical thermodynamics of Model ${\cal M}_{\rm I}$.
Left: condensate $O$ along the hairy
branch.  Right: grand-potential density relative to the RN-AdS$_4$
branch.  The hairy solution approaches the critical point
continuously, with
$O\propto(T_c-T)^{1/2}$ and
$\Delta{\cal F}\propto(T_c-T)^2$. The black and red dots label $T=T_c$ and $T=0$, respectively.
}
\label{fig:num_MF_thermo}
\end{figure}

\begin{figure}[h!]
\begin{center}
\includegraphics[width=0.43\textwidth]{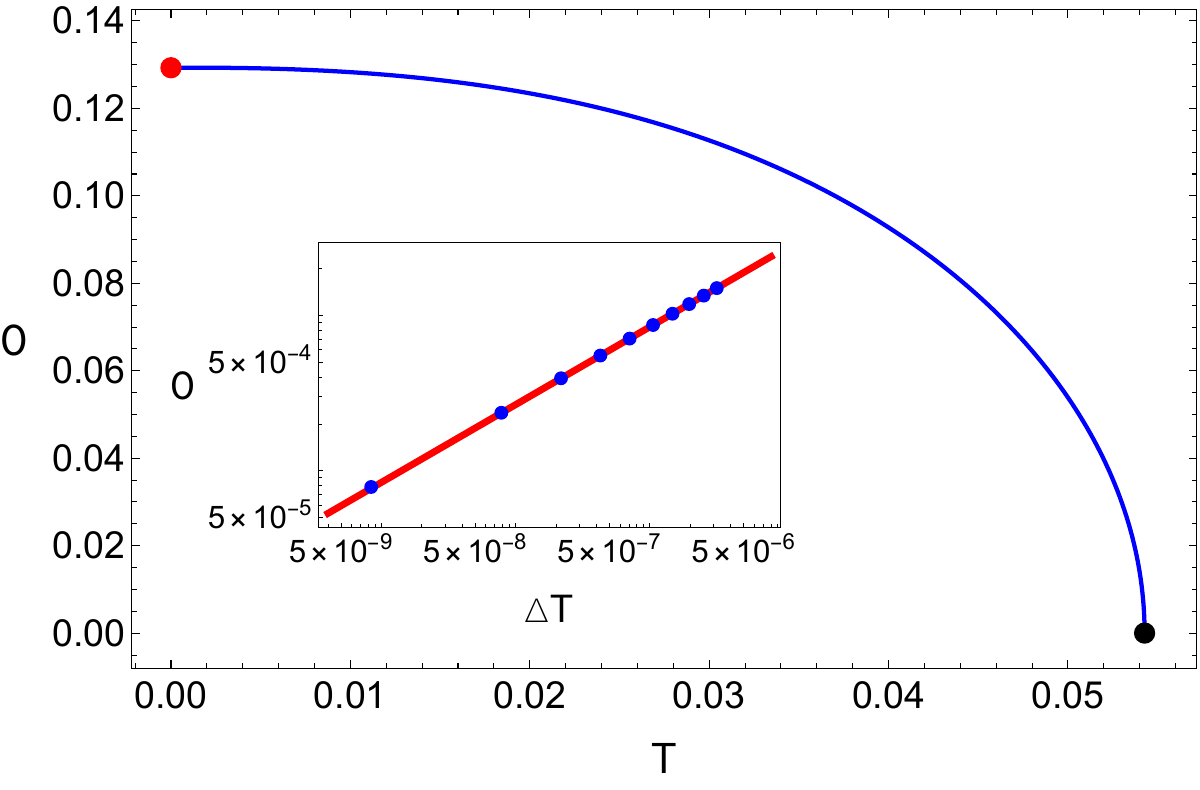}
\hspace{0.04\textwidth}
\includegraphics[width=0.43\textwidth]{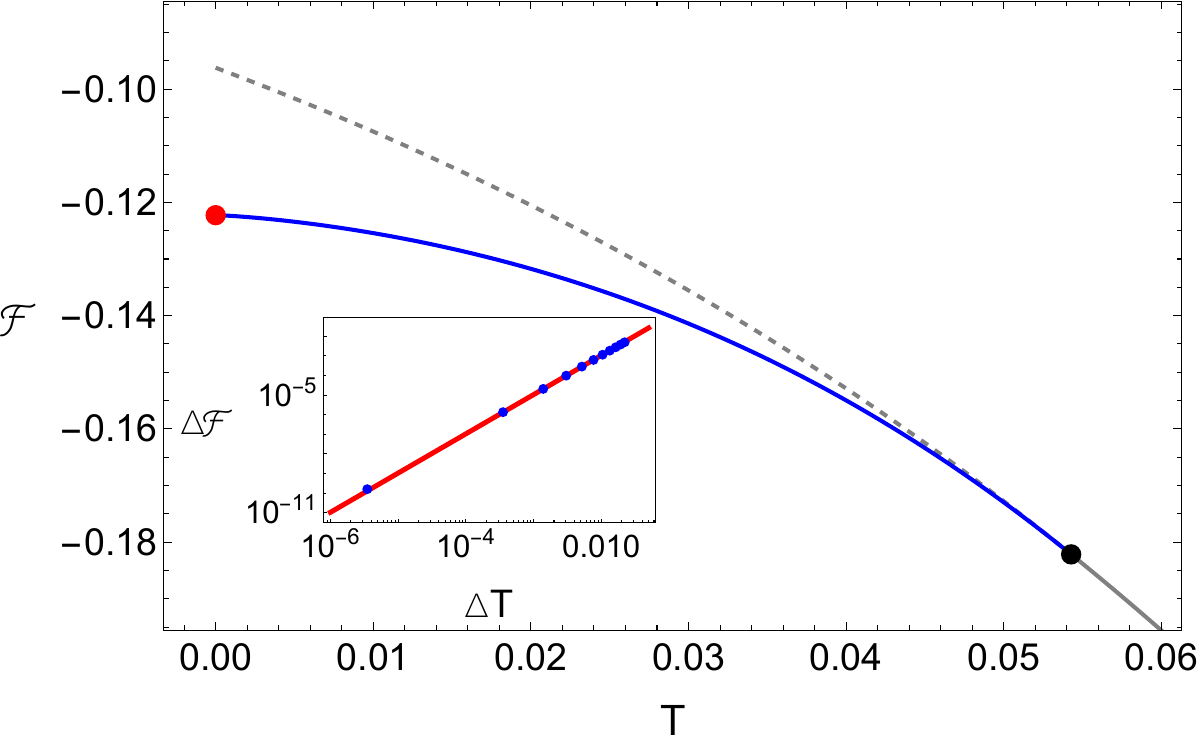}
\end{center}
\vspace{-0.5cm}
\caption{\small
Same as Fig.~\ref{fig:num_MF_thermo}, but for Model
${\cal M}_{\rm II}$.  The two models share the same critical
temperature and critical powers, while the nonlinear amplitudes are
different.
}
\label{fig:num_MR_thermo}
\end{figure}

Figures~\ref{fig:num_MF_thermo} and
\ref{fig:num_MR_thermo} show the corresponding fully nonlinear
solutions.  In both models the hairy branch joins the RN-AdS$_4$
branch continuously at $T=T_c$.  The condensate vanishes as
$O\propto\Delta T^{1/2}$, while the
grand-potential difference behaves as
$\Delta{\cal F}\propto\Delta T^2$, in agreement with the perturbative analysis.

Although the two theories have identical linearized scalar dynamics
at the critical RN solution, their nonlinear coefficients are
different.  This accounts for the different amplitudes in
Eq.~\eqref{eq:num_condensate_prediction}, while leaving the critical
powers unchanged.

\subsubsection{Interior matching}
\label{subsec:num_interior}

We next continue the nonlinear hairy solutions through the event
horizon and test the interior matching derived in
Sec.~\ref{subsec:critical_interior_matching}. Combining
Eq.~\eqref{eq:RN_Kasner_temperature_scaling} with
Eq.~\eqref{eq:num_common_critical_data}, the analytic prediction is
\begin{align}
\begin{split}
{\cal M}_{\rm I}:~~~
1-p_t
\simeq
97.366\,\Delta T
;~~~~~~~
{\cal M}_{\rm II}:~~~
1-p_t
\simeq
2481.7\,\Delta T
.
\label{eq:num_pt_temperature}
\end{split}
\end{align}
Note that their coefficients are determined
entirely from the exterior amplitude relation and the transmission
coefficient $B_-$, without fitting the nonlinear interior solution.

Numerically, we find
\begin{align}
\begin{split}
{\cal M}_{\rm I}:~~~
1-p_t
\simeq
97.330\,\Delta T
;
~~~~~~
{\cal M}_{\rm II}:~~~
1-p_t
\simeq
2476.5\,\Delta T .
\end{split}
\label{eq:num_pt_temperature-1}
\end{align}
which are also in good agreement with \eqref{eq:num_pt_temperature}.

\begin{figure}[h!]
\begin{center}
\includegraphics[width=0.43\textwidth]{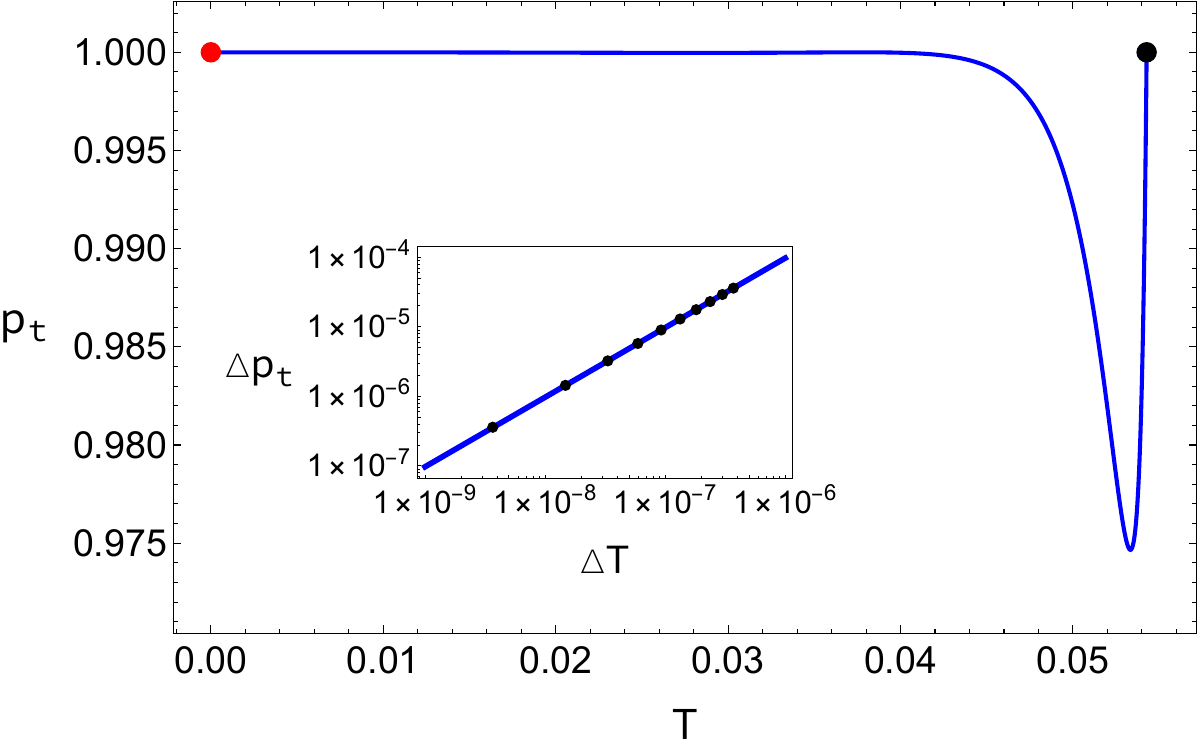}
~~~~~~~
\includegraphics[width=0.425\textwidth]{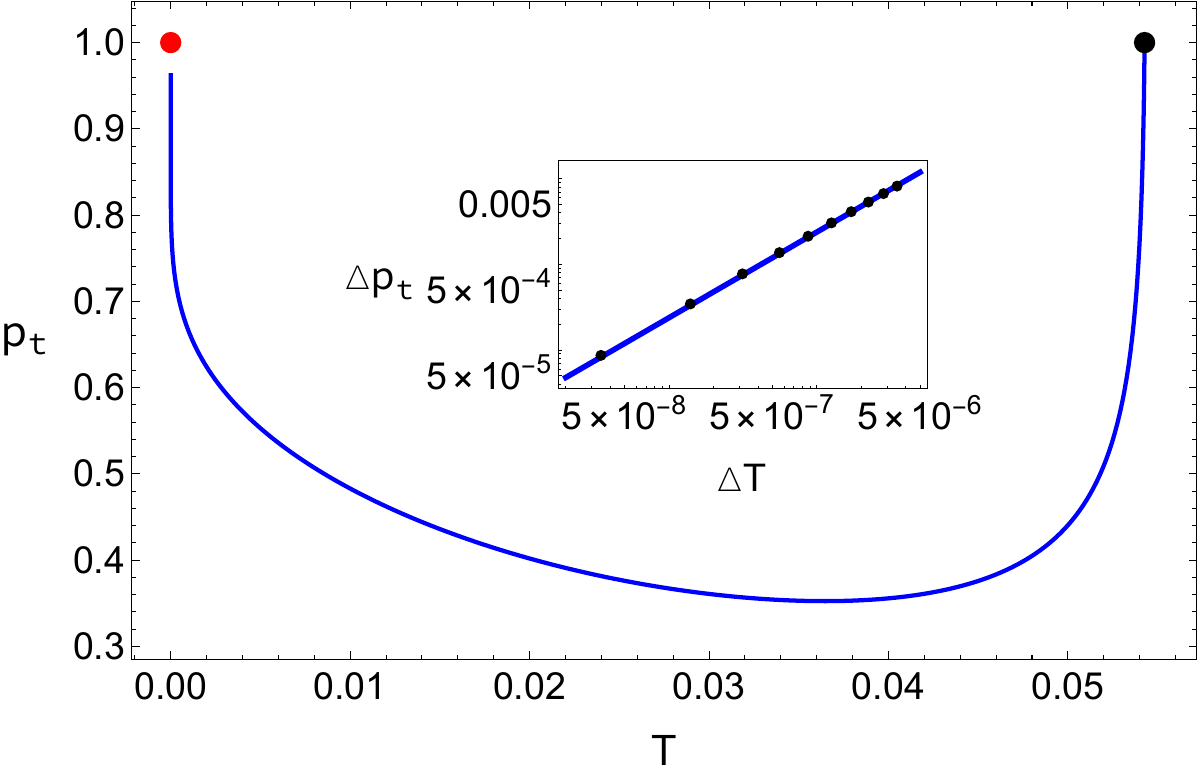}
\end{center}
\vspace{-0.5cm}
\caption{\small 
Kasner exponent $p_t$ along the hairy branch for Model ${\cal M}_{\rm I}$ ({\em left}) and Model ${\cal M}_{\rm II}$ ({\em right}).
The insets show $\Delta p_t\equiv1-p_t$ as a function of
$\Delta T=T_c-T$ near the critical point, where the dots are numerical data, while the lines are the analytic predictions in ~\eqref{eq:num_pt_temperature}. 
}
\label{fig:num_MF_Kasner}
\end{figure}

Fig.~\ref{fig:num_MF_Kasner} 
show the numerical first-Kasner exponent.
For both models
$p_t\rightarrow 1$
as
$T\rightarrow T_c^-$,
and the near-critical data approach the linear behavior predicted in
Eq.~\eqref{eq:num_pt_temperature}.  The numerical solutions therefore
confirm not only the critical power
$1-p_t\propto T_c-T$, but also the model-dependent absolute
amplitude inherited from the exterior bifurcation.

Taken together, the exterior and interior numerical results exhibit
a simple hierarchy.  The nonlinear theory first determines the
relation between $\epsilon$ and the boundary distance
$\Delta T$.  The critical RN zero mode then transmits this amplitude
through the would-be Cauchy horizon, fixing the ER collapse and the
first Kasner epoch.  Consequently, the two models have identical
leading matching at fixed $\epsilon$, while their different
temperature--amplitude relations produce different numerical
coefficients when the same observables are expressed in terms of
$T_c-T$.

\section{Infrared control of low-temperature interior scaling}  
\label{sec:low_temperature_scaling}
The previous sections focused on the near-critical limit $T\to T_c$, where a universal interior scaling law emerges. 
We now consider a different asymptotic limit along the hairy branch, $T\to0$.  
We show that the
relevant information is encoded by the IR endpoint of the full
nonlinear solution, similar to the one-horizon case in \cite{Gao:2026tck}.

For rotating BTZ, thermodynamic competition with a horizonless scalar
soliton complicates a direct zero-temperature black-hole limit.  We
therefore restrict the analysis to the two scalarized charged
black-brane models shown in \eqref{eq:critical_models}.
They have the same quadratic couplings, critical RN--AdS$_4$ geometry,
and critical zero mode.  Consequently, they share the same
near-critical scaling powers and the same ER matching at fixed
bifurcation amplitude.  Their nonlinear completions nevertheless
select different zero-temperature IR endpoints.  

\subsection{Power-law scaling in Model ${\cal M}_{\rm I}$ at low temperature}
\label{subsec:lowT_finite_scalar}

For Model ${\cal M}_{\rm I}$ in \eqref{eq:critical_models}, the zero-temperature IR  solution approaches an AdS$_2\times\mathbb{R}^2$ throat with a finite scalar field, similar to the geometry in \cite{Iqbal:2011ae}. The details are discussed in Appendix \ref{app:finite_scalar_endpoint}. 

For the scalar field, the approach to the extremal throat contains two distinct radial
deformations \eqref{app:eq:finite_scalar_deformation}. The first is the regular analytic deformation of the
coupled throat, while the second is the noninteger scalar irrelevant
mode. 
The distinction between these two modes will be important below: the
analytic mode controls the leading smooth thermal response, whereas
the noninteger mode provides the perturbation that is transmitted to
the inner horizon.

We first list the parameters at zero temperature used below. According to Appendix \ref{app:finite_scalar_endpoint}, for the model considered here, 
\begin{equation}
m_{\rm IR}^2
=
19.073,
\qquad
L_2^2
=
0.17283,
\qquad
\delta_{\rm IR}
=
1.3832.
\label{eq:lowT_IR_numbers}
\end{equation} 
Integrating from the
AdS$_2$ throat to the asymptotic boundary gives 
\begin{align}
z_0
&=
3.5161,
~~~~~~~
\chi_*
=
0.16596,
~~~~~~~
f_2
=
-\frac{V_*}{z_0^2}
=
0.46801,
\nonumber\\
c_{\rm a}
&=
-0.42425,
~~~~~~~
A_{\rm IR}
=
-0.087219,
~~~~~~~
\left. O\right|_{T=0}
=
0.018301\,, 
\label{eq:lowT_zeroT_data}
\end{align}
for parameters in \eqref{app:eq:extremal_expansion} and 
\eqref{app:eq:finite_scalar_deformation}. 
These quantities are obtained directly from the zero-temperature
solution and provide the matching data for the low-temperature
analysis below.

These zero temperature parameters are useful to determine the low temperature behavior. 
From the thermodynamic relation
\eqref{eq:app_grand_first_law}, mixed derivatives at fixed $\mu$
give
\begin{equation}
\left(
\frac{\partial\phi_2}{\partial T}
\right)_{\mu,J}
=
\left(
\frac{\partial s}{\partial J}
\right)_{\mu,T}.
\label{eq:lowT_Maxwell_relation}
\end{equation}
The condensate consequently has
the low-temperature expansion
\begin{equation}
O(T)
=
\phi_{2,0}
+
A_1 T
+
{\cal O}(T^{\delta_{\rm IR}}),
\qquad
A_1
=
\left.
\frac{\partial s_0}{\partial J}
\right|_{\mu=1,J=0}=
-\frac{8\pi}{z_0^3}
\left.
\frac{\partial z_0}{\partial J}
\right|_{\mu=1,J=0}, 
\label{eq:lowT_condensate_expansion}
\end{equation}
where we have used $s_0=4\pi/z_0^2$.

The required derivative is obtained from the linear response of the
zero-temperature solution to an infinitesimal scalar source, keeping $\mu=1$.  Numerically,
\begin{equation}
\left.
\frac{\partial z_0}{\partial J}
\right|_{\mu=1,J=0}
=
-0.22070,
\qquad
A_1
=
0.12761.
\label{eq:lowT_linear_response_data}
\end{equation}
Using the zero-temperature value in
Eq.~\eqref{eq:lowT_zeroT_data}, we finally obtain
\begin{equation}
O
=
0.018301
+
0.12761\,T
+
{\cal O}(T^{\delta_{\rm IR}}).
\label{eq:lowT_condensate_result}
\end{equation}
Numerically, we find that $O\simeq 0.018337 + 0.12606 T$ in  Fig.~\ref{fig:num_MF_thermo} at low temperature, in good agreement with the analytic prediction. 

\subsubsection{Transmission of the scalar IR mode to the singularity}
\label{subsec:lowT_IR_to_singularity}

We now determine how the two IR deformations of the scalar field in
Eq.~\eqref{app:eq:finite_scalar_deformation} are continued through the
near-extremal throat. Closely related construction has been used for charged magnetic brane  \cite{DHoker:2010onp}. 

At small but nonzero temperature, the leading near-extremal throat is 
\begin{equation}
f
\simeq
f_2
\left(
(z_0-z)^2-(z_0-z_h)^2
\right)\,,
\label{eq:lowT_nearextremal_f}
\end{equation}
where $z_0$ is the extremal horizon, 
$z_h$ is the event horizon at low temperature, 
while the inner horizon is located at $z_-=2z_0-z_h$,
respectively.  The Hawking temperature is 
\begin{equation}
T = \frac{f_2 e^{-\chi_*/2}}{2\pi} (z_0-z_h).
\label{eq:lowT_gammaT}
\end{equation}

\vspace{-0.2cm}
\paragraph{Analytic IR mode.}

We first consider the solution of the scalar field inside black hoel which approaches the analytic term 
$c_{\rm a}(z_0-z_h)$ in   \eqref{app:eq:finite_scalar_deformation} near the outer horizon. 
It is convenient to solve the system in coordinate 
\begin{equation}
x
\equiv
\frac{z_0-z}{z_0-z_h},
\end{equation}
where $x=\pm1$ are outer and inner horizon,respectively. 
Writing its finite-temperature continuation as
$\delta\phi_{\rm a}=(z_0-z_h)\varphi_{\rm a}(x)$, the scalar equation
expanded to first order in $(z_0-z_h)$ becomes
\begin{equation}
\frac{d}{dx}
\left[
\left(
x^2-1
\right)
\frac{d\varphi_{\rm a}}{dx}
\right]
-
\delta_{\rm IR}
\left(
\delta_{\rm IR}+1
\right)
\varphi_{\rm a}
=
-\frac{
4\partial_\phi V|_*
}{
V_*z_0
}
x .
\label{eq:lowT_analytic_throat_equation}
\end{equation}
Here we have used
$L_2^2=-1/V_*$ and
$\delta_{\rm IR}(\delta_{\rm IR}+1)
=m_{\rm IR}^2L_2^2$.

Using Eq.~\eqref{app:eq:finite_scalar_deformation}, the solution that matches the
analytic zero-temperature deformation is simply
$
\varphi_{\rm a}(x)
=
c_{\rm a}x .
$ 
Hence
\begin{equation}
\delta\phi_{\rm a}
=
c_{\rm a}(z_0-z) .
\label{eq:lowT_analytic_continuation}
\end{equation}
This continuation is manifestly regular at both event and inner horizon. 
The analytic IR mode therefore produces no logarithmic
component at the inner horizon.  Notice that this conclusion follows
from the coupled near-extremal throat expansion rather than from an assumption about regularity.

\vspace{-0.2cm}
\paragraph{Noninteger irrelevant mode.}

The solution approaching noninteger deformation in   \eqref{app:eq:finite_scalar_deformation} behaves differently.  At leading order
its finite-temperature continuation obeys
\begin{equation}
\frac{d}{dx}
\left[
\left(
x^2-1
\right)
\frac{d\varphi}{dx}
\right]
-
\delta
\left(
\delta+1
\right)
\varphi
=
0 .
\label{eq:lowT_legendre_equation}
\end{equation}
Matching to the zero-temperature behavior
$A_{\rm IR}(z_0-z)^{\delta}$ selects the solution regular at the
event horizon,
\begin{equation}
\varphi(x)
=
A_{\rm IR}
(z_0-z_h)^{\delta}
\frac{
P_{\delta}(x)
}{
c_{\delta}
},
\qquad
c_\delta
=
\frac{
2^\delta
\Gamma\left(\delta+\frac12\right)
}{
\sqrt{\pi}\,
\Gamma(\delta+1)
}.
\label{eq:lowT_legendre_solution}
\end{equation}
Near the inner horizon, $x\rightarrow-1$, 
form
\begin{equation}
\frac{
P_\delta(x)
}{
c_\delta
}
=
a_\delta
+
b_\delta
\log(1+x)
+\cdots ,
\qquad
b_\delta
=
\frac{
\sin(\pi\delta)
}{
\pi c_\delta
}.
\label{eq:lowT_inner_legendre}
\end{equation}
For an integer Legendre index $\delta=1$, one has
$P_1(x)=x$ and $b_1=0$, illustrating why an analytic
deformation does not generate a logarithm at the inner horizon.
The actual analytic term in Eq.~\eqref{eq:lowT_analytic_continuation},
however, is a particular solution of the coupled inhomogeneous
equation \eqref{eq:lowT_analytic_throat_equation}, rather than the
noninteger scalar eigenmode. For the noninteger
irrelevant exponent $\delta_{\rm IR}$, 
\begin{equation}
b_{\delta_{\rm IR}}
=
-0.25945.
\end{equation}

The physical logarithmic amplitude transmitted to the inner horizon
is therefore
\begin{equation}
{\cal B}_{\rm IR}
\equiv
A_{\rm IR}
b_{\delta_{\rm IR}}
(z_0-z_h)^{\delta_{\rm IR}}
=
0.92178\,
T^{\delta_{\rm IR}},
\label{eq:lowT_inner_seed}
\end{equation}
Thus the mode that is subleading in the value of the scalar near the
outer horizon is precisely the mode that controls the singular
response at the inner horizon.

\vspace{-0.2cm}
\paragraph{Near-extremal separation of scales.}
We now show even at low temperature, the discussion  on the matching before and after collapse in Sec. \ref{sec:general_mechanism} still applies. 
From \eqref{eq:lowT_nearextremal_f}, 
$
z_--z_h
=
2(z_0-z_h)
=
\mathcal O(T),
$
and the inner-horizon slope scales as
\begin{equation}
f'(z_-)
=
2f_2(z_0-z_h)
+
\mathcal O((z_0-z_h)^2)
=
\mathcal O(T).
\label{eq:lowT_inner_slope}
\end{equation}

For every nonzero temperature, the inner horizon of the near-extremal throat remains 
simple.  Applying the local matching of
Sec.~\ref{sec:general_mechanism} at fixed $T$, with $d=2$ and
$Z_-=1$, gives
\begin{equation}
\Delta z_{\rm ER}
=
\frac{
z_-
}{
4
}
{\cal B}_{\rm IR}^2
\left[
1+o(1)
\right]
=
\mathcal O
\left(
T^{2\delta_{\rm IR}}
\right).
\label{eq:lowT_ER_width}
\end{equation}
The ER layer therefore shrinks parametrically faster than the
separation between the two horizons,
\begin{equation}
\frac{
\Delta z_{\rm ER}
}{
z_--z_h
}
=
\mathcal O
\left(
T^{2\delta_{\rm IR}-1}
\right)
\rightarrow
0,
\qquad
2\delta_{\rm IR}-1
=
1.7664.
\label{eq:lowT_scale_separation}
\end{equation}

This separation also controls the local simple-zero approximation.
Expanding around the inner horizon,
\begin{equation}
f(z)
=
f'_-
\left(
z-z_-
\right)
+
\frac12 f''_-
\left(
z-z_-
\right)^2
+\cdots ,
\end{equation}
with $f'_-={\cal O}(T)$ and $f''_-={\cal O}(1)$, the relative correction across
the ER layer behaves as
\begin{equation}
\frac{
f''_-\,\Delta z_{\rm ER}
}{
f'_-
}
=
\mathcal O
\left(
T^{2\delta_{\rm IR}-1}
\right)
\rightarrow0 .
\label{eq:lowT_simple_zero_control}
\end{equation}
Hence, although the two horizons merge on the scale of the full
near-extremal throat, the nonlinear ER region resolves the inner
horizon as an increasingly accurate simple zero.

The same double scaling is visible in the quantities entering the ER
normal form of Appendix~\ref{app:ER_normal_form}.  Since
${\cal B}_{\rm IR}={\cal O}(T^{\delta_{\rm IR}})$ and
$f'_-={\cal O}(T)$, Eq.~\eqref{eq:general_momentum_matching} gives
\begin{equation}
\Pi_\star
=
\mathcal O
\left(
T^{\delta_{\rm IR}+1}
\right).
\end{equation}
Moreover,
\begin{equation}
{\cal C}_-
=
\mathcal O(T^{-1}),
\qquad
{\cal G}_c
=
{\cal C}_- \Pi_\star^2
=
\mathcal O
\left(
T^{2\delta_{\rm IR}+1}
\right),
\qquad
\Lambda_-
=
\mathcal O(T).
\label{eq:lowT_ER_double_scaling}
\end{equation}
It follows that
\begin{equation}
\frac{
{\cal G}_c
}{
\Lambda_-
}
=
\mathcal O
\left(
T^{2\delta_{\rm IR}}
\right),
\end{equation}
in agreement with
Eq.~\eqref{eq:lowT_ER_width}.  
More generally, this argument shows that the simple-Cauchy-horizon
matching remains parametrically controlled in the near-extremal limit
provided
$\delta_{\rm IR}>1/2$.\footnote{
If $\delta_{\rm IR}\leq1/2$, the ER scale is no longer parametrically
smaller than the horizon separation, and the simple-zero boundary-layer
analysis must be replaced by a genuinely near-extremal matching problem. It is interesting to explore the matching in holographic quantum phase transitions  \cite{Iqbal:2011aj}.}

\paragraph{Kasner epoch.}

We now use the Kasner matching of
Sec.~\ref{subsec:general_ER_to_Kasner}.  Replacing the critical
logarithmic amplitude by ${\cal B}_{\rm IR}$ gives 
\begin{equation}
|v|
\simeq 
\frac{
4
}{
|{\cal B}_{\rm IR}|
}
,
\qquad
|v|^{-1}
\simeq
0.23045\,
T^{\delta_{\rm IR}}
.
\label{eq:lowT_v_scaling}
\end{equation}
Thus the post-collapse Kasner epoch is driven to the
large-$|v|$ regime as $T\rightarrow0$. Note that according to Appendix \ref{subsec:Kasner_dynamics}, there exists only one Kasner regime at extremally low temperature. 

Using the large-$|v|$ relations derived in
Sec.~\ref{subsec:general_ER_to_Kasner}, we obtain 
\begin{align}
1-p_t
&\simeq
0.84968\,
T^{2\delta_{\rm IR}}
,~~~~
p_x
=
p_y
\simeq
0.42484\,
T^{2\delta_{\rm IR}}
,
~~~~~~
|p_\phi|
\simeq
1.3036\,
T^{\delta_{\rm IR}}
\label{eq:lowT_Kasner_predictions}
\end{align}
with $\delta_{\rm IR}$ given in \eqref{eq:lowT_IR_numbers}. 

\begin{figure}[h!]
\begin{center}
\includegraphics[width=0.6\textwidth]{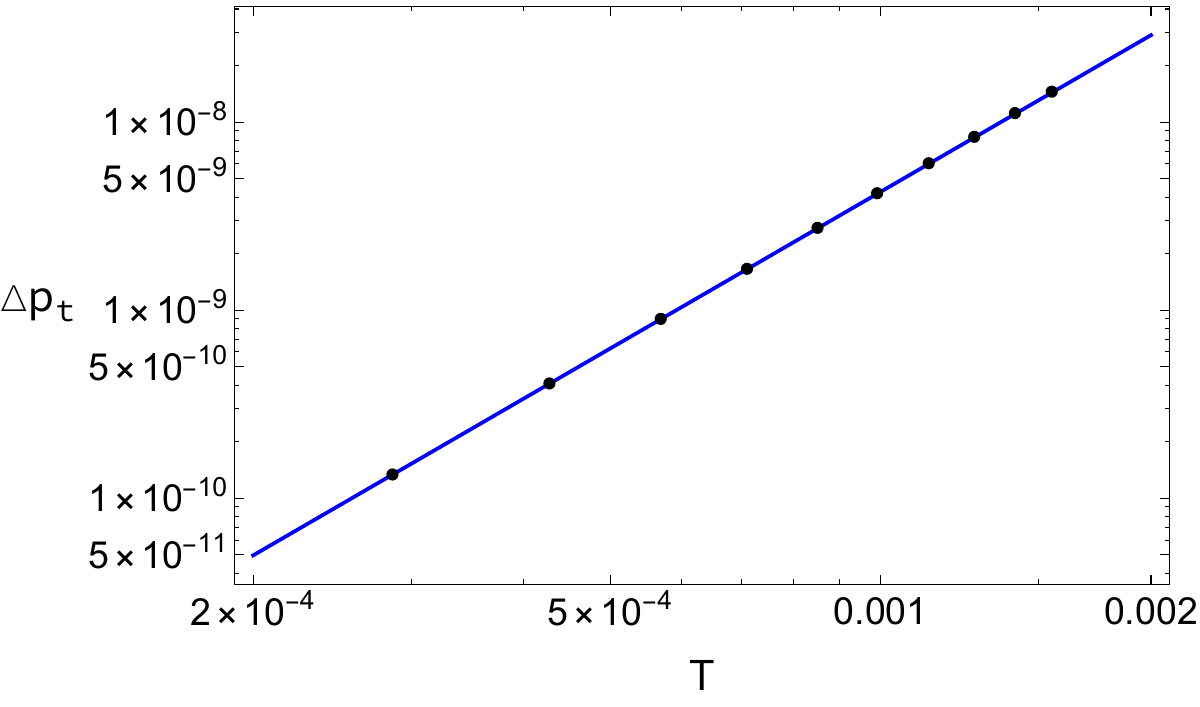}
\end{center}
\vspace{-0.5cm}
\caption{\small Log-log plot of the zero-temperature scaling of
$\Delta p_t \equiv 1-p_t$ for Model ${\cal M}_{\rm I}$.
The black dots are the numerical data, while the blue solid line is the analytic prediction \eqref{eq:lowT_Kasner_predictions}.
The good agreement verifies the predicted low-temperature scaling.
}
\label{fig:num_dpt-lowT1_Kasner}
\end{figure}

Fitting the first Kasner exponent in the low-temperature regime of the fully backreacted solution gives
\begin{align}
    1-p_t \simeq 0.88780\, T^{2.7717}\,.
\end{align}
As shown in Fig. \ref{fig:num_dpt-lowT1_Kasner}, the numerical value matches well with \eqref{eq:lowT_Kasner_predictions},  
which provides a direct numerical confirmation that the approach of the first Kasner epoch
to the limiting geometry is controlled by the leading irrelevant
IR deformation.

\subsection{Inverse-log scaling in Model ${\cal M}_{\rm II}$ at low temperature}
\label{subsec:lowT_running}

We next consider Model ${\cal M}_{\rm II}$ in
Eq.~\eqref{eq:critical_models} where the scalar
does not approach a finite IR fixed point. The
source-free zero-temperature solution develops the running-scalar
IR geometry derived in
Appendix~\ref{app:running_scalar_endpoint}. Its complete asymptotic
form is given in Eq.~\eqref{app:eq:running_sep_solution}. In
particular, the scalar grows linearly, the redshift becomes
exponentially large, and the scalar velocity $z\phi'$ diverges in  IR; see Eq.~\eqref{app:eq:running_sep_velocity} in which the parameter $C$ is determined by the global source-free
zero-temperature solution, 
\begin{align}
\rho_0
&=
0.36679,
~~~~~~~~
C
=
0.57865.
\label{eq:lowT_running_C_number}
\end{align}

At small but nonzero temperature the zero-temperature running-scalar IR geometry is
cut off by a narrow region around the event horizon
$z=z_h$. As $T\to0$, the event horizon moves
to $z_h\to\infty$.  Introducing the stretched coordinate 
\begin{equation}
z
=
z_h+\frac{\sigma}{z_h},
\label{eq:lowT_running_layer_coordinate}
\end{equation}
we write
\begin{align}
\phi
&=
Cz_h+\frac{w(\sigma)}{z_h},
~~~~~~~
f
=
\frac{F(\sigma)}{z_h^2},
~~~~~~~~
\chi
=
\frac{C^2z_h^2}{4}
-\frac{1}{12}\log z_h
+\chi_\infty
+\Xi(\sigma).
\label{eq:lowT_running_layer_ansatz}
\end{align}

Substituting the above ansatz
\eqref{eq:lowT_running_layer_ansatz} into \eqref{eq:chi_eom}-\eqref{eq:phi_eom}, 
and expanding at fixed $\sigma$ as $z_h\to\infty$, and matching \eqref{eq:lowT_running_layer_ansatz} to the zero-temperature IR solution
\eqref{app:eq:running_sep_solution}, we obtain the
leading solution
\begin{equation}
w(\sigma)
=
C\sigma+\frac{1}{12C},
\qquad
\Xi(\sigma)
=
\frac{C^2\sigma}{2},
\qquad
F(\sigma)
=
\frac{16}{C^2}
\left[
1-\exp\left(\frac{C^2\sigma}{4}\right)
\right].
\label{eq:lowT_running_layer_solution}
\end{equation}
Note that a closely related limiting procedure for low temperature black hole  was studied in Einstein--dilaton gravity ~\cite{Gursoy:2008za}.

The event horizon is at $\sigma=0$.  In particular,
\begin{equation}
f'_h
=
-\frac{4}{z_h}
+\mathcal O(z_h^{-3}),
\qquad
\chi_h
=
\frac{C^2z_h^2}{4}
-\frac{1}{12}\log z_h
+\chi_\infty
+\mathcal O(z_h^{-2}).
\label{eq:lowT_running_horizon_data}
\end{equation}

The Hawking temperature therefore behaves as
\begin{equation}
T
=
T_{\rm R}\,
z_h^{-23/24}
\exp\left(
-\frac{C^2z_h^2}{8}
\right)
\left[
1+\mathcal O(z_h^{-2})
\right],
\qquad
T_{\rm R}
=
\frac{1}{\pi}e^{-\chi_\infty/2}.
\label{eq:lowT_running_T_zh}
\end{equation}
Since $s=4\pi/z_h^2$, equivalently it can be written as
\begin{equation}
T
=
T_{\rm R}\,
\left(\frac{s}{4\pi}\right)^{23/48}
\exp\left(
-\frac{\pi C^2}{2s}
\right)
\left[
1+\mathcal O(s)
\right].
\label{eq:lowT_running_T_s}
\end{equation}
The low-temperature scale is therefore generated by an essential
singularity rather than by an ordinary power law.

On the interior side of the same boundary layer, the scalar velocity
approaches
\begin{equation}
v_1
=
Cz_h
\left[
1+\mathcal O(z_h^{-2})
\right].
\label{eq:lowT_running_v1}
\end{equation}
Hence $|v_1|\rightarrow\infty$ as $T\rightarrow0$.  
Note that the method to obtain the above relation is different from the ER collapse mechanism introduced in Sec. \ref{sec:general_mechanism}. 
Using the
large-$|v|$ Kasner relations of
Sec.~\ref{subsec:general_ER_to_Kasner}, we obtain
\begin{align}
1-p_t
&\simeq
\frac{16}{C^2z_h^2}
=
\frac{4}{\pi C^2}s
,~~~~~~
p_x
=
p_y
=
\frac{2}{\pi C^2}s,
~~~~~~
|p_\phi|
=
\frac{2\sqrt{2}}{|C|\sqrt{\pi}}\,
s^{1/2}.
\label{eq:lowT_running_Kasner_s}
\end{align}
For the value of $C$ above,
\begin{equation}
1-p_t
\simeq 
3.80260\,s.
\label{eq:lowT_running_pt_s_number}
\end{equation}
The first Kasner velocity is already parametrically large at low
temperature, so the electric wall discussed in
Sec.~\ref{subsec:Kasner_dynamics} is asymptotically irrelevant and no
subsequent electric Kasner inversion occurs sufficiently close to
$T=0$.

To express the result directly in terms of temperature, define
\begin{equation}
X
\equiv
\frac{C^2z_h^2}{8}
=
\frac{\pi C^2}{2s},
\qquad
\alpha_{\rm R}
\equiv
\frac{23}{48}.
\label{eq:lowT_running_X_alpha}
\end{equation}
Eq.~\eqref{eq:lowT_running_T_zh} takes the form
\begin{equation}
T
=
T_0
X^{-\alpha_{\rm R}}
e^{-X}
\left[
1+\mathcal O(X^{-1})
\right],
\label{eq:lowT_running_T_X}
\end{equation}
where $T_0$ is a constant fixed by the zero-temperature solution, $T_0
=
T_{\rm R}
\left(\frac{8}{C^2}\right)^{-23/48}$. From \eqref{eq:lowT_running_T_zh}, for this model we have $T_0=0.034169$. 
Thus 
\begin{equation}
X
=
\ell
-
\alpha_{\rm R}\log\ell
+
\mathcal O(1),~~~~~~\text{where}~~~ \ell
\equiv\log(T_0/T). 
\label{eq:lowT_running_X_ell}
\end{equation}
Since $v_1^2=8X+\mathcal O(1)$, the Kasner exponents obey
\begin{align}
1-p_t
\simeq
\frac{
2
}{
\ell-\frac{23}{48}\log\ell
},
~~~~~~
p_x
=
p_y
\simeq
\frac{
1
}{
\ell-\frac{23}{48}\log\ell
},
~~~~~~
|p_\phi|
\simeq
\frac{
2
}{
\sqrt{
\ell-\frac{23}{48}\log\ell
}
}.
\label{eq:lowT_running_Kasner_temperature}
\end{align}
From \eqref{eq:lowT_running_X_ell}, we have low
temperature relation
\begin{equation}
1-p_t
\simeq
\frac{
2
}{
\log (1/T)
}
\qquad
(T\rightarrow0).
\label{eq:lowT_running_pt_leading}
\end{equation}
The running-scalar endpoint therefore produces an inverse-logarithmic Kasner scaling rather than the power-law infrared scaling found for
Model ${\cal M}_{\rm I}$.  

\begin{figure}[h!]
\begin{center}
\includegraphics[width=0.58\textwidth]{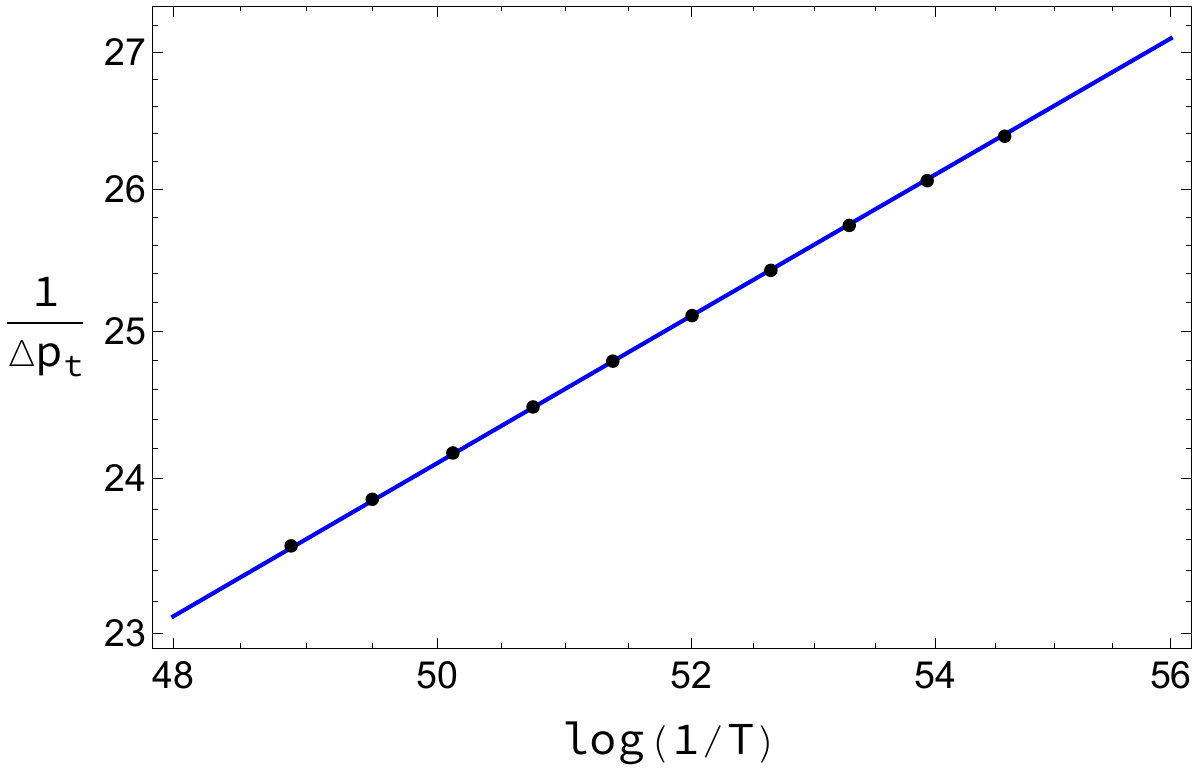}
\end{center}
\vspace{-0.5cm}
\caption{\small
Zero-temperature scaling of $\Delta p_t\equiv1-p_t$ for Model ${\cal M}_{\rm II}$.
The black dots are the numerical data and the blue solid line is the analytic prediction.
The linear behavior of $1/\Delta p_t$ versus $-\log T$ confirms the predicted asymptotic scaling, with slope $1/2$. 
}
\label{fig:num_dpt-M2_Kasner}
\end{figure}

In Fig. \ref{fig:num_dpt-M2_Kasner}, we show the low-temperature scaling of
$\Delta p_t$ for Model ${\cal M}_{\rm II}$.
The numerical data for $1/\Delta p_t$ become asymptotically linear in
$\log (1/T)$, with slope approaching $1/2$, in agreement with the analytic
prediction $\Delta p_t\simeq 2/\log (1/T)$.  Moreover, the numerical fitting gives
\begin{align}
    \frac{1}{1-p_t} \simeq 0.49555 \log (1/T)\,,~~~~~
\end{align}
which is in good agreement with the analytical result \eqref{eq:lowT_running_pt_leading}. 

\section{Discussion}
\label{sec:condis}

Our main result is a transmission mechanism by which linear critical
data at the onset of an exterior instability determine a nonlinear
reorganization of the interior. Along a continuous hairy branch
approaching a two-horizon reference solution, the event-horizon-regular
zero mode develops a logarithmic component at the reference Cauchy
horizon, and the transmitted amplitude
${\cal B}_-=\epsilon B_-$ controls the resulting nonlinear
Einstein--Rosen layer. Although the exterior scalar amplitude vanishes
at criticality, the would-be Cauchy horizon converts it into a
parametrically enhanced radial gradient. Its backreaction becomes
order one only within an increasingly narrow region, producing an
increasingly sharp Einstein--Rosen collapse. If a
scalar-kinetic-dominated regime follows, the same transmitted amplitude
carries the exterior critical scaling into the first Kasner epoch.

The logarithmic near-horizon structure and the 
ER scaling with respect to ${\cal B}_-$ are determined by
the local Cauchy-horizon geometry, whereas the transmission coefficient
$B_-$ retains information about the global propagation of the critical
zero mode. Later Kasner transitions can instead probe the large-field
completion of the matter sector. This distinguishes the present
two-horizon problem from the direct transmission to a Kasner regime found for single-horizon black holes~\cite{Gao:2026tck, Liu:2021hap,Gao:2023zbd,Caceres:2023zhl,Zhao:2026mkx}: here the
would-be Cauchy horizon introduces a singular intermediate layer before
the first Kasner epoch. The same framework interprets the 
Kasner scalings observed numerically in~\cite{Shao:2025fki, Shao:2025apr, Li:2025mhy}. 

The $T\to0$ limit gives distinct scaling laws. For the charged models studied here, a 
finite-scalar AdS$_2\times\mathbb R^2$ endpoint transmits a noninteger
irrelevant mode and gives
$1-p_t^{(1)}\propto T^{2\delta_{\rm IR}}$, whereas a running-scalar
endpoint produces an essential temperature--redshift relation and the
inverse-logarithmic behavior
$1-p_t^{(1)}\simeq 2/\log(1/T)$. Thus identical finite-temperature 
critical data need not imply identical low-temperature interiors.

There are some open questions for further exploration. 
Our construction concerns stationary, symmetry-preserving solutions. 
Time-dependent perturbations may produce mass inflation,
while nonaxisymmetric or superradiant modes may introduce mode mixing
and additional scales. It would be interesting to study whether the stationary critical layer survives
as an intermediate scaling regime, or even an attractor, in the fully dynamical evolution.


Additionally, several exceptional limits may lead to new universality classes. For an extremal reference Cauchy horizon, the radial kinetic coefficient has a higher-order zero, so the repeated-root argument underlying the
logarithmic channel must be reconsidered. Likewise, if the leading transmission coefficient vanishes, the nonlinear scale must be generated by a subleading mode or by higher-order perturbation theory. It would be useful to determine whether such zeros occur only under
fine tuning or can be protected by a symmetry, and whether they produce new critical powers or crossover functions.

Finally, it would also be interesting to determine whether the singular zero-mode transmission and the resulting ER layer admit a direct boundary signature, in light of recent evidence that
inner horizons and regions beyond them are encoded in the analytic structure of holographic correlators~\cite{Ceplak:2025dds, Dodelson:2025jff, AliAhmad:2026wem}. 
Recent work suggests that black-hole interiors can be probed by timelike entanglement entropy \cite{Anegawa:2024kdj,Li:2026bof} and
by high-frequency correlators and asymptotic quasinormal spectra, e.g.  
\cite{Afkhami-Jeddi:2025wra,Jia:2026ryl,Giombi:2026kdz, Grozdanov:2026cut,
Arnaudo:2026tcy,Hartnoll:2026vhu}. In our setting, an
order-parameter quasinormal mode softens to the static onset mode at
a continuous transition \cite{Amado:2009ts}, whose normalized radial
continuation fixes ${\cal B}_-$. It is natural
to ask whether this transmission data and the associated
Einstein--Rosen crossover leave identifiable signatures in timelike
entanglement or boundary spectral data. A complementary direction is to formulate the near-critical scaling
in terms of the interior response coefficients introduced by the
Iyer-Wald first law of interior dynamics
\cite{Xiong:2026zkb}. 
\subsection*{Acknowledgments}
We would like to thank Song He, 
Xin-Meng Wu, Ya-Wen Sun, Jun-Kun Zhao for useful discussions. This work was supported by the National Natural Science Foundation of China Grants No. 12375041, 12447134 and 12575046. H-D.~L.~also acknowledges the support from 
the Postdoctoral Innovation Project of Shandong Province SDCX-ZG-202503036.

\vspace{0.95cm}
\appendix

\section{Derivation of the explicit equations in Cauchy horizon collapse}
\label{app:ER_normal_form}

In this appendix we derive the equations in  \eqref{eq:ER_Psi_equation} during Einstein--Rosen bridge collapse 
used in
Sec.~\ref{subsec:general_ER}.  
The derivation applies to the stationary metric introduced in
Eq.~\eqref{eq:general_stationary_metric} for a large class of 
Einstein--matter systems under the assumptions \eqref{eq:app_ER_regular_sources} and \eqref{eq:app_ER_remainders}. 

The Einstein equations in the ansatz of
Sec.~\ref{sec:general_mechanism} can be written in the form
\begin{align}
\chi'
&=
\frac{zZ}{d}\phi'^2
+
{\cal R}_\chi,
~~~~~~~
f'
=
\frac{d+1}{z}f
+
\frac12 f\chi'
+
{\cal S}
+
{\cal R}_f .
\label{eq:app_ER_radial}
\end{align}
Here ${\cal S}$ denotes the regular matter contribution that remains
finite at the reference Cauchy horizon, while
${\cal R}_\chi$ and ${\cal R}_f$ collect the remaining matter, or geometric terms, which vanish in the rotating and charged 
examples considered in Sec. \ref{sec:rotating_BTZ} and \ref{sec:charged_RN}.

We further require
\begin{equation}
Z_->0,
\qquad
{\cal S}_->0
\label{eq:app_ER_regular_sources}
\end{equation}
at the reference Cauchy horizon with the subscript indicating their value at $z\to z_-$.  The positivity of
${\cal S}_-$ is an additional dynamical condition for the ER completion; it is verified explicitly for the charged 
and rotating examples.

We define
\begin{equation}
{\cal Y}\equiv e^{-\chi/2}
.
\label{eq:app_ER_y}
\end{equation}
Using Eq.~\eqref{eq:general_ER_geometry}, the longitudinal metric
coefficient is
\begin{equation}
{\cal G}
=
-\frac{f\,{\cal Y}^2}{z^2}.
\label{eq:app_ER_G}
\end{equation}
In this section we mainly use $\mathcal{Y}$ and $\mathcal{G}$ in place of $f$ and $\chi$,  
while the dynamics of scalar field is described by  $\Phi_\phi$.
The canonical momentum defined in
Eq.~\eqref{eq:general_radial_momentum} can then be written as
\begin{equation}
\Pi_\phi
=
\frac{Zf{\cal Y}}{z^d}\phi'
=
-Zz^{2-d}\frac{{\cal G}}{{\cal Y}}\phi'.
\label{eq:app_ER_flux}
\end{equation}

It is useful to rewrite Eq.~\eqref{eq:app_ER_radial} directly in
terms of ${\cal Y}$, ${\cal G}$, and $\Pi_\phi$. 
Eqs.~\eqref{eq:app_ER_y}, \eqref{eq:app_ER_G},
\eqref{eq:app_ER_flux}, and \eqref{eq:app_ER_radial} give 
\begin{align}
\begin{split}
{\cal Y}'
&=
-\frac{z^{2d-3}}{2d\, Z}\,\Pi_\phi^2\,
\frac{{\cal Y}^3}{{\cal G}^2}
-\frac12\,{\cal Y}\,{\cal R}_\chi,
\\
{\cal G}'
&=
\frac{d-1}{z}{\cal G}
-\frac{{\cal Y}^2}{z^2}{\cal S}
-\frac{z^{2d-3}}{2d\, Z}\,\Pi_\phi^2\,
\frac{{\cal Y}^2}{\cal G}
-\frac12\,{\cal G}\,{\cal R}_\chi
-\frac{{\cal Y}^2}{z^2}{\cal R}_f \,.
\label{eq:app_ER_exact}
\end{split}
\end{align}

To simplify the above equations in the nonlinear region, we now estimate the scaling of the terms on the right-hand sides. As discussed in 
Sec.~\ref{subsec:erc}, the onset
of nonlinear backreaction occurs over $\Delta z
=
{\cal O}(|{\cal B}_-|^2)$. The outer matching gives  
\begin{equation}
\Pi_\star =
{\cal O}(|{\cal B}_-|),
\qquad
\phi =
{\cal O}({\cal B}_-\log {\cal B}_-)= o(1),
\label{eq:app_ER_known_scaling}
\end{equation}
On a fixed interval 
of the
stretched boundary-layer coordinate we further require 
\begin{align}
&Z=Z_-+o(1),
\qquad
{\cal S}={\cal S}_-+o(1),
\qquad
z=z_-+{\cal O}({\cal B}_-^2),  
\\
&|{\cal B}_-|^2{\cal R}_\chi=o(1),
\qquad
{\cal R}_f=o(1),
\qquad
\Pi_\phi'={\cal O}(1),
\label{eq:app_ER_remainders}
\end{align}
The last condition in Eq.~\eqref{eq:app_ER_remainders} implies
\begin{equation}
\Delta\Pi_\phi
=
\Pi_\phi' \De z
=
{\cal O}(|{\cal B}_-|^2)
=
o(|{\cal B}_-|).
\end{equation}
Thus the scalar momentum is conserved to leading order,
$
\Pi_\phi
=
\Pi_\star[1+o(1)].
$ 
Using Eq.~\eqref{eq:general_momentum_matching}, we have 
\begin{equation}
\Pi_\star
=
\frac{
Z_-{\cal Y}_-f'(z_-)
}{
z_-^d
}
{\cal B}_-
+
o(|{\cal B}_-|).
\label{eq:app_ER_matching}
\end{equation}

The scaling of $\mathcal{G}$ at the boundary layer, i.e., at the beginning of the nonlinear region, is also important for the discussion. On the inter-horizon side of the reference Cauchy horizon,
${\cal G}>0$ and
\begin{equation}
{\cal G}(z)
=
\frac{{\cal Y}_-^2f'(z_-)}{z_-^2}(z_--z)
+
{\cal O}\!\left((\De z)^2\right),
\label{eq:app_ER_outer_G}
\end{equation}
At the boundary layer, $z_--z=\De z=\mathcal{O}(\calB^{\,^2})$, so we have
\be 
\mathcal{G}=\mathcal{O}(\calB^{\,^2})
\label{eq:G-scaling}
\ee 
Note that this is not the scaling of $\mathcal{G}$ throughout the nonlinear region. Generally, $\mathcal{G}$ decreases monotonically towards 
the first Kasner epoch, so it would not be larger than $\mathcal{O}(\calB^{\,^2})$. 


We now apply the scaling~\eqref{eq:app_ER_known_scaling} and
\eqref{eq:G-scaling} and condition \eqref{eq:app_ER_remainders} to the equations 
Eq.~\eqref{eq:app_ER_exact}. In $\mathcal{Y}'$, the second term is $o(\calB^{\,-2})$, which is much smaller than the scalar-gradient term $\mathcal{O}(\calB^{\,-2})$. In $\mathcal{G}'$, the regular source term $\mathcal{Y}^2\mathcal{S}/z^2$ is $\mathcal{O}(1)$. The geometric term
$(d-1){\cal G}/z$ is ${\cal O}(|{\cal B}_-|^2)$, and $\mathcal{Y}^2 \mathcal{R}_f/z^2$ is $o(1)$. Since $\mathcal{G}$ is generally monotonically decreasing while keeps positive, the term $\mathcal{G}\,\mathcal{R}_\chi/2$ is smaller than $\mathcal{O}(1)$. The term containing scalar-gradient is non-negligible, since it is $\mathcal{O}(1)$ at the boundary layer, and will increase as $\mathcal{G}$ decrease within the nonlinear region.

The leading equations~\eqref{eq:app_ER_exact} therefore
simplifies to 
\begin{align}
{\cal Y}'
&\simeq
-\alpha_-\Pi_\star^2
\frac{{\cal Y}^3}{{\cal G}^2},
~~~~~~~~~
{\cal G}'
\simeq
-\alpha_-{\cal Y}^2
\left(
\frac{1}{\cal C}_-
+
\frac{\Pi_\star^2}{\cal G}
\right),
\label{eq:app_ER_reduced}
\end{align}
where
\begin{equation}
\alpha_-
=
\frac{z_-^{2d-3}}{2d\,Z_-},
\qquad
{\cal C}_-
\equiv
\frac{
\alpha_-z_-^2
}{{\cal S}_-}=
\frac{
z_-^{2d-1}
}{
2d Z_-{\cal S}_-
}
>0 .
\label{eq:app_ER_constants}
\end{equation}
The last inequality of \eqref{eq:app_ER_constants} is a natural assumption. It is satisfied 
for the examples considered in Sec. \ref{sec:rotating_BTZ} and \ref{sec:charged_RN} where  
${\cal S}_-= f'(z_-) >0$.


The two equations in
Eq.~\eqref{eq:app_ER_reduced} can be further simplified as 
\begin{equation}
\frac{d\log {\cal Y}}{d{\cal G}}
=
\frac{{\cal G}_c}{
{\cal G}({\cal G}+{\cal G}_c)
},
\qquad \text{where}~~~
{\cal G}_c
\equiv
{\cal C}_-\,\Pi_\star^2.
\label{eq:app_ER_y_equation}
\end{equation}
Matching to the reference geometry on the outer side,
${\cal Y}\rightarrow {\cal Y}_-$ for ${\cal G}\gg{\cal G}_c$, gives
\begin{equation}
{\cal Y}
\simeq
{\cal Y}_-
\frac{\cal G}{{\cal G}+{\cal G}_c}.
\label{eq:app_ER_y_solution}
\end{equation}

Substituting Eq.~\eqref{eq:app_ER_y_solution} back into
Eqs.~\eqref{eq:app_ER_flux} and
\eqref{eq:app_ER_reduced}, one obtains
\begin{align}
{\cal G}'
&\simeq
-\Lambda_-
\frac{\cal G}{{\cal G}+{\cal G}_c},
~~~~~~~~~~~~
\phi'
\simeq
-{\cal K}_-
\frac{\Pi_\star}{{\cal G}+{\cal G}_c},
\label{eq:app_ER_normal_form}
\end{align}
with
\begin{equation}
{\cal K}_-
=
\frac{z_-^{d-2}{\cal Y}_-}{Z_-},
\qquad
\Lambda_-
=
\frac{\alpha_-{\cal Y}_-^2}{{\cal C}_-}=
\frac{
{\cal Y}_-^2f'(z_-)
}{
z_-^2
}
=
-{\cal G}'(z_-) ,
\label{eq:app_ER_dictionary}
\end{equation}
where we have used \eqref{eq:app_ER_constants}. 
 Eq.~\eqref{eq:app_ER_normal_form} is precisely the equations \eqref{eq:ER_Psi_equation} used in Sec.~\ref{subsec:general_ER}.

Combining the coefficients 
\eqref{eq:app_ER_dictionary} with the relation
\eqref{eq:app_ER_matching} reproduces the collapse width, exponential
rate, and post-collapse scalar gradient quoted in
Sec.~\ref{subsec:general_ER}, i.e. 
\begin{align}
\Gamma_{\rm ER}
\equiv
\frac{\Lambda_-}{{\cal G}_c}
\simeq
\frac{2d}{Z_-z_-{\cal B}_-^2},
~~~~~~~~~
\phi'_{\rm aft}
\equiv
-\frac{{\cal K}_-\Pi_\star}{{\cal G}_c}
\simeq
-\frac{2d}{Z_-z_-{\cal B}_-}.
\label{eq:app_ER_final_amplitudes}
\end{align}
 In particular, the dependence on the
reference redshift ${\cal Y}_-$ and on the Cauchy-horizon slope $f'(z_-)$ cancels from the final leading amplitudes.

The rotating and charged models considered in Sec. \ref{sec:rotating_BTZ} and \ref{sec:charged_RN} provide
explicit realizations of the above conditions.  
In both cases, we have 
\be
Z(z)=1, ~~~{\cal R}_\chi={\cal R}_f=0. 
\label{eq:relation1}
\ee
Additionally, the regular sources are
\begin{align}
\begin{split}
{\cal S}_{\rm BTZ}
&=
\frac{V(\phi)}{z}
+
\frac{J^2z^3}{2},
\qquad
d=1,
\label{eq:app_ER_explicit_sources}
\\
{\cal S}_{\rm RN}
&=
\frac{z^3E^2}{4G(\phi)}
+
\frac{V(\phi)}{2z},
\qquad
d=2,
\end{split}
\end{align}
where $J$ and $E$ are the conserved rotational and electric fluxes,
respectively, with $N'=-Jz{\cal Y}$ in the rotating case.  For finite fluxes
and regular small-field couplings,
${\cal S}={\cal S}_-+o(1)$ and $\Pi_\phi'=O(1)$ across the boundary
layer.  Moreover, at the reference Cauchy horizon,
$
{\cal S}_-
=
f'(z_-)
>
0 .
$ 
Thus both examples satisfy the conditions used above and realize the same leading evolutions.

\section{Holographic renormalization and thermodynamics}
In this appendix we collect the holographic renormalization and thermodynamics for the models we studied in the main text. 

\subsection{Rotating black holes}
\label{sec:thermo-rotating}
We regulate the geometry at $z=\varepsilon$, denote the induced metric by $h_{ab}$, and
take the outward normal $n^a$ to point toward the conformal boundary.  The renormalized action is
\begin{equation}
 \begin{split}
 S_{\rm ren}=\lim_{\varepsilon\to0}\Bigg\{S_{\rm bulk}
 &-\int_{z=\varepsilon}\dd^2x\sqrt{-h}
 \left(2-2K-\phi n^a\partial_a\phi-\frac38\phi^2-\frac{\kappa}{4}\,\sqrt\varepsilon\phi^2\right)\,.
 \end{split}
 \label{eq:completeBoundaryAction}
\end{equation}
The Gibbons--Hawking term is $2K$ with $K=\nabla_{\mu}n^{\mu}$. The remaining terms are gravitational and scalar counterterms required for renormalization, with the boundary conditions chosen to define the grand canonical ensemble at fixed double-trace coupling $\kappa$.

The renormalized on-shell action gives the energy and angular momentum densities per unit length along the spatial $x$ direction, with $x\sim x+2 \pi$,
\begin{equation}
 \mathcal E
 =-f_{(2)}+\frac54\alpha\eta-\frac14\kappa\alpha^2\,,\qquad
 \mathcal J=J\,,
\end{equation}
where the variables $f_{(2)}, \al,\eta,J$ are defined in the UV expansion \eqref{eq:UV-expansion}.
The on-shell action also gives the free energy density with the statistical relation at fixed $\kappa$,
\begin{equation}
 \mathcal F
 =\mathcal E-Ts-\Omega \mathcal J\,,
 \qquad\dd\mathcal F=-s\,\dd T-\mathcal J\,\dd \Omega \,,
 \label{eq:rota-free-energy}
\end{equation}
where the temperature $T$, entropy density $s$, and angular velocity $\Omega$ are given in \eqref{eq:rota-horizon-quantities}.
The preferred saddle at  $(T,\Omega)$ thus minimizes $\mathcal F$.

For a hairy black hole saddle, combining \eqref{eq:rota-free-energy} with \eqref{eq:rota-horizon-quantities} gives
\begin{equation}
 \mathcal F
 =f_{(2)}-\frac58\alpha\eta-\frac14\kappa\alpha^2
 +\Omega J\,.
 \label{eq:F-BH}
\end{equation}
For the rotating BTZ black hole \eqref{eq:BTZ-metric-func},
\begin{align}
 \mathcal E_{\rm BTZ}
 =\frac{1+\Omega^2}{z_+^2},~~~
 \mathcal J_{\rm BTZ}=\frac{2\Omega}{z_+^2},~~~
 \mathcal F_{\rm BTZ}=
 -\frac{4\pi^2T^2}{1-\Omega^2}\,.
 \label{eq:BTZ-thermodynamics}
\end{align}
For the thermal AdS \eqref{eq:ads-metric} with temperature $T$, although its local shift is $N=0$, $\Omega$ enters through the twisted Euclidean boundary-torus identification. We have
\begin{equation}
 \mathcal E_{\rm AdS}=-1,\qquad
 \mathcal J_{\rm AdS}=0,\qquad s_{\rm AdS}=0,\qquad
 \mathcal F_{\rm AdS}=-1\,.
 \label{eq:AdS-thermodynamics}
\end{equation}
Comparison with Eq.~\eqref{eq:BTZ-thermodynamics} gives the Hawking-Page transition temperature,
\begin{equation}
 T_{\rm HP}(\Omega)=\frac{\sqrt{1-\Omega^2}}{2\pi}.
\end{equation}
A regular boson star with $s=\mathcal J=0$, similar to the thermal $\AdS$, can also be considered in grand canonical ensemble with $\mathcal F_{\rm BS}=\mathcal E_{\rm BS}$, which is obtained numerically in this paper.

\subsection{Charged black holes}
\label{app:holo-ren}

We collect here the thermodynamic relations needed for the charged
black holes of Sec.~\ref{sec:charged_RN}. For a flat boundary geometry
and $m^2=-2$, the renormalized action in standard quantization is
\begin{align}
S_{\rm ren}
=
\lim_{\varepsilon\to0}
\left[
S_{\rm bulk}
-
\int_{z=\varepsilon} d^3x\,\sqrt{-h}
\left(
2K+4+\frac12\phi^2
\right)
\right],
\label{eq:app_ren_action}
\end{align}
where $h_{ab}$ is the induced metric at $z=\varepsilon$ and $K$ is
the trace of the extrinsic curvature. 
Using the asymptotic expansion
\eqref{eq:UV_expansion}, with the boundary time normalized by
$\chi_0=0$, the finite Dirichlet action $S_{\rm ren}$ yields the free energy density,
\begin{equation}
\mathcal{F}=\frac{T S_{\rm ren}}{V_2}
=
f_3-\phi_1\phi_2 .
\label{eq:app_ID}
\end{equation}
where $V_2\equiv\int \dd x \dd y$ is the spatial volume.

In standard quantization the source and expectation value are
$J=\phi_1$ and
$O=\phi_2$, respectively.  In alternative
quantization they are instead
$J=\phi_2$ and
$O=\phi_1$, after the corresponding Legendre
transformation.  For the source-free boundary conditions used in the
main text, both quantizations give 
\begin{equation}
\mathcal F=f_3 .
\label{eq:app_sourcefree_omega}
\end{equation}

The radially conserved charge introduced in
Sec.~\ref{sec:charged_RN} gives the identity
\begin{equation}
3f_3
-
2\phi_1\phi_2
+
\mu\rho
+
Ts
=
0 .
\label{eq:app_radial_identity}
\end{equation}
Together with the renormalized boundary stress tensor, this yields
the quantum statistical relation
\begin{equation}
\mathcal F
=
\mathcal E
-
Ts
-
\mu\rho .
\label{eq:app_QSR}
\end{equation}

Allowing the scalar source to vary, the grand potential satisfies
\begin{equation}
d\mathcal F
=
-s\,dT
-\rho\,d\mu
-
O\,dJ .
\label{eq:app_grand_first_law}
\end{equation}
Thus, at fixed scalar source,
\begin{equation}
d\epsilon
=
T\,ds
+
\mu\,d\rho,
\qquad
d\mathcal F
=
-s\,dT
-\rho\,d\mu .
\label{eq:app_first_law}
\end{equation}
These relations are the thermodynamic identities used in
Secs.~\ref{subsec:num_exterior} and
\ref{subsec:lowT_finite_scalar}.

\section{Relation between the horizon and Kasner shifts}
\label{app:shift_relation}

For the rotating solutions in Einstein--scalar theory of Sec.~\ref{sec:rotating_BTZ},
the radial equations admit the conserved charge
\begin{align}
Q
=
\frac{e^{-\chi/2}}{z}
\left(f'-f\chi'\right)
-
\frac{2e^{\chi/2}}{z}NN',
\qquad
Q'=0 .
\label{app:eq:rotating_radial_charge}
\end{align}
The equation for the shift function also implies that
$e^{\chi/2}N'/z$ is radially conserved.  Equation
\eqref{app:eq:rotating_radial_charge} can therefore be written as
\begin{align}
Q
=
\frac{e^{\chi/2}}{z}
\left(fe^{-\chi}-N^2\right)' ,
\end{align}
showing that
$d(fe^{-\chi}-N^2)/dN$ is constant along the radial flow.

Using the asymptotic normalization
$f(0)=1$, $\chi(0)=N(0)=0$, together with
$f(z_+)=0$ and $N(z_+)=N_h$ at the event horizon, one obtains the
exact relation
\begin{align}
fe^{-\chi}
=
(N-N_h)
\left(
N-\frac{1}{N_h}
\right).
\label{app:eq:shift_identity}
\end{align}

In the Kasner regime described in Sec.~\ref{sec:rotating_BTZ},
$fe^{-\chi}\rightarrow0$ and $N\rightarrow N_K$.  Equation
\eqref{app:eq:shift_identity} then gives
\begin{align}
N_K=N_h
\qquad\text{or}\qquad
N_K=\frac{1}{N_h}.
\end{align}
For a rotating solution, the conserved rotational flux implies that
$N(z)$ is monotonic, excluding the first possibility.  Hence
\begin{align}
N_K=\frac{1}{N_h}.
\label{app:eq:NK_Nh}
\end{align}
With the convention $N_h=-\Omega$ used in the main text, this becomes
\begin{align}
N_K=-\frac{1}{\Omega}.
\end{align}

\section{Large-field control of the deep-interior dynamics}
\label{subsec:Kasner_dynamics}
In charged black holes, the matching discussed in Sec. \ref{sec:inte} fixes the first Kasner epoch but not, in general,
the later evolution.  As emphasized in
Sec.~\ref{subsec:general_scope}, once $|\phi|$ is no longer small the
radial dynamics can probe the full nonlinear form of the matter 
couplings.  In the present charged model this dependence can be made
explicit through the large-field behavior of $G(\phi)$ \cite{Cai:2023igv}.

Introduce 
\begin{equation}
\rho=\log z,
\qquad
v=\frac{d\phi}{d\rho}=z\phi',
\qquad
H=-\frac{e^{-\chi/2}f}{z^3},
\qquad
{\cal R}=-\frac{\dot H}{H},
\label{eq:electric_transition_variables}
\end{equation}
where a dot denotes $d/d\rho$.  A Kasner epoch corresponds to
$v\simeq{\rm const}$ and ${\cal R}\simeq0$.  

For the polynomial potentials considered here, 
$V(\phi)$ grows at most polynomially in $\phi$ and 
we therefore neglect $V(\phi)$ in the following leading-order 
analysis. Einstein  equations
give
\begin{equation}
{\cal R}
=
\frac{
zE^2e^{-\chi/2}
}{
4G(\phi)H
}.
\label{eq:electric_R}
\end{equation}
Defining 
\begin{equation}
\xi(\phi)
\equiv
\frac{\partial_\phi G}{G},
\label{eq:lambda_eff_def}
\end{equation}
the leading flow reduces to
\begin{align}
\dot v
&=
{\cal R}\left(v+2\xi\right),
~~~~~~~~
\dot {\cal R}
=
{\cal R}
\left[
1-\frac{v^2}{4}
-\xi v
+{\cal R}
\right].
\label{eq:electric_flow_full}
\end{align}
Thus the fate of a later Kasner epoch is controlled by the
large-field limit of $\xi$.

\vspace{-0.3cm}
\paragraph{Polynomial growth.}

The two charged models studied in the main text belong to this class, from \eqref{eq:model1},  $G(\phi)\sim g_2\phi^2$ at large field.  More
generally, suppose at large $|\phi|$, 
\begin{equation}
G(\phi)
\simeq
c|\phi|^q,
\qquad
q>1,
\label{eq:Kasner_G_polynomial}
\end{equation}
one has $\xi\rightarrow0$.  Along a Kasner plateau,
$
{\cal R}
\simeq
z^{\,1-v_K^2/4}
/
(\log z)^q
.
$
${\cal R}$ is therefore irrelevant for $|v_K|\geq2$ and grows
for $|v_K|<2$.  In the latter case
Eq.~\eqref{eq:electric_flow_full} becomes
\begin{align}
\dot v
&=
{\cal R}v,
~~~~~~~~
\dot {\cal R}
=
{\cal R}
\left(
1-\frac{v^2}{4}+{\cal R}
\right),
\end{align}
with first integral
\begin{equation}
{\cal R} 
=
C_{\rm tr}v-1-\frac{v^2}{4}.
\label{eq:electric_first_integral}
\end{equation}
where $C_{\rm tr}$ is an integration constant along the transition. 
If ${\cal R}$ region connects two Kasner plateaus with
velocities $v_1$ and $v_2$, then ${\cal R}\rightarrow0$ on both sides and
we have 
\begin{equation}
v_1v_2=4.
\label{eq:electric_Kasner_inversion}
\end{equation}
An unstable epoch with $|v_1|<2$ is therefore mapped to one with
$|v_2|>2$.  This is the 
Kasner inversion.

Near the critical point,
Eq.~\eqref{eq:RN_ER_temperature_scaling} gives
$|v_1|\rightarrow\infty$.  Hence there is always a neighborhood of
$T_c$ in which the first post-collapse Kasner epoch is always stable. 
Numerically we have found that for $g_2=g_4=1$, there exists Kasner inversion in hairy black hole phase.

\vspace{-0.3cm}
\paragraph{Asymptotically exponential growth.}

A different continuation with the same quadratic coupling around the
RN solution is
\begin{equation}
G(\phi)
=
\cosh(c\phi),
\qquad
c^2=2g_2 .
\label{eq:G_cosh}
\end{equation}
When $\phi\to 0$, 
$G(\phi)=1+g_2\phi^2+\mathcal O(\phi^4)$, so this theory has the same
linear reference-background zero-mode problem and therefore the same leading
Cauchy-horizon matching.

At large field, 
$
\xi(\phi)
\rightarrow
\sigma c,
~
\sigma=\operatorname{sgn}(\phi).
$
${\cal R}$ growth rate on a Kasner plateau is then
\begin{equation}
\frac{\dot {\cal R}}{{\cal R}}
\simeq
1-\frac{v_K^2}{4}
-\sigma c v_K .
\label{eq:cosh_wall_exponent}
\end{equation}
When $\dot {\cal R}\simeq 0$, we have the marginal velocities
\begin{equation}
v_{\sigma,\pm}
=
-2\sigma c
\pm
2\sqrt{1+ c^2}. 
\label{eq:cosh_marginal_velocities}
\end{equation}
Thus ${\cal R}$ grows for
$v_{\sigma,-}<v_K<v_{\sigma,+}$. For $v$ outside this region, the system stays in the same Kasner epoch. 

For constant asymptotic $\xi=\sigma c$, introducing
$w=v+2\sigma c$ reduces
Eq.~\eqref{eq:electric_flow_full} to the polynomial form with
$1$ replaced by $1+ c^2$.  The two Kasner plateaus connected by
the transition therefore satisfy
\begin{equation}
\left(
v_1+2\sigma c
\right)
\left(
v_2+2\sigma c
\right)
=
4\left(1+ c^2\right).
\label{eq:cosh_Kasner_transition}
\end{equation}
Unlike the polynomial inversion, this transformation need not change
the sign of $p_t$; it is therefore more naturally described as a
Kasner transition.

Equations~\eqref{eq:electric_Kasner_inversion} and
\eqref{eq:cosh_Kasner_transition} illustrate the separation between
the two scales emphasized in Sec.~\ref{subsec:general_scope}.
Gauge couplings with the same quadratic expansion around $\phi=0$
can produce the same critical zero mode, the same ER collapse, and
the same first-Kasner matching, while their later deep-interior
evolution is different.  The former is controlled by small-field
critical data; the latter probes the large-field completion of
$G(\phi)$.

For gauge couplings in  \eqref{eq:Kasner_G_polynomial} and \eqref{eq:G_cosh}, the scalar velocity $v$ becomes sufficiently large both near the phase transition and in the low-temperature limit, so that no further Kasner transition occurs in these regimes. Away from the transition, however, $v$ may enter the range in which a Kasner inversion is triggered. The model studied in the main text belongs to the class described by Eq.~\eqref{eq:Kasner_G_polynomial}. By contrast, if the scalar potential grows superexponentially in $\phi$, additional Kasner transitions may occur even arbitrarily close to the phase transition~\cite{Gao:2026tck}.  

\section{Zero-temperature infrared geometries}
\label{app:zero_temperature}

In this appendix we derive the two distinct zero-temperature IR geometries relevant for the hairy black-brane solutions in Sec. \ref{sec:low_temperature_scaling}.

\subsection{Extremal IR geometry with AdS$_2$  $\times$ $\mathbb{R}^2$}
\label{app:finite_scalar_endpoint}

We first consider the zero-temperature solution in Model ${\cal M}_{\rm I}$ that approaches a
regular extremal horizon at a finite radial position $z=z_0$, with
$\phi(z_0)=\phi_*$ finite.  The leading near horizon behavior is
\begin{equation}
f
=
f_2 \, (z_0-z)^2+\cdots,
\qquad
\phi
=
\phi_*+\cdots,
\qquad
\chi
=
\chi_*+\cdots,
\qquad
\psi
=
\psi_1 (z_0-z)+\cdots ,
\label{app:eq:extremal_expansion}
\end{equation}
where $f_2, \phi_*,\chi_*, \psi_1$ are constants. 

Substituting this expansion into
Eqs.~\eqref{eq:f_eom} and \eqref{eq:phi_eom} gives, at leading order,
\begin{equation}
\frac{z_0^4E^2}{G_*}
+
2V_*
=
0,
\qquad
\partial_\phi V|_*
-
\frac{z_0^4E^2}{2}
\frac{\partial_\phi G|_*}{G_*^2}
=
0 ,
\label{app:eq:extremal_balance}
\end{equation}
where a star denotes evaluation at $\phi=\phi_*$.  Eliminating the
electric flux between these two equations yields
\begin{equation}
\left.
\frac{d}{d\phi}
\left[
V(\phi)G(\phi)
\right]
\right|_{\phi=\phi_*}
=
0 ,
\label{app:eq:attractor_condition}
\end{equation}
which is the attractor condition. 
A regular charged throat
requires $G_*>0$ and $V_*<0$.

At the next order, the Einstein equation gives
$ 
f_2
=
-\frac{V_*}{z_0^2}>0 .
$ 
The resulting near-horizon geometry is
AdS$_2\times\mathbb{R}^2$, with
$
L_2^2
=
-\frac{1}{V_*}.
$

Now we consider scalar perturbations around this throat.  The
homogeneous scalar fluctuation is governed by the effective IR mass
\begin{equation}
m_{\rm IR}^2
=
\frac{
\partial_\phi^2(VG)|_*
}{
G_*
}.
\label{app:eq:IR_mass}
\end{equation}
For a mode $\delta\phi\propto (z_0-z)^\delta$, the leading scalar equation
then gives
$
\delta(\delta+1)
=
m_{\rm IR}^2L_2^2 .
$ 
The positive root is the irrelevant exponent
$\delta_{\rm IR}$ quoted in
Sec.~\ref{subsec:lowT_finite_scalar}.

The full regular expansion also contains an analytic deformation
associated with the coupled variation of the throat geometry.
Consequently, close to the extremal horizon at zero temperature, the scalar takes the form
\begin{equation}
\phi
=
\phi_*
+
c_{\rm a}(z_0-z)
+
A_{\rm IR}(z_0-z)^{\delta_{\rm IR}}
+\cdots .
\label{app:eq:finite_scalar_deformation}
\end{equation}
Expanding the coupled radial equations to first order in $z_0-z$ fixes
the coefficient 
\begin{equation}
c_{\rm a}
=
\frac{
4V_{,\phi,*}
}{
z_0
\left(
-2V_*-m_{\rm IR}^2
\right)
}.
\label{app:eq:analytic_mode}
\end{equation}
By contrast, the amplitude $A_{\rm IR}$ in \eqref{app:eq:finite_scalar_deformation} is not fixed by the local
near-horizon analysis.  It is determined by integrating the extremal
solution to the asymptotic AdS$_4$ region and imposing the UV 
sourceless boundary condition for the scalar field.

\subsection{Running-scalar IR geometry}
\label{app:running_scalar_endpoint}

We next consider a zero-temperature IR endpoint for which the scalar does
not approach a finite value, but instead runs to large field as $z\rightarrow\infty$.  We introduce
\begin{equation}
H
\equiv
\frac{f e^{-\chi/2}}{z^3},
\qquad
P
\equiv
\frac{f e^{-\chi/2}\phi'}{z^2}.
\label{app:eq:HP_definition}
\end{equation}
Einstein equations become
\begin{align}
H'
&=
e^{-\chi/2}
\left[
\frac{E^2}{4G(\phi)}
+
\frac{V(\phi)}{2z^4}
\right],
~~~~~~~~~
P'
=
e^{-\chi/2}
\left[
\frac{\partial_\phi V}{z^4}
-
\frac{E^2}{2}
\frac{\partial_\phi G}{G(\phi)^2}
\right],
\label{app:eq:HP_equations}
\end{align}
together with the exact relation
\begin{equation}
\frac{P}{H}
=
z\phi'
\label{app:eq:PH_identity}
\end{equation}
whenever $H\neq0$.

If the source terms in Eq.~\eqref{app:eq:HP_equations} are integrable
in IR, $H$ and $P$ approach constants.  When
$H\rightarrow H_0\neq0$, Eq.~\eqref{app:eq:PH_identity} implies
$z\phi'\rightarrow a$.  The zero-temperature conserved-charge
condition then gives $a^2=12$, leading to the local logarithmic branch
$\phi\simeq\pm2\sqrt{3}\log z$.  This is a consistent IR 
solution, but it is not the branch selected by the source-free
zero-temperature interpolation of Model ${\cal M}_{\rm II}$.

The physical branch instead approaches the degenerate limit
$H\rightarrow0$.  For
$G(\phi)=1+g_2\phi^2$ and $V(\phi)=-6-\phi^2$, assuming when $z\to\infty$, 
\begin{equation}
\phi
=
Cz+\frac{b}{z}+\cdots,
\qquad
f
=
\frac{F}{z^2}+\cdots ,
\label{app:eq:running_sep_ansatz}
\end{equation}
with $C\neq0$.  The equation
$\chi'=z\phi'^2/2$ then gives
\begin{equation}
\chi
=
\frac{C^2z^2}{4}
-
Cb\log z
+
\chi_\infty
+\mathcal O(z^{-2}).
\label{app:eq:running_sep_chi}
\end{equation}
This geometry reminds us the IR geometry for insulating phase discussed in \cite{Liu:2018spp}. 
Substituting these expansions into \eqref{app:eq:HP_equations}, the leading
order-$z$ terms cancel only if
$
C^4
=
E^2/(2g_2).
$ 
The leading scalar equation then fixes
$
F
=
16/C^2,
$
while the next order of \eqref{app:eq:HP_equations} gives
$
b
=
(
1-\frac{1}{2g_2}
)/(2C).
$

For Model ${\cal M}_{\rm II}$, $g_2=0.6$, then 
$b=1/(12C)$.  The physical running-scalar endpoint therefore has the
asymptotic form in  $z\to\infty$, 
\begin{align}
\begin{split}
\phi
&=
Cz
+
\frac{1}{12Cz}
+
\mathcal O(z^{-3}),
~~~~~~~~~
f
=
\frac{16}{C^2z^2}
\left[
1+\mathcal O(z^{-2})
\right].
\\
\chi
&=
\frac{C^2z^2}{4}
-
\frac{1}{12}\log z
+
\chi_\infty
+
\mathcal O(z^{-2}),
\label{app:eq:running_sep_solution}
\end{split}
\end{align}
Fixing the boundary condition $\chi_0=0$ and $E=-\rho_0$, we obtain
$
C^4
=
\frac{\rho_0^2}{2g_2}.
$

Eq.~\eqref{app:eq:running_sep_solution} indeed belongs to the
degenerate branch: $H$ is exponentially suppressed and tends to zero
as $z\rightarrow\infty$.  At the same time,
\begin{equation}
z\phi'
=
Cz
-
\frac{1}{12Cz}
+
\mathcal O(z^{-3})
~~~\rightarrow~~~
\pm\infty ,
\label{app:eq:running_sep_velocity}
\end{equation}
so the scalar velocity does not approach the finite value of the
logarithmic branch.  The endpoint is also curvature singular, as can be seen from the Ricci scalar 
\begin{equation}
R
=
-2C^2z^2
+
\mathcal O(1).
\label{app:eq:running_sep_curvature}
\end{equation}
The exponentially suppressed lapse $e^{-\chi}$ and the divergent scalar velocity 
are used 
for 
the low-temperature
continuation constructed in Sec.~\ref{subsec:lowT_running}.

\end{document}